\documentclass[12pt]{iopart}

\usepackage{iopams}
\usepackage{setstack}   
\usepackage[latin9]{inputenc}
\usepackage{nicefrac}
\usepackage{graphicx}
\usepackage{amsthm}   
\usepackage[all]{xy}  
\usepackage[margin=10pt,font=small,labelfont=bf]{caption}

  \theoremstyle{plain}
  \newtheorem{thm}{Theorem}[section]
  \theoremstyle{definition}
  \newtheorem{defn}[thm]{Definition}
  \theoremstyle{plain}
  \newtheorem{cor}[thm]{Corollary}
  \theoremstyle{remark}
  \newtheorem*{rem*}{Remark}
  \theoremstyle{plain}
  \newtheorem{prop}[thm]{Proposition}
  \theoremstyle{plain}
  \newtheorem{lem}[thm]{Lemma}

\date{\today}

\def\at{\Big|}

\def\ind{{\rm Ind}}
\def\res{{\rm Res}}

\eqnobysec   

\begin{document}

\title{The Isospectral Fruits of Representation Theory: Quantum Graphs and
Drums}

\author{Ram Band$^{1}$, Ori Parzanchevski$^{2}$ and Gilad Ben-Shach$^{1,3}$}

\address{$^{1}$ Department of Physics of Complex Systems\\
 The Weizmann Institute of Science, Rehovot 76100, Israel}
\ead{rami.band@weizmann.ac.il}

\address{$^{2}$ Institute of mathematics\\
 Hebrew University, Jerusalem 91904, Israel}
\ead{parzan@math.huji.ac.il}

\address{$^{3}$ Department of Physics and Department of Mathematics
and Statistics\\
 McGill University, 3600 rue University, Montreal, QC, Canada, H3A
2T8} \ead{gilad@ben-shach.com}

\begin{abstract}
We present a method which enables one to construct isospectral
objects, such as quantum graphs and drums. One aspect of the method
is based on representation theory arguments which are shown and
proved. The complementary part concerns techniques of assembly which
are both stated generally and demonstrated. For that purpose,
quantum graphs are grist to the mill. We develop the intuition that
stands behind the construction as well as the practical skills of
producing isospectral objects. We discuss the theoretical
implications which include Sunada's theorem of isospectrality
\cite{Sunada} arising as a particular case of this method. A gallery
of new isospectral examples is presented and some known examples are
shown to result from our theory.
\end{abstract}

\submitto{\JPA}

\section{Introduction}

\label{sec:introduction}In 1966, Marc Kac asked his famous question,
{}``Can one hear the shape of a drum?''~\cite{Kac}. This question
can be rephrased as {}``does the Laplacian on every planar domain
with Dirichlet boundary conditions have a unique spectrum?''. Ever
since the time when Kac posed this fascinating question, physicists
and mathematicians alike have attacked the problem from various
angles. Attempts were made both to reconstruct the shape of an
object from its spectrum, and to find different objects that are
isospectral, i.e., have the same spectrum. The interested reader can
find an elaborate summary of these efforts in
\cite{Kac}-\cite{Carlson99}. In 1985, Sunada presented a theorem
that describes a method to construct isospectral Riemannian
manifolds \cite{Sunada}. Buser and later Berard expanded on this
theorem, and offered a proof based on the concept of
transplantation, as summarized by Brooks
\cite{Buser,Berard,Brooks-Sun}. Over the years, several pairs of
isospectral objects were found, but these were not planar domains,
and therefore did not serve as an exact answer to Kac's question. In
1992, by applying an extension of Sunada's theorem, Gordon, Webb and
Wolpert were able to finally answer Kac's question as it related to
drums, presenting the first pair of isospectral two-dimensional
planar domains \cite{Gordon1,Gordon2}. Buser et al. later obtained a
set of seventeen isospectral families of planar domains, both
Neumann and Dirichlet isospectral \cite{BuserConway}. Jakobson et
al. and Levitin et al. extended the choice of boundary conditions by
considering objects with alternating boundary conditions, and found
sets of four planar domains that are mutually isospectral
\cite{Jakobson,Levitin}. In the late 1990's, Gutkin and Smilansky
reposed and answered Kac's question as it applies to quantum graphs
\cite{gutkinus}. Recently, Band et al. presented a pair of
isospectral quantum graphs \cite{BSS06}, whose construction was
generalized to the method described in this paper.

We begin by reviewing the terminology and the relevant definitions
for quantum graphs. In section~\ref{sec:basic_example}, we rederive
the graphs constructed in~\cite{BSS06}, to help the reader gain an
intuitive understanding of the method. Once the reader is familiar
with the notions used, we formalize a theorem (section
\ref{sec:algebra}) along with a corollary, which together form the
crux of the construction method. With the theorem in hand, we return
to the basic example presented in section~\ref{sec:basic_example},
and show that the isospectral pair can be expanded indefinitely $-$
section \ref{sec:extending_example}. After describing the assembly
process rigorously in section \ref{sec:rigorous}, we devote section
\ref{sec:further_investigation} to further investigating the
theoretical implications of the theory. Finally, in sections
\ref{sec:gallery_of_graphs}, \ref{sec:drums} we demonstrate how to
apply the construction to other types of objects, and present a
variety of examples of graphs, drums, and manifolds.

\section{Quantum graphs}

\label{sec:graphs_intro} A \emph{graph} $\Gamma$ consists of a
finite set of vertices $V=\left\{ v_{i}\right\} $ and a finite set
$E=\left\{ e_{j}\right\} $ of edges connecting the vertices. Each
edge $e$ can be identified with a pair of vertices $\left\{
v_{i},v_{k}\right\} $. We denote by $E_{v}$ the set of all edges
incident to the vertex $v$. The degree (valency) of the vertex is
$d_{v}=\left|E_{v}\right|$. This becomes a \emph{metric graph} if
each edge is assigned a finite length $l_{e}>0$. It is then possible
to identify an edge $e$ with a finite segment $\left[0,l_{e}\right]$
of the real line having the natural coordinate $x_{e}$ along it. In
this context, a \emph{function on the graph} is a vector
$f=\left(f\big|_{e_{1}},\ldots,f\big|_{e_{|E|}}\right)$ of functions
$f\big|_{e_{j}}:\left[0,l_{e_{j}}\right]\rightarrow\mathbb{C}$ on
the edges. Notice that in general it is not required that for $v\in
V$ and $e,e'\in E_{v}$ the functions $f\big|_{e}$ and $f\big|_{e'}$
agree on $v$.

To obtain a \emph{quantum graph}, we consider the following Hilbert
space:
$L^{2}\left(\Gamma\right)=\bigoplus_{j=1}^{\left|E\right|}L^{2}\left(\left[0,l_{e_{j}}\right]\right)$
with the inner product: $\left\langle f,g\right\rangle
=\sum_{j=1}^{\left|E\right|}\int_{0}^{l_{e_{j}}}\overline{f\big|_{e_{j}}}\cdot
g\big|_{e_{j}}\, dx_{e_{j}}$. The operator which draws our interest
is the negative Laplacian: $-\bigtriangleup
f=\left(-f''\at_{e_{1}},\ldots,-f''\at_{e_{\left|E\right|}}\right)$.
The domain of definition for this operator is the Sobolev space,
$W^{2,2}\left(\Gamma\right)$, the space of all functions $f$ such
that $f\big|_{e_{j}}\in
W^{2,2}\left(\left[0,l_{e_{j}}\right]\right)$ for all $1\leq
j\leq\left|E\right|$. In addition we require the functions to obey
certain boundary conditions stated a priori: for a vertex $v\in V$,
we consider homogeneous boundary conditions which involve the
function's values and derivatives at the vertex, of the form
$A_{v}\cdot f\at_{v}+B_{v}\cdot f'\at_{v}=0$. Here $A_{v}$ and
$B_{v}$ are $d_{v}\times d_{v}$ complex matrices, $f\at_{v}$ is the
vector $\left(f\at_{e_{1}}(v),\ldots,f\at_{e_{d_{v}}}(v)\right)^{T}$
of the vertex values of the function along each edge incident to
$v$, and
$f'\at_{v}=\left(f'\at_{e_{1}}(v),\ldots,f'\at_{e_{d_{v}}}(v)\right)^{T}$
is the vector of outgoing derivatives of $f$ taken at the vertex.
Before stating the boundary conditions, the graph is merely a
collection of independent edges with functions defined separately on
each edge. The connectivity of the graph is manifested through the
boundary conditions, which are local in nature: we relate the values
of the function and its derivatives on each vertex, but no relation
is assumed between those values on different vertices, or along the
edges. In summary, a quantum graph is a metric graph equipped with a
differential operator and with homogeneous differential boundary
conditions at the vertices. One can generalize the metric Laplacian
by including a potential or a magnetic flux defined on the edges.
However, these generalizations will not be addressed here, and the
interested reader is referred to the reviews \cite{GS06,Kuch08}.

 A standard choice of boundary condition
which we adopt is the so
called \emph{Neumann} boundary condition%
\footnote{This boundary condition is sometimes referred to as
\emph{Kirchhoff}
condition in the literature.%
}:\\
 $\bullet\quad$ $f$ agrees on the vertices: $\forall v\in V\quad\forall e,e'\in E_{v}\,\,:\quad f\at_{e}\left(v\right)=f\at_{e'}\left(v\right)$.
\\
 $\bullet\quad$ The sum of outgoing derivatives at each vertex is
zero: $\forall v\in V$ :$\sum\limits _{e\in
E_{v}}f'\at_{e}\left(v\right)=0$.

It is worth noting that a Neumann vertex of valency two can be added
at (or removed from) any point along an edge without changing the
eigenspaces of the Laplacian, and thus, from a spectral point of
view, without really changing the graph. Thus, loops (edges
connecting a vertex to itself) and parallel edges (edges with the
same endpoints) can be eliminated by the introduction of such
{}``dummy'' vertices $-$ we shall occasionally exploit this to
simplify notation by assuming, without loss of generality, that we
are dealing with graphs with no loops or parallel edges. A possible
choice for matrices that correspond to the Neumann boundary
conditions is:\[ A_{v}=\left(\begin{array}{cccc}
1 & -1\\
 & \ddots & \ddots\\
 &  & 1 & -1\\
0 & \cdots & 0 & 0\end{array}\right)\qquad
B_{v}=\left(\begin{array}{cccc}
0 & 0 & \cdots & 0\\
\vdots & \vdots & \ddots & \vdots\\
0 & 0 & \cdots &  0\\
1 & 1 & \cdots &  1\end{array}\right)\,.\]
 For a vertex of degree one, the Neumann boundary condition is expressed
by the matrices $A_{v}=\left(0\right),\,\, B_{v}=\left(1\right)$ and
means that the derivative of the function equals zero at that
vertex. Another useful boundary condition for a degree one vertex is
the \emph{Dirichlet} boundary condition, which means that the
function vanishes at that vertex: $A_{v}=\left(1\right),\,\,
B_{v}=\left(0\right)$. It can be seen that the Neumann condition
renders the Laplacian self-adjoint, which guarantees that its
spectrum is real. In general, Kostrykin and Schrader provide
necessary and sufficient conditions that ensure the self-adjointness
of the Laplacian: for every $v\in V$ the $d_{v}\times2d_{v}$ matrix
$\left(A_{v}\, B_{v}\right)$ must be of maximal rank $d_{v}$, and
the matrix $A_{v}\cdot B_{v}^{\dagger}$ must be self-adjoint
\cite{KosSch99}. For a quantum graph $\Gamma$ we shall denote by
$\Phi_{\Gamma}\left(\lambda\right)$ the space of complex functions
on $\Gamma$ (i.e., satisfying the boundary conditions at the
vertices) which are eigenfunctions of $\Gamma$'s (negative)
Laplacian with eigenvalue $\lambda$:
\begin{equation}
\Phi_{\Gamma}\left(\lambda\right)=\left\{ f\in
W^{2,2}\left(\Gamma\right)\,|\,-\bigtriangleup f=\lambda f\right\}
\,. \end{equation}
 We define the \emph{spectrum} of $\Gamma$ to be the function
 \begin{equation}
\sigma_{\Gamma}:\lambda\mapsto\dim_{\mathbb{C}}\Phi_{\Gamma}\left(\lambda\right)\,,
\end{equation}
 which assigns to each eigenvalue its multiplicity. Two quantum graphs
$\Gamma$ and $\Gamma'$ are said to be \emph{isospectral} if their
spectra coincide, that is $\sigma_{\Gamma}\equiv\sigma_{\Gamma'}$.

\medskip{}

Quantum graphs play an important role in the study of quantum chaos.
This connection was first revealed by the work of Kottos and
Smilansky \cite{KS97,KS99}. They show that the spectral statistics
of quantum graphs follow the predictions of random-matrix theory
very closely. They propose a derivation of a trace formula for
quantum graphs and point out its similarity to the famous Gutzwiller
trace formula \cite{Gutz71,Gutz90} for chaotic Hamiltonian systems.
The trace formula for a quantum graph connects the spectrum of the
graph's Laplacian to the total length of the graph and the lengths
of its periodic orbits. The main result in the field of
isospectrality of quantum graphs is that of Gutkin and Smilansky,
\cite{gutkinus}, where they use the trace formula to show that under
certain conditions a quantum graph can be heard, meaning that it can
be recovered from the spectrum of its Laplacian. The necessary
conditions include the graph being simple and its edges having
rationally independent lengths. When these conditions are not
satisfied, isospectral quantum graphs indeed arise. An early example
appears in \cite{Roth}, in which Roth obtains isospectrality
exploiting a spectral trace formula. vonBelow \cite{vonBelow} uses
the connection between spectra of discrete graphs and spectra of
equilateral quantum graphs to turn isospectral discrete graphs into
isospectral quantum graphs. In \cite{Oren} isospectrality of
weighted discrete graphs provides isospectral quantum graphs whose
edges vary in length. A wealth of examples is constructed in
\cite{gutkinus,Tashkent}, using an analogy of the isospectral drums
obtained by Buser et al.\ \cite{BuserConway}. A recent example is
the pair of isospectral dihedral graphs presented in \cite{BSS06}.
Their construction was generalized to obtain the more complete
theory which is presented in this paper.

\section{A basic example}

\label{sec:basic_example}

\begin{figure}[!h]
\hfill{}
\begin{minipage}[b][1\totalheight]{0.4\columnwidth}
\begin{center}
\includegraphics[width=0.6\textwidth]{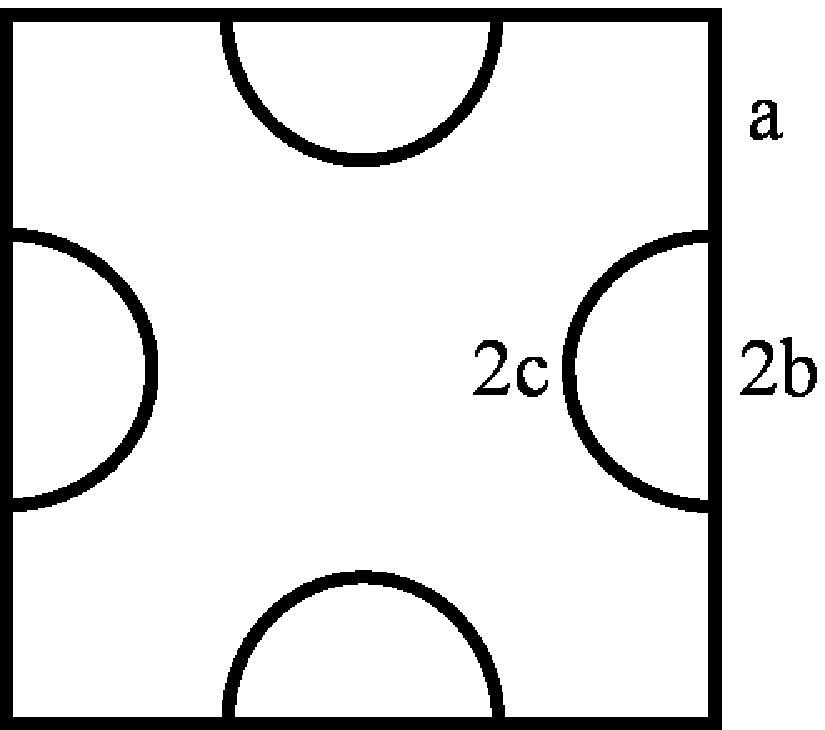}
\end{center}
\vspace{0pt}
\begin{center}
(a)  \;\;
\end{center}
\end{minipage} \hfill{}
\begin{minipage}[b][1\totalheight]{0.4\columnwidth}
\begin{center}
\includegraphics[width=0.72\textwidth]{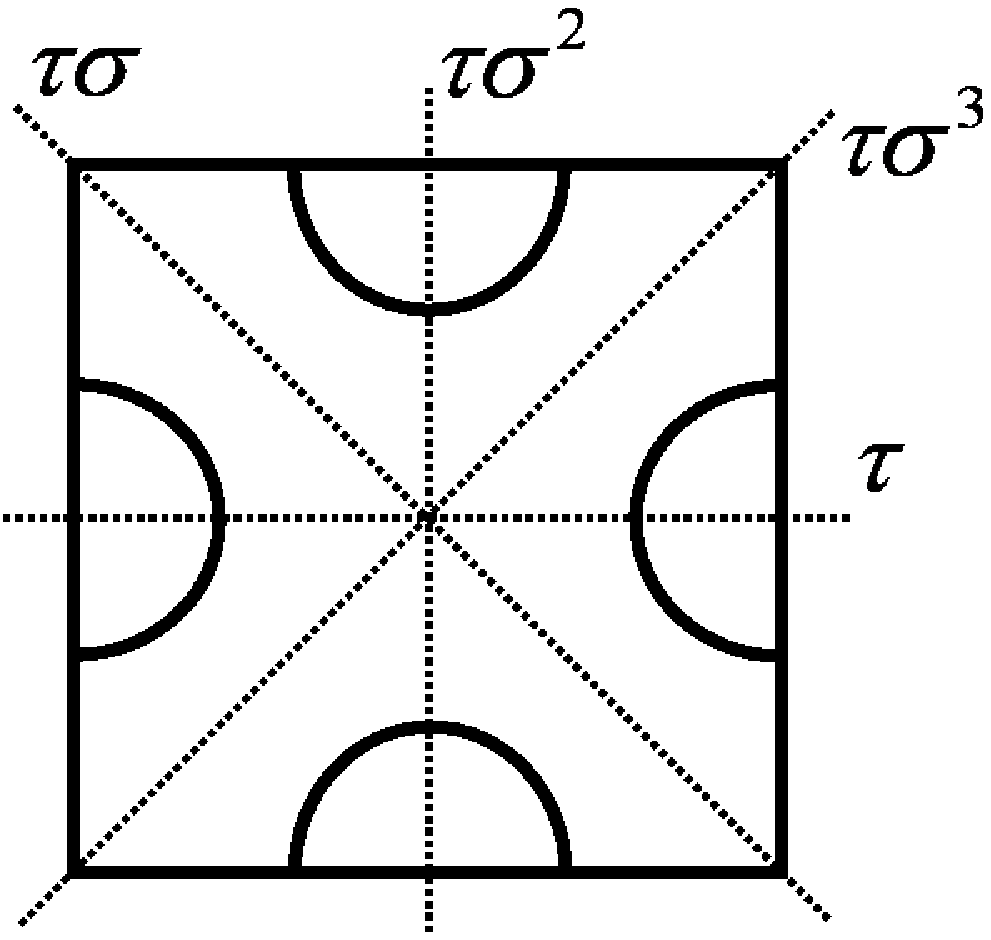}
\end{center}
\vspace{-8pt}
\begin{center}
(b)  \;\;
\end{center}
\end{minipage} \hfill{}

 \caption{(a) A graph that obeys the dihedral symmetry of the
square. The lengths of some edges are marked. (b) The same graph,
showing the axes of the reflection elements in $D_{4}$.}

\label{fig:full_square}
\end{figure}

Let $\Gamma$ be the graph given in figure \ref{fig:full_square}(a).
The lengths of the edges are determined by the parameters $a,b,c$
and we impose Neumann boundary conditions at all vertices.
$G=D_{4}$, the dihedral group of the square, is the symmetry group
of $\Gamma$. $G$ consists of the identity, three rotations and four
reflections. Let $\tau$ denote the reflection of $\Gamma$ along the
horizontal axis and $\sigma$ the rotation of $\Gamma$
counterclockwise by $\nicefrac{\pi}{2}$. The axes of the reflection
elements in $G$ are shown in figure \ref{fig:full_square}(b). We can
describe $G$ and two of its subgroups $H_{1},H_{2}\leq G$ by:
\begin{eqnarray*}
G & = & \left<\sigma, \tau\right> = \{e,\,\sigma,\,\sigma^{2},\,\sigma^{3},\,\tau,\,\tau\sigma,\,\tau\sigma^{2},\,\tau\sigma^{3}\}\\
H_{1} & = & \left<\tau, \tau\sigma^2\right> = \{e,\,\tau,\,\tau\sigma^{2},\,\sigma^{2}\}\\
H_{2} & = & \left<\tau\sigma, \tau\sigma^3\right> =
\{e,\,\tau\sigma,\,\tau\sigma^{3},\,\sigma^{2}\}\end{eqnarray*}
 Consider the following one dimensional representations of $H_{1},H_{2}$,
respectively: \begin{eqnarray} R_{1} & : & \left\{
\begin{array}{llll}
e\mapsto\left(1\right), & \tau\mapsto\left(-1\right), & \tau\sigma^{2}\mapsto\left(1\right), & \sigma^{2}\mapsto\left(-1\right)\end{array}\right\} \label{eq:r1_rep}\\
R_{2} & : & \left\{ \begin{array}{llll} e\mapsto\left(1\right), &
\tau\sigma\mapsto\left(1\right), &
\tau\sigma^{3}\mapsto\left(-1\right), &
\sigma^{2}\mapsto\left(-1\right)\end{array}\right\}
\label{eq:r2_rep}\end{eqnarray}

\begin{figure}[!h]

\begin{centering}
\hfill{}%
\begin{minipage}[c][1\totalheight]{0.3\columnwidth}%
\begin{center}
\includegraphics[scale=0.4]{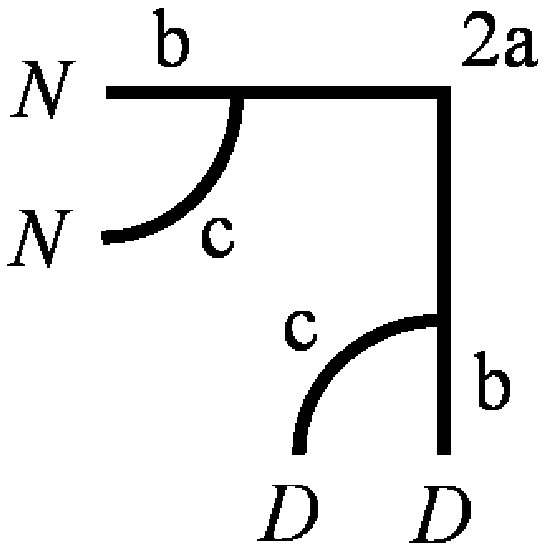}
\par\end{center}

\begin{center}
(a)
\par\end{center}%
\end{minipage}\hfill{}%
\begin{minipage}[c][1\totalheight]{0.3\columnwidth}%
\begin{center}
\includegraphics[scale=0.4]{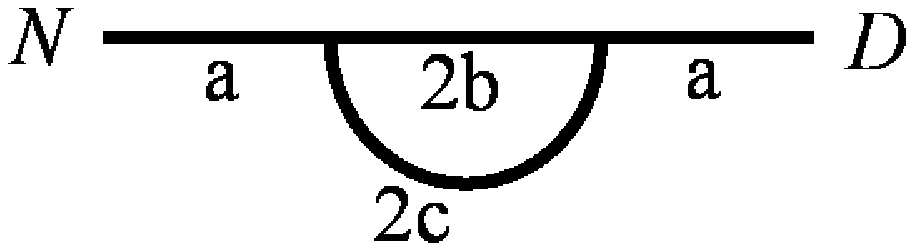}
\par\end{center}

\bigskip{}

\begin{center}
(b)
\par\end{center}%
\end{minipage}\hfill{}
\par\end{centering}


\caption{The isospectral graphs (a) $\nicefrac{\Gamma}{R_{1}},\,$
(b) $\nicefrac{\Gamma}{R_{2}}$. Neumann boundary conditions are
assumed if nothing else is specified. $D$ stands for Dirichlet
boundary conditions and $N$ for Neumann. }

\label{fig:dihedral_pair}
\end{figure}

We will use these representations to construct two graphs denoted by
$\nicefrac{\Gamma}{R_{1}},\,\nicefrac{\Gamma}{R_{2}}$ (figure
\ref{fig:dihedral_pair}) which will be found to be isospectral.

We now explain the process of building the quotient graph
$\nicefrac{\Gamma}{R_{1}}$. Let $\lambda\in\mathbb{C}$ and
$\tilde{f}\in\Phi_{\Gamma}(\lambda)$ be a function which transforms
according to the representation $R_{1}$, i.e.:
\begin{equation}
\forall g\in H_{1},\quad
g\tilde{f}=\rho_{R_{1}}(g)\tilde{f}.\end{equation}
 In the l.h.s., the action of the group $H_{1}$ on $\tilde{f}$ is by $\forall x\in\Gamma,\,\,(g\tilde{f})(x)=\tilde{f}(g^{-1}x)$.
We use the transformation law of $\tilde{f}$ in order to deduce its
properties: we know that $\tau\tilde{f}=-\tilde{f}$, which means
that $\tilde{f}$ is an anti-symmetric function with respect to the
horizontal reflection. We deduce that $\tilde{f}$ vanishes on the
fixed points of $\tau$ (marked with diamonds in figure
\ref{fig:intuitive_quotient1}(a)). In a similar manner, we see that
$\tilde{f}$ is symmetric with respect to the vertical reflection,
since $\tau\sigma^{2}\tilde{f}=\tilde{f}$, and therefore the
derivative of $\tilde{f}$ vanishes at the corresponding fixed points
(the squares in figure \ref{fig:intuitive_quotient1}(a)).
Furthermore, it is enough to know the restriction of $\tilde{f}$ to
the first quadrant (the bold subgraph in figure
\ref{fig:intuitive_quotient1}(a)) in order to deduce $\tilde{f}$ on
the whole graph, using the known action of the reflections:
\begin{equation}
\tau\tilde{f}=-\tilde{f}\,,\quad\tau\sigma^{2}\tilde{f}=\tilde{f}\,.\label{eq:R1_isotip_reflec}\end{equation}
 Our construction process is now complete. The quotient graph $\nicefrac{\Gamma}{R_{1}}$
is the subgraph which lies in the first quadrant, with the boundary
conditions of Dirichlet and Neumann in the appropriate locations, as
was found for $\tilde{f}$ (figure \ref{fig:intuitive_quotient1}(b)).

\begin{center}
\begin{figure}[!h]
\begin{centering}
\hfill{}%
\begin{minipage}[c][1\totalheight][t]{0.4\columnwidth}%
\begin{center}
\includegraphics[width=0.6\textwidth]{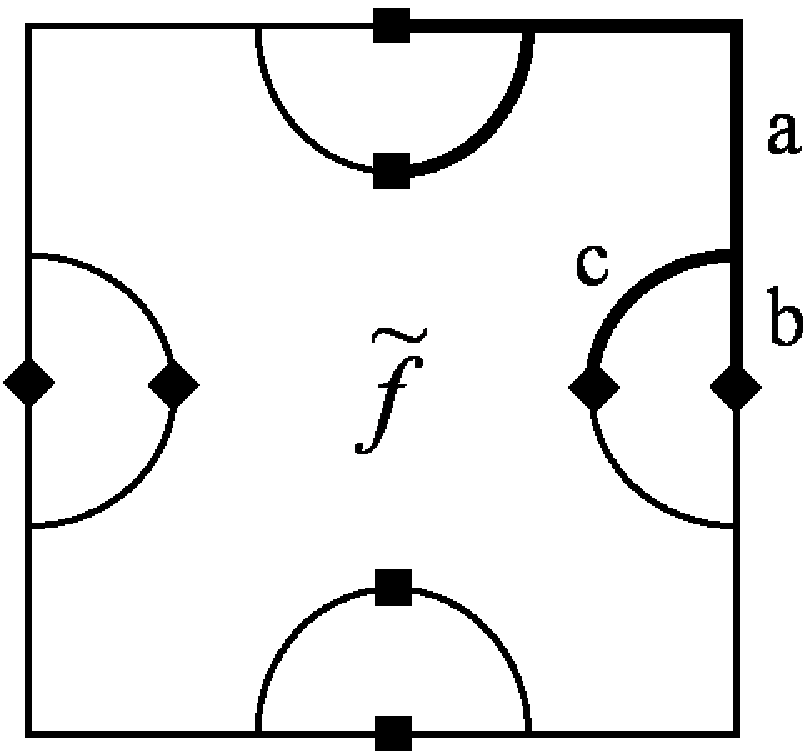}
\par\end{center}

\begin{center}
(a)
\par\end{center}%
\end{minipage}\hfill{}%
\begin{minipage}[c][1\totalheight][t]{0.35\columnwidth}%
\begin{center}
\includegraphics[width=0.6\textwidth]{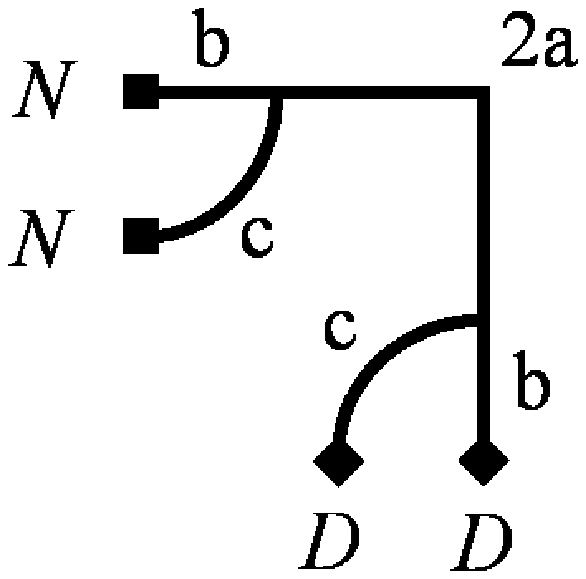}
\par\end{center}

\begin{center}
(b)
\par\end{center}%
\end{minipage}\hfill{}
\par\end{centering}

\caption{(a) The information we have on $\tilde{f}$ which transforms
according to $R_1$. Diamonds mark the vertices at which the function
vanishes and squares the vertices with zero derivative. (b) The
quotient graph $\nicefrac{\Gamma}{R_{1}}$ which encodes this
information. \textit{D} stands for Dirichlet boundary conditions and
\textit{N} for Neumann.}

\label{fig:intuitive_quotient1}
\end{figure}

\par\end{center}

From this example, we conclude that the construction of the quotient
graph is motivated by an encoding scheme. We choose a fundamental
domain for the action of $G$ on $\Gamma$, i.e., a minimal subgraph
from which the entire graph can be reached by the action of the
group. We take this domain to be the quotient graph
$\nicefrac{\Gamma}{R_{1}}$. We encode a function
$\tilde{f}\in\Phi_{\Gamma}(\lambda)$, which transforms according to
the representation $R_{1}$, by a function
$f\in\Phi_{\nicefrac{\Gamma}{R_{1}}}(\lambda)$. The encoding is
described by the map
$\Psi:\Phi_{\Gamma}^{R_{1}}(\lambda)\rightarrow\Phi_{\nicefrac{\Gamma}{R_{1}}}(\lambda)$,
where $\Phi_{\Gamma}^{R_{1}}(\lambda)$ is the space of all functions
$\tilde{f}\in\Phi_{\Gamma}(\lambda)$ that transform according to the
representation $R_{1}$, and $\Psi$ is just the restriction map to
the fundamental domain. An important observation is that given
$f\in\Phi_{\nicefrac{\Gamma}{R_{1}}}(\lambda)$, it is possible to
construct a unique function
$\tilde{f}\in\Phi_{\Gamma}^{R_{1}}(\lambda)$ (using
(\ref{eq:R1_isotip_reflec})), whose restriction to the fundamental
domain is $f$ (this is the decoding process). It follows that $\Psi$
is invertible and thus is an isomorphism:
\begin{equation}
\Phi_{\Gamma}^{R_{1}}(\lambda)\cong\Phi_{\nicefrac{\Gamma}{R_{1}}}(\lambda)
\end{equation}

The quotient $\nicefrac{\Gamma}{R_{2}}$ is constructed similarly.
For any $\lambda\in\mathbb{C}$, we consider
$\tilde{f}\in\Phi_{\Gamma}^{R_{2}}(\lambda)$, which means that
$\tilde{f}$ transforms in the following way: \begin{equation}
\tau\sigma\tilde{f}=\tilde{f}\,,\quad\tau\sigma^{3}\tilde{f}=-\tilde{f}\label{eq:R2_isotip_reflec}\end{equation}
 and carry on with the same arguments as above. We do not specify
the details of this process but rather summarize it in figure
\ref{fig:intuitive_quotient2}.

\begin{center}
\begin{figure}[!h]
 \hfill{}%
\begin{minipage}[c][1\totalheight][t]{0.4\columnwidth}%
\begin{center}
\includegraphics[width=0.6\textwidth]{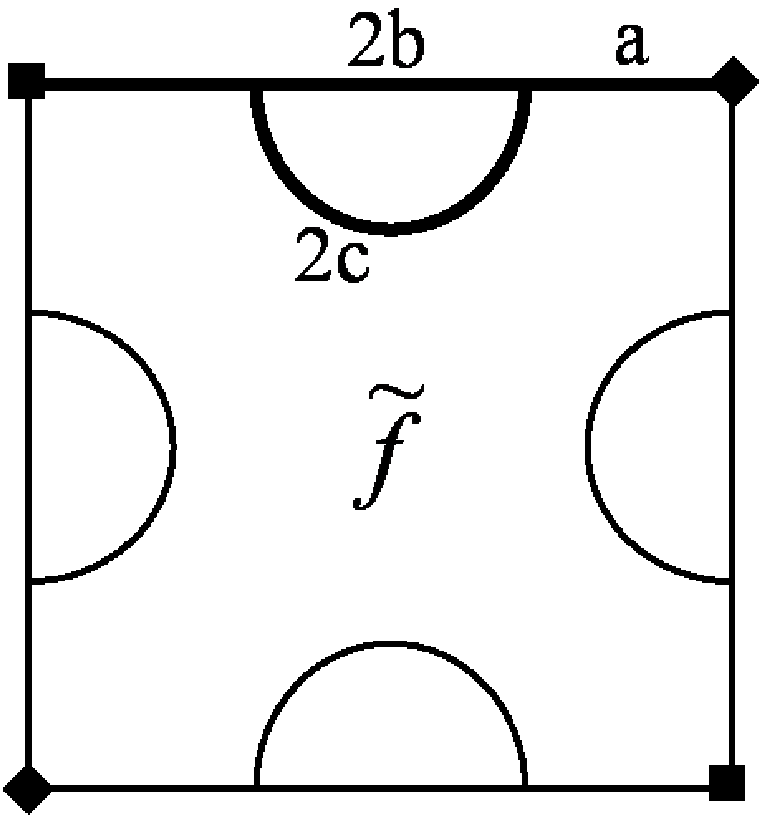}
\par\end{center}

\begin{center}
(a)
\par\end{center}%
\end{minipage}\hfill{}%
\begin{minipage}[c][1\totalheight][t]{0.35\columnwidth}%
\begin{center}
\includegraphics[width=0.9\textwidth]{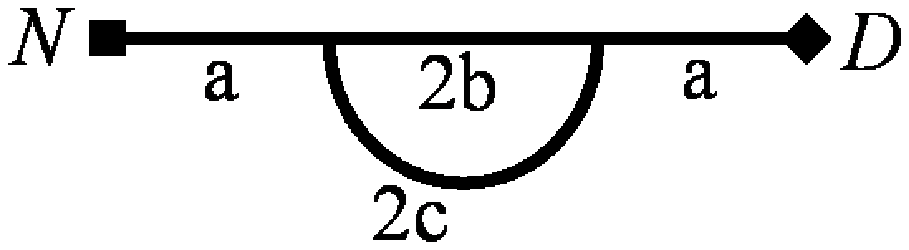}
\par\end{center}

\medskip{}

\begin{center}
(b)
\par\end{center}%
\end{minipage}\hfill{}

\caption{(a) The information we have on
$\tilde{f}\in\Phi_{\Gamma}^{R_{2}}(\lambda)$. Diamonds mark the
vertices at which the function vanishes and squares the vertices
with zero derivative. (b) The quotient graph
$\nicefrac{\Gamma}{R_{2}}$ which encodes this information.
\textit{D} stands for Dirichlet boundary conditions and \textit{N}
for Neumann.}

\label{fig:intuitive_quotient2}
\end{figure}

\par\end{center}

The isospectrality of $\nicefrac{\Gamma}{R_{1}}$ and
$\nicefrac{\Gamma}{R_{2}}$ is a direct consequence of corollary
\ref{cor:sunadapair} which appears in the next section.

\section{Representation theory and isospectrality}

\label{sec:algebra} Having informally exposed some of the relations
between representations and quantum graphs, we begin a more precise
examination, culminating in a general theorem on isospectrality
\footnote{See \ref{appendix} for a short review of the Algebra used
in this section.}. Let $\Gamma$ be a graph that obeys a certain
finite symmetry group $G$; this means that the action of $G$
preserves the lengths of the edges and the connectivity and boundary
conditions at the vertices. We do not assume that $G$ is the maximal
symmetry group of $\Gamma$. For every $\lambda\in\mathbb{C}$,
$\Phi_{\Gamma}\left(\lambda\right)$, the vector space of all
$\lambda$-eigenfunctions of the Laplacian on $\Gamma$, is a carrier
space of some representation of $G$. This follows from the Laplacian
commuting with the symmetry group: $\forall g\in
G,\,\,-\bigtriangleup(gf)=g\left(-\bigtriangleup
f\right)=\lambda(gf)$, which renders
$\Phi_{\Gamma}\left(\lambda\right)$ closed under the action of $G$.
Let $R$ denote the representation carried by
$\Phi_{\Gamma}\left(\lambda\right)$, and decompose it into the
irreducible representations of $G$. Such a decomposition allows us
to present $\Phi_{\Gamma}\left(\lambda\right)$ as some direct sum of
carrier spaces of the irreducible representations of $G$. Denote by
$S_{1},\ldots,S_{r}$ the irreducible representations of $G$ and
assume that $S_{i}$ appears $n_{i}$ times in $R$, i.e.,
$R\cong\bigoplus_{i=1}^{r}n_{i}\, S_{i}$. Then
\begin{eqnarray}
\nonumber
\Phi_{\Gamma}\left(\lambda\right)&=&\left(V_{1}^{S_{1}}\oplus\ldots\oplus
V_{n_{1}}^{S_{1}}\right)\oplus\ldots\oplus\left(V_{1}^{S_{r}}\oplus\ldots\oplus
V_{n_{r}}^{S_{r}}\right)\\
&=&\bigoplus _{i=1}^{r}\:
\Phi_{\Gamma}^{S_{i}}\left(\lambda\right)\,,\label{eq:isotypic-decomp}\end{eqnarray}
 where, for each $i\in\left\{ 1..r\right\} $, $\left\{ V_{j}^{S_{i}}\right\} _{j=1}^{n_{i}}$
are carrier spaces of the irreducible representation $S_{i}$ (one
carrier space for each copy of $S_{i}$ in $R$), and their direct sum
in $\Phi_{\Gamma}\left(\lambda\right)$ is denoted by
$\Phi_{\Gamma}^{S_{i}}\left(\lambda\right)$.
$\Phi_{\Gamma}^{S_{i}}\left(\lambda\right)$ is called the
$S_{i}$-\emph{isotypic component} of
$\Phi_{\Gamma}\left(\lambda\right)$; it is the vector space
consisting of all $\lambda$-eigenfunctions which transform under the
action of $G$ according to the representation $S_{i}$. Denoting by
$\mathbf{1}_{G}$ the trivial representation of $G$,
$\Phi_{\Gamma}^{\mathbf{1}_{G}}\left(\lambda\right)$ is called the
trivial component of $\Phi_{\Gamma}\left(\lambda\right)$, and is the
space of all $\lambda$-eigenfunctions which are invariant under the
action of $G$.

We pause the algebraic discussion for the purpose of reexamining the
example from the previous section. Recall that $R_1$ is a
representation of the group $H_1$, and that we have constructed a
quotient graph $\nicefrac{\Gamma}{R_{1}}$, such that the following
isomorphisms were established:
\begin{equation}
\forall\lambda\in\mathbb{C},\quad\Psi:\Phi_{\Gamma}^{R_{1}}\left(\lambda\right)\overset{\cong}{\longrightarrow}\Phi_{\nicefrac{\Gamma}{R_{1}}}\left(\lambda\right)\label{eq:isomorphic_eigenspaces}\end{equation}

 By definition, the dimension of $\Phi_{\nicefrac{\Gamma}{R_{1}}}\left(\lambda\right)$
is $\sigma_{\nicefrac{\Gamma}{R_{1}}}(\lambda)$, the multiplicity of
$\lambda$ in the spectrum of $\nicefrac{\Gamma}{R_{1}}$. In analogy,
we denote by $\sigma_{\Gamma}^{R_{1}}\left(\lambda\right)$ the
dimension of $\Phi_{\Gamma}^{R_{1}}\left(\lambda\right)$. But what
does this dimension tell us? It can be thought of as the spectrum
that will be observed by someone who can only see functions which
transform according to $R_{1}$. We call $\sigma_{\Gamma}^{R_{1}}$
the $R_{1}$-\emph{spectrum} of $\Gamma$, and note in particular that
it is a subspectrum of $\Gamma$:
$\sigma_{\Gamma}^{R_{1}}\leq\sigma_{\Gamma}$. In terms of dimensions
(\ref{eq:isomorphic_eigenspaces}) gives \begin{equation}
\sigma_{\Gamma}^{R_{1}}\equiv\sigma_{\nicefrac{\Gamma}{R_{1}}}\,.\label{eq:sigma_sigma}\end{equation}
 We thus identify the role of the quotient graph $\nicefrac{\Gamma}{R_{1}}$
as having the same spectrum as the $R_{1}$-spectrum of the original
graph. This will be the characterizing property of all our quotient
graphs, and it is therefore time to generalize the discussion.

\medskip{}

As we have defined the $R_{1}$-spectrum by
$\sigma_{\Gamma}^{R_{1}}(\lambda)=\dim\Phi_{\Gamma}^{R_{1}}(\lambda)$,
we now define the $S$-spectrum of $\Gamma$, for any irreducible
representation $S$ (not necessarily one dimensional), as
\begin{equation} \sigma_{\Gamma}^{S}(\lambda):=\frac{{\rm
dim}\Phi_{\Gamma}^{S}(\lambda)}{{\rm
dim}S}\,,\label{eq:irrep_spectrum}\end{equation}
 which can be interpreted as the number of copies of $S$ of which
$\Phi_{\Gamma}^{S}\left(\lambda\right)$ consists. This is also equal
to the number of copies of $S$ in $\Phi_{\Gamma}(\lambda)$ (the
corresponding $n_{i}$ in (\ref{eq:isotypic-decomp})). Using the
orthogonality relations of irreducible characters, we may rewrite
(\ref{eq:irrep_spectrum}) as
$\sigma_{\Gamma}^{S}\left(\lambda\right)=\left\langle
\chi_{S},\chi_{\Phi_{\Gamma}\left(\lambda\right)}\right\rangle
_{G}$. We use this equality to generalize the definition:

\begin{defn}
For any representation $R$, the $R$-spectrum of $\Gamma$
is\begin{equation} \sigma_{\Gamma}^{R}(\lambda):=\left\langle
\chi_{R},\chi_{\Phi_{\Gamma}\left(\lambda\right)}\right\rangle
_{G}\,.\label{def:rep_spectrum}\end{equation}

\end{defn}
Note that (\ref{eq:irrep_spectrum}) holds for irreducible
representations and is not true for all $R$ (in fact,
$\Phi^{R}_{\Gamma}(\lambda)$ is not even defined when $R$ is
reducible). Equipped with the notion of the $R$-spectrum of
$\Gamma$, we can now define what is a $\nicefrac{\Gamma}{R}$ graph.

\begin{defn}
\label{def:quotient-definition}A \emph{$\nicefrac{\Gamma}{R}$-graph}
is any quantum graph $\Gamma'$ whose spectrum is equal to the
$R$-spectrum of $\Gamma$: \[
\sigma_{\Gamma'}\equiv\sigma_{\Gamma}^{R}\,.\]

\end{defn}
Since all $\nicefrac{\Gamma}{R}$-graphs have the same spectrum by
definition, we allow ourselves, by abuse of language, to refer to
\emph{the spectrum of $\nicefrac{\Gamma}{R}$}. The following
isospectrality theorem then follows:

\begin{thm}
\label{thm:mainthm}Let $\Gamma$ be a quantum graph equipped with an
action of a group $G$, $H$ a subgroup of $G$, and $R$ a
representation of $H$. Then $\nicefrac{\Gamma}{R}$ is isospectral to
$\nicefrac{\Gamma}{\ind_{H}^{G}R}$.
\end{thm}
\begin{proof}
\begin{eqnarray*}
\forall\lambda\in\mathbb{C}\qquad\sigma_{\Gamma/R}(\lambda)=\sigma_{\Gamma}^{R}(\lambda) & = & \langle\chi_{R}\,,\,\chi_{\Phi_{\Gamma}(\lambda)}\rangle_{H}\\
 & = & \langle\chi_{\ind_{H}^{G}R}\,,\,\chi_{\Phi_{\Gamma}(\lambda)}\rangle_{G}\\
 & = & \sigma_{\Gamma}^{\ind_{H}^{G}R}(\lambda)=\sigma_{\Gamma/{\ind_{H}^{G}R}}(\lambda)\end{eqnarray*}
 where moving to the second line we have used the Frobenius Reciprocity
Theorem.
\end{proof}
\begin{cor}
\label{cor:sunadapair}If $G$ acts on $\Gamma$ and $H_{1},H_{2}$ are
subgroups of $G$ with corresponding representations $R_{1},R_{2}$
such that $\ind_{H_{1}}^{G}R_{1}\cong\ind_{H_{2}}^{G}R_{2}$, then
$\nicefrac{\Gamma}{R_{1}}$ and $\nicefrac{\Gamma}{R_{2}}$ are
isospectral.
\end{cor}

\begin{rem*}
This corollary is in fact equivalent to the theorem, as can be seen
by taking $H_{2}=G$, $R_{2}=\ind_{H_{1}}^{G}R_{1}$.
\end{rem*}
The isospectrality of the pair of graphs constructed in section
\ref{sec:basic_example} (figure \ref{fig:dihedral_pair}) is obtained
from the above corollary. Returning to that example, one first
notices that the two graphs presented there obey definition
\ref{def:quotient-definition}. This is true since we have shown that
\[
\forall\lambda\in\mathbb{C}\quad\Phi_{\nicefrac{\Gamma}{R_{1}}}(\lambda)\cong\Phi_{\Gamma}^{R_{1}}(\lambda)\]
 from which follows $\sigma_{\nicefrac{\Gamma}{R_{1}}}\equiv\sigma_{\Gamma}^{R_{1}}$
and therefore the first graph presented in section
\ref{sec:basic_example} can be honestly called a
$\nicefrac{\Gamma}{R_{1}}$-graph. The same goes for the second
graph, $\nicefrac{\Gamma}{R_{2}}$. The representations $R_{1},R_{2}$
used in the construction satisfy the condition
$\ind_{H_{1}}^{G}R_{1}\cong\ind_{H_{2}}^{G}R_{2}$ and this enables
us to apply corollary \ref{cor:sunadapair} and conclude that
$\nicefrac{\Gamma}{R_{1}}$ and $\nicefrac{\Gamma}{R_{2}}$ are
isospectral.

\section{Extending the basic example}

\label{sec:extending_example}It is clear that theorem
\ref{thm:mainthm} and corollary \ref{cor:sunadapair} would yield
isospectral examples only when the required quotient graphs indeed
exist. We saw the existence of two such quotients in section
\ref{sec:basic_example}. A proof of the existence of the quotient of
any graph by any representation, along with a rigorous construction
technique, are given in \cite{PB08}. In this section, we present key
examples which enable us to gain insight into this method, as well
as to understand the procedure implemented in \cite{PB08}.

We return to the basic example brought forth in section
\ref{sec:basic_example}, wishing to extend it by discovering more
quantum graphs which are isospectral to the pair of graphs
$\nicefrac{\Gamma}{R_{1}},\nicefrac{\Gamma}{R_{2}}$ in figure
\ref{fig:dihedral_pair}. Corollary \ref{cor:sunadapair} offers a
method by which this can be achieved: find a subgroup $H_{3}\leq G$
and a representation $R_{3}$ of it such that
$\ind_{H_{1}}^{G}R_{1}\cong\ind_{H_{2}}^{G}R_{2}\cong\ind_{H_{3}}^{G}R_{3}$.
Then, $\nicefrac{\Gamma}{R_{3}}$ is isospectral to
$\nicefrac{\Gamma}{R_{1}}$ and $\nicefrac{\Gamma}{R_{2}}$. Such a
subgroup and a representation indeed exist: \begin{eqnarray}
H_{3} & = & \{e,\,\sigma,\,\sigma^{2},\,\sigma^{3}\}\nonumber \\
R_{3} & : & \left\{ \begin{array}{llll} e\mapsto\left(1\right), &
\sigma\mapsto\left(i\right), & \sigma^{2}\mapsto\left(-1\right), &
\sigma^{3}\mapsto\left(-i\right)\end{array}\right\}
\label{eq:r3_rep}\end{eqnarray}

\begin{figure}[!h]
\begin{centering}
\hfill{}%
\begin{minipage}[c][1\totalheight][t]{0.4\columnwidth}%

\begin{center}
\includegraphics[scale=0.5]{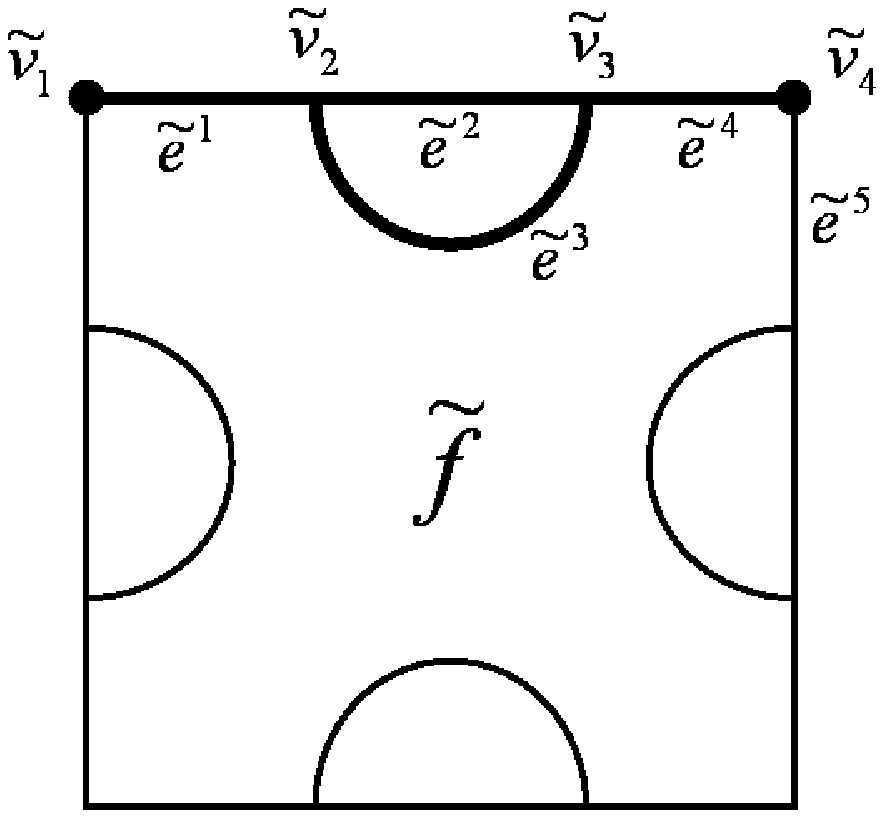}
\par\end{center}

\begin{center}
(a)
\par\end{center}%
\end{minipage}\hfill{}%
\begin{minipage}[c][1\totalheight][t]{0.5\columnwidth}%
\begin{center}
\begin{minipage}[c][1\totalheight]{0.4\columnwidth}%
\begin{center}
\includegraphics[scale=0.5]{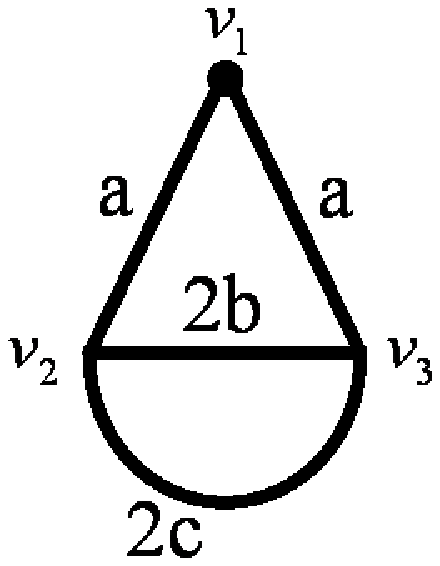}
\par\end{center}%
\end{minipage}%
\begin{minipage}[c][1\totalheight]{0.58\columnwidth}%
\begin{center}
\footnotesize $A_{v_{1}}=\left(\begin{array}{cc}
1 & -i\\
0 & 0\end{array}\right)$
\par\end{center}

\begin{center}
\footnotesize $B_{v_{1}}=\left(\begin{array}{cc}
0 & 0\\
1 & i\end{array}\right)$
\par\end{center}%
\end{minipage}
\par\end{center}

\begin{center}
(b)
\par\end{center}%
\end{minipage}\hfill{}
\par\end{centering}

\caption{(a) The information we have on
$\tilde{f}\in\Phi_{\Gamma}^{R_{3}}\left(\lambda\right)$. There is a
factor of $i$ between the values and the clockwise derivatives of
$\tilde{f}$ at the marked vertices $\tilde{v}_{1},\tilde{v}_{4}$.
(b) The quotient graph $\nicefrac{\Gamma}{R_{3}}$ which encodes this
information. The only non-Neumann boundary condition is at $v_{1}$
and it is specified by the matrices $A_{v_{1}},\, B_{v_{1}}$.}

\label{fig:intuitive_quotient3}
\end{figure}

We use the intuitive approach obtained from the basic example in
order to construct the quotient $\nicefrac{\Gamma}{R_{3}}$. Let
$\tilde{f}$ be a function that transforms according to the
representation $R_{3}$. The action of the rotation element $\sigma$
on $\tilde{f}$ is given by \begin{equation}
\sigma\,\tilde{f}=i\,\tilde{f}.\label{eq:sigma_action}\end{equation}
 This means that knowing the values of $\tilde{f}$ on a quarter of
the graph (for example, the quarter marked in bold in figure
\ref{fig:intuitive_quotient3}(a)) enables us to deduce the values of
$\tilde{f}$ on the whole of $\Gamma$. We therefore take this
subgraph to be our quotient graph and check what boundary conditions
we should impose on it. From (\ref{eq:sigma_action}) we obtain
\begin{eqnarray}
\tilde{f}\at_{\tilde{e}^{1}}(\tilde{v}_{1}) & = & i\,\tilde{f}\at_{\tilde{e}^{5}}(\tilde{v}_{4})\label{eq:R3_bc1}\\
\tilde{f}'\at_{\tilde{e}^{1}}(\tilde{v}_{1}) & = &
i\,\tilde{f}'\at_{\tilde{e}^{5}}(\tilde{v}_{4})\,.\label{eq:R3_bc2}\end{eqnarray}
 These equations suggest that we should identify vertices $\tilde{v}_{1},\tilde{v}_{4}$.
They merge to become a single vertex $v_{1}$ in the quotient (figure
\ref{fig:intuitive_quotient3}(b)). Equation (\ref{eq:R3_bc1}) gives
the boundary condition for the values at this new vertex. The
boundary condition for the derivatives at $v_{1}$ is obtained from
(\ref{eq:R3_bc2}), by recalling that $\tilde{f}$ obeys Neumann
boundary conditions at the vertex $\tilde{v}_{4}$ of $\Gamma$:
\begin{equation}
\tilde{f}'\at_{\tilde{e}^{4}}(\tilde{v}_{4})+\tilde{f}'\at_{\tilde{e}^{5}}(\tilde{v}_{4})=0\,.\label{eq:R3_bc3}\end{equation}
 Therefore, the boundary conditions at $v_{1}$ in $\nicefrac{\Gamma}{R_{3}}$
are described by the matrices given in figure
\ref{fig:intuitive_quotient3}(b), and this concludes the
construction of $\nicefrac{\Gamma}{R_{3}}$. The isomorphisms
\begin{equation}
\forall\lambda\in\mathbb{C},\quad\Psi:\Phi_{\Gamma}^{R_{3}}\left(\lambda\right)\overset{\cong}{\longrightarrow}\Phi_{\nicefrac{\Gamma}{R_{3}}}\left(\lambda\right)\label{eq:iso-R3}\end{equation}
 can be easily deduced from the construction process. Taking dimensions
gives
$\sigma_{\Gamma}^{R_{3}}~\equiv~\sigma_{\nicefrac{\Gamma}{R_{3}}}$,
which proves the validity of this quotient.

We already obtained an isospectral triple, but this does not cause
us to stagnate and we go further with our isospectral quest. By
theorem \ref{thm:mainthm}, any $\nicefrac{\Gamma}{R}$-graph, where
$R\cong\ind_{H_{1}}^{G}R_{1}$, would be isospectral to our three
graphs. A simple calculation (See \ref{appendix}) shows that $R$ is
the single two dimensional irreducible representation of $G=D_{4}$.
By a choice of basis we can describe $R$ as a matrix representation.
Such a representation is:
\begin{equation}
\fl \left\{ \begin{array}{cccc}
e\mapsto{\scriptsize\left(\begin{array}{cc} 1 & 0\\
0 & 1\end{array}\right)}, & \sigma\mapsto{\scriptsize\left(\begin{array}{cc}0 & 1\\
-1 & 0\end{array}\right)}, & \sigma^{2}\mapsto{\scriptsize\left(\begin{array}{cc}-1 & 0\\
0 & -1\end{array}\right)}, & \sigma^{3}\mapsto{\scriptsize\left(\begin{array}{cc}0 & -1\\
1 & 0\end{array}\right)},\\
\\\tau\mapsto{\scriptsize\left(\begin{array}{cc}-1 & 0\\
0 & 1\end{array}\right)}, & \tau\sigma\mapsto{\scriptsize\left(\begin{array}{cc}0 & -1\\
-1 & 0\end{array}\right)}, & \tau\sigma^{2}\mapsto{\scriptsize\left(\begin{array}{cc}1 & 0\\
0 & -1\end{array}\right)}, & \tau\sigma^{3}\mapsto{\scriptsize\left(\begin{array}{cc}0 & 1\\
1 & 0\end{array}\right)}\end{array}\right\}
\label{eq:repform1}\end{equation}

 We now construct the quotient $\nicefrac{\Gamma}{R}$.
 The graph obtained, which is shown in figure \ref{fig:intuitive_quotient4}(b) is the same as
 $\nicefrac{\Gamma}{R_1}$ (figure \ref{fig:dihedral_pair}(a)) and we elaborate
 on this phenomenon later on. The construction process of this ``new''
 graph is now presented for didactic reasons. The procedure is similar to those already described, although a slight
complication arises from $R$ not being one dimensional.

\begin{center}
\begin{figure}[!h]
\begin{centering}
\begin{minipage}[c][1\totalheight]{0.63\columnwidth}%
\begin{center}
\includegraphics[width=0.9\columnwidth]{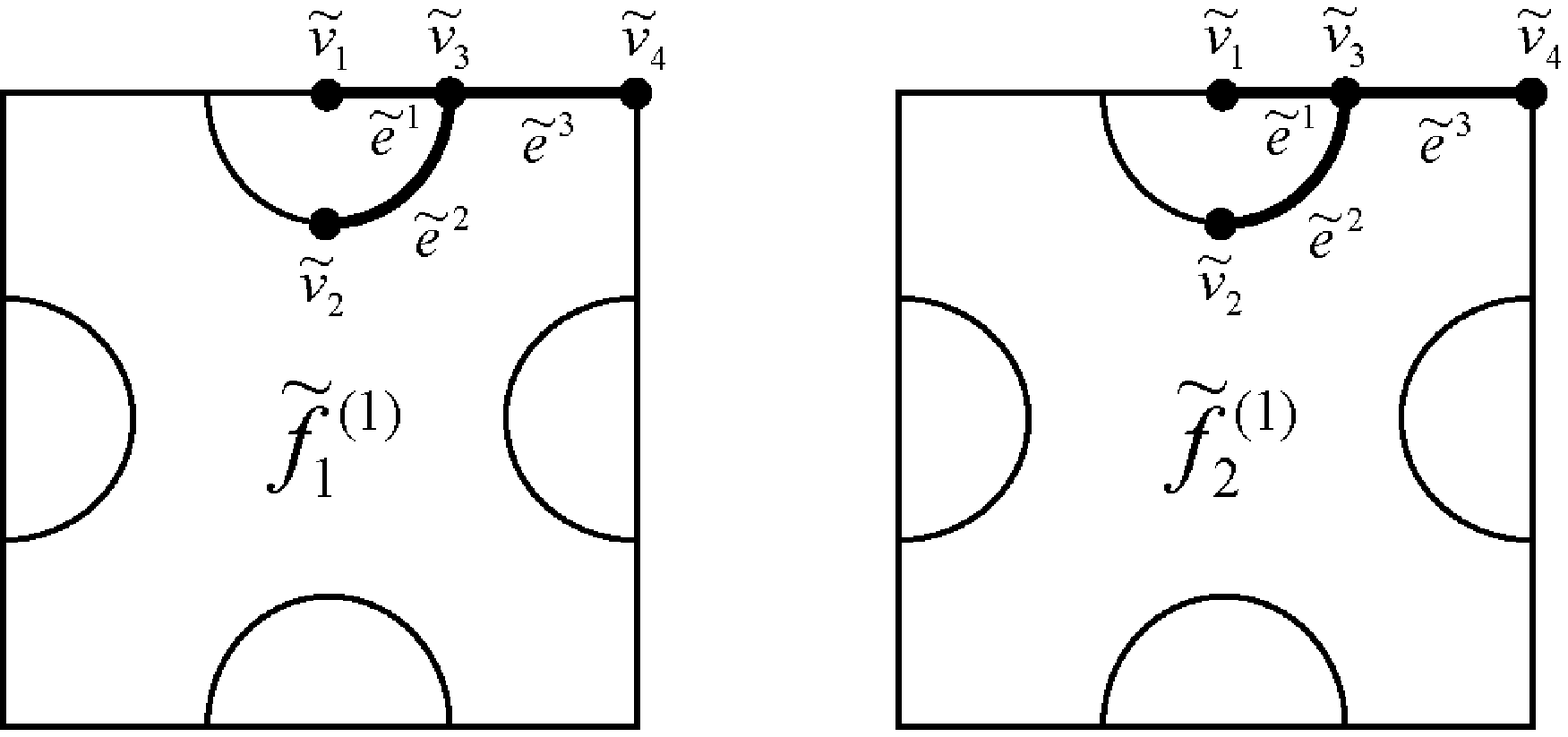}
\par\end{center}

\begin{center}
(a)
\par\end{center}%
\end{minipage}\hfill{}%
\begin{minipage}[c][1\totalheight]{0.35\columnwidth}%
\begin{center}
\includegraphics[scale=0.5]{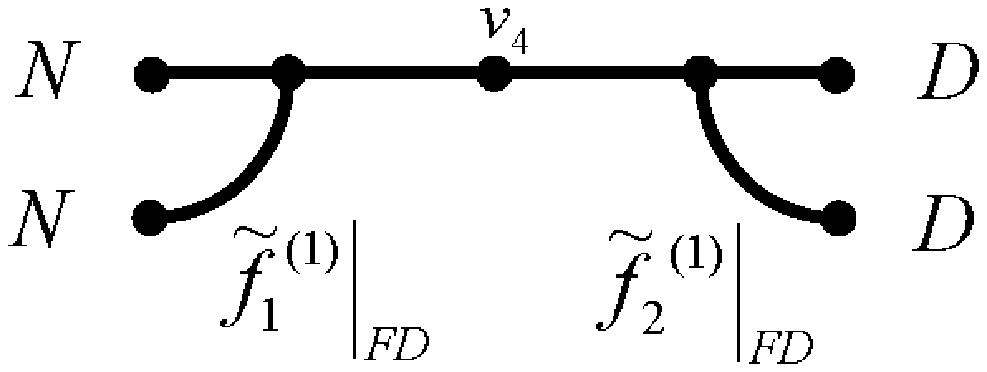}
\par\end{center}

\begin{center}
(b)
\par\end{center}%
\end{minipage}
\par\end{centering}

\caption{(a) Two copies of the graph $\Gamma$ which assist in the
visualization of the functions
$\tilde{f}_{1}^{(1)},\tilde{f}_{2}^{(1)}$ on $\Gamma$. The
fundamental domain is marked in bold. (b) The quotient graph
$\nicefrac{\Gamma}{R}$ formed as two copies of the fundamental
domain.}

\label{fig:intuitive_quotient4}
\end{figure}

\par\end{center}

We consider two functions
$\tilde{f}_1^{(1)},\tilde{f}_2^{(1)}\in\Phi_{\Gamma}(\lambda)$ (for
some $\lambda\in\mathbb{C}$) that transform according to the matrix
representation (\ref{eq:repform1}). It follows that
$\tilde{f}_1^{(1)},\tilde{f}_2^{(1)}\in\Phi_{\Gamma}^{R}(\lambda)$,
and that $\tilde{f}_{1}^{(1)}$ and $\tilde{f}_{2}^{(1)}$ form a
basis for a carrier space of the representation $R$, which we denote
by $V_{1}^{R}$. We may proceed in this manner, choosing
$\tilde{f}_{1}^{\left(2\right)},\tilde{f}_{2}^{\left(2\right)},\ldots$
such that each pair,
$\left\{\tilde{f}_{1}^{\left(i\right)},\tilde{f}_{2}^{\left(i\right)}\right\}$,
transforms according to (\ref{eq:repform1}), and is linearly
independent of the previous ones. Therefore, each pair,
$\left\{\tilde{f}_{1}^{\left(i\right)},\tilde{f}_{2}^{\left(i\right)}\right\}$,
spans a different carrier space of $R$, which we denote by
$V_{i}^{R}$ and we get that
$\Phi_{\Gamma}^{R}\left(\lambda\right)=\bigoplus\limits
_{i=1}^{n}V_{i}^{R}$. The number of carrier spaces is
\begin{equation}
n=\frac{\dim\Phi_{\Gamma}^{R}\left(\lambda\right)}{\dim
R}=\left\langle
\chi_{R},\chi_{\Phi_{\Gamma}^{R}\left(\lambda\right)}\right\rangle
_{G}=\left\langle
\chi_{R},\chi_{\Phi_{\Gamma}\left(\lambda\right)}\right\rangle
_{G}=\sigma_{\Gamma}^{R}(\lambda)\,\label{eq:dim_equality1}\end{equation}
 (recall that $R$ is irreducible). We wish to construct a quotient
$\nicefrac{\Gamma}{R}$ such that \begin{equation}
\dim\Phi_{\nicefrac{\Gamma}{R}}(\lambda)=n\,,\label{eq:dim_equality2}\end{equation}
 which means that $\sigma_{\nicefrac{\Gamma}{R}}(\lambda)=\sigma_{\Gamma}^{R}(\lambda)$.
If this holds for every $\lambda$, then $\nicefrac{\Gamma}{R}$ is
indeed the desired quotient graph (definition
\ref{def:quotient-definition}). If we again relate the construction
process to the encoding technique, we see that we can achieve
(\ref{eq:dim_equality2}) if each carrier space $V_{i}^{R}$ is
encoded by a single function
$f^{(i)}\in\Phi_{\nicefrac{\Gamma}{R}}(\lambda)$, in a manner that
$\left\{ f^{(i)}\right\}_{i=1}^{n} $ is a basis for
$\Phi_{\nicefrac{\Gamma}{R}}\left(\lambda\right)$. We demonstrate
this idea by encoding $V_{1}^{R}={\rm Span}\left\{
\tilde{f}_{1}^{(1)},\tilde{f}_{2}^{(1)}\right\} $. This is done by
thinking of the basis functions
$\tilde{f}_{1}^{(1)},\tilde{f}_{2}^{(1)}$ as residing on two copies
of the graph $\Gamma$ (figure \ref{fig:intuitive_quotient4}(a)).
Knowing the values of $\tilde{f}_{1}^{(1)}$ and
$\tilde{f}_{2}^{(1)}$ on a fundamental domain for the action of $G$
on $\Gamma$ (e.g., the bold subgraphs in figure
\ref{fig:intuitive_quotient4}(a)) allows one to deduce the values of
$\tilde{f}_{1}^{(1)}$ and $\tilde{f}_{2}^{(1)}$ on the whole graph,
using the known action of the group (\ref{eq:repform1}). Therefore,
the quotient graph is the union of these two copies of the
fundamental domain. Its boundary conditions can be concluded from
(\ref{eq:repform1}), which gives the relations between the values of
$\tilde{f}_{1}^{(1)},\tilde{f}_{2}^{(1)}$ and between their
derivatives:
\begin{eqnarray}
\tau\sigma^{2}\,\tilde{f}_{1}^{(1)}=\tilde{f}_{1}^{(1)} & \quad\Rightarrow\quad & \left(\tilde{f}_{1}^{(1)}\right)'\at_{\tilde{e}^{1}}(\tilde{v}_{1})=\left(\tilde{f}_{1}^{(1)}\right)'\at_{\tilde{e}^{2}}(\tilde{v}_{2})=0\label{eq:fourth_dihedral1}\\
\tau\sigma^{2}\,\tilde{f}_{2}^{(1)}=-\tilde{f}_{2}^{(1)} & \quad\Rightarrow\quad & \tilde{f}_{2}^{(1)}\at_{\tilde{e}^{1}}(\tilde{v}_{1})=\tilde{f}_{2}^{(1)}\at_{\tilde{e}^{2}}(\tilde{v}_{2})=0\label{eq:fourth_dihedral2}\\
\tau\sigma^{3}\,\tilde{f}_{1}^{(1)}= \tilde{f}_{2}^{(1)}&&
.\label{eq:fourth_dihedral3}\end{eqnarray}
 We recall that $\tilde{f}_{1}^{(1)}$ satisfies Neumann boundary
conditions at $\tilde{v}_{4}$: \begin{eqnarray}
\left(\tau\sigma^{3}\tilde{f}_{1}^{(1)}\right)\at_{\tilde{e}^{3}}(\tilde{v}_{4})=\tilde{f}_{1}^{(1)}\at_{\tau\sigma^{3}\tilde{e}^{3}}(\tilde{v}_{4})=\tilde{f}_{1}^{(1)}\at_{\tilde{e}^{3}}(\tilde{v}_{4})\label{eq:v4_temp_values}\\
\left(\tau\sigma^{3}\tilde{f}_{1}^{(1)}\right)'\at_{\tilde{e}^{3}}(\tilde{v}_{4})=\left(\tilde{f}_{1}^{(1)}\right)'\at_{\tau\sigma^{3}\tilde{e}^{3}}(\tilde{v}_{4})=-\left(\tilde{f}_{1}^{(1)}\right)'\at_{\tilde{e}^{3}}(\tilde{v}_{4})\quad.\label{eq:v4_temp_deriv}\end{eqnarray}
 Plugging (\ref{eq:fourth_dihedral3}) into (\ref{eq:v4_temp_values}),
(\ref{eq:v4_temp_deriv}), gives the following relations:
\begin{eqnarray}
&&\tilde{f}_{2}^{(1)}\at_{\tilde{e}^{3}}(\tilde{v}_{4})  =  \tilde{f}_{1}^{(1)}\at_{\tilde{e}^{3}}(\tilde{v}_{4})\label{eq:v4_temp_values2}\\
&&
\left(\tilde{f}_{2}^{(1)}\right)'\at_{\tilde{e}^{3}}(\tilde{v}_{4})
 =
-\left(\tilde{f}_{1}^{(1)}\right)'\at_{\tilde{e}^{3}}(\tilde{v}_{4})\quad.\label{eq:v4_temp_deriv2}\end{eqnarray}
 Equations (\ref{eq:v4_temp_values2}) and (\ref{eq:v4_temp_deriv2})
motivate us to glue the two sub-graphs at the vertex $\tilde{v}_{4}$
and supplement the new graph with Neumann boundary conditions at
this vertex, which we denote by $v_4$. The relations
(\ref{eq:fourth_dihedral1}), (\ref{eq:fourth_dihedral2}) give
Neumann and Dirichlet boundary conditions, respectively, on the
other vertices, and this fully describes the quotient graph
$\nicefrac{\Gamma}{R}$ (figure \ref{fig:intuitive_quotient4}(b)).

We now explain how this encoding enables us to prove that
$\sigma_{\nicefrac{\Gamma}{R}}\equiv\sigma_{\Gamma}^{R}$, and ensure
the validity of our $\nicefrac{\Gamma}{R}$. Given
$\tilde{f}_{1}^{(1)},\tilde{f}_{2}^{(1)}$ as above, we form a
function $f^{(1)}\in\Phi_{\nicefrac{\Gamma}{R}}(\lambda)$ whose left
half is equal to the restriction of $\tilde{f}_{1}^{(1)}$ to the
fundamental domain, and whose right half is the restriction of
$\tilde{f}_{2}^{(1)}$ to the fundamental domain. The considerations
above apply for every carrier space $V_{i}^{R}$, and we can encode
its basis $\left\{ \tilde{f}_{1}^{(i)},\tilde{f}_{2}^{(i)}\right\} $
by a function $f^{(i)}\in\Phi_{\nicefrac{\Gamma}{R}}(\lambda)$ in
the same way as was done for $i=1$.

The set $\bigcup\limits _{i=1}^{n}\left\{
\tilde{f}_{1}^{(i)},\tilde{f}_{2}^{(i)}\right\} $ forms a basis for
$\Phi_{\Gamma}^{R}(\lambda)$, and is therefore linearly independent.
We now show that $\left\{ f^{(i)}\right\} _{i=1}^{n}$ are linearly
independent as well, which gives
$\frac{1}{2}\dim\Phi_{\Gamma}^{R}(\lambda)\leq\dim\Phi_{\nicefrac{\Gamma}{R}}\left(\lambda\right)$,
i.e.,
$\sigma_{\Gamma}^{R}\left(\lambda\right)\leq\sigma_{\nicefrac{\Gamma}{R}}\left(\lambda\right)$.
Assume that $\sum_{i=1}^{n}c_{i}f^{(i)}=0$, so we have for the
restrictions of $\left\{
\tilde{f}_{1}^{(i)},\tilde{f}_{2}^{(i)}\right\} _{i=1}^{n}$ on the
fundamental domain that
$\sum_{i=1}^{n}c_{i}\tilde{f}_{1}^{(i)}\at_{FD}=\sum_{i=1}^{n}c_{i}\tilde{f}_{2}^{(i)}\at_{FD}=0$.
Using the known action of the group (\ref{eq:repform1}), which
linearly relate the values of $\left\{
\tilde{f}_{1}^{(i)},\tilde{f}_{2}^{(i)}\right\} _{i=1}^{n}$
everywhere on $\Gamma$ to their values on the fundamental domain, we
obtain
$\sum_{i=1}^{n}c_{i}\tilde{f}_{1}^{(i)}=\sum_{i=1}^{n}c_{i}\tilde{f}_{2}^{(i)}=0$,
hence, $\forall i\;\; c_{i}=0$.

In order to show the opposite inequality
$\frac{1}{2}\dim\Phi_{\Gamma}^{R}(\lambda)\geq\dim\Phi_{\nicefrac{\Gamma}{R}}\left(\lambda\right)$
we employ the decoding process, which turns a function $g$ on
$\nicefrac{\Gamma}{R}$ into a pair of functions $\left\{
\tilde{g}_1,\tilde{g}_2\right\} $ on $\Gamma$ which transform
according to (\ref{eq:repform1}). This decoding is done in the
following way: first, we set $\tilde{g}_1\at_{FD}$ to be equal to
the restriction of $g$ to the left half of $\nicefrac{\Gamma}{R}$
and $\tilde{g}_2\at_{FD}$ to the restriction of $g$ to its right
half. Next, since $\tilde{g}_1,\tilde{g}_2$ should transform
according to the matrix representation (\ref{eq:repform1}), we know
how to express the values of each of $\tilde{g}_1$ and $\tilde{g}_2$
on $\Gamma$ as a linear combination of their values on the
fundamental domain. We now show that starting from any set of
linearly independent functions $\left\{ g^{(i)}\right\} _{i=1}^{m}$
on $\nicefrac{\Gamma}{R}$, and performing the decoding process on
each of them to obtain the set $\left\{
\tilde{g}_{1}^{(i)},\tilde{g}_{2}^{(i)}\right\} _{i=1}^{m}$ of
functions on $\Gamma$, the latter set is linearly independent as
well. Assume that
$\sum_{i=1}^{m}c_{1}^{(i)}\tilde{g}_{1}^{(i)}+c_{2}^{(i)}\tilde{g}_{2}^{(i)}=0$.
Since $\left\{ \tilde{g}_{j}^{\left(i\right)}\right\} $ transform
according to (\ref{eq:repform1}), we can apply the elements of $G$
to this relation in order to find others; for example, from
$\tau\sigma^{3}$ we obtain
$\sum_{i=1}^{m}c_{2}^{(i)}\tilde{g}_{1}^{(i)}+c_{1}^{(i)}\tilde{g}_{2}^{(i)}=0$.
Since $R$ is irreducible, the matrices in (\ref{eq:repform1})
additively span $M_{2}\left(\mathbb{C}\right)$ (\cite{Curtis},
section 3C). Therefore, by applying a suitable combination of $G$'s
elements we obtain
\[
\sum_{i=1}^{m}c_{1}^{(i)}\tilde{g}_{1}^{(i)}=\sum_{i=1}^{m}c_{1}^{(i)}\tilde{g}_{2}^{(i)}=\sum_{i=1}^{m}c_{2}^{(i)}\tilde{g}_{1}^{(i)}=\sum_{i=1}^{m}c_{2}^{(i)}\tilde{g}_{2}^{(i)}=0\]
 so that $\sum_{i=1}^{m}c_{1}^{(i)}g^{(i)}=\sum_{i=1}^{m}c_{2}^{(i)}g^{(i)}=0$,
and therefore $\forall i\; c_{1}^{(i)}=c_{2}^{(i)}=0$. Hence
$\frac{1}{2}\dim\Phi_{\Gamma}^{R}(\lambda)=\dim\Phi_{\nicefrac{\Gamma}{R}}\left(\lambda\right)$,
which gives
$\sigma_{\Gamma}^{R}(\lambda)=\sigma_{\nicefrac{\Gamma}{R}}(\lambda)$,
and finishes the proof (see (\ref{eq:dim_equality1}),
(\ref{eq:dim_equality2})).\\

As was already mentioned, the ``new'' graph $\nicefrac{\Gamma}{R}$
we have obtained is in fact one of the graphs that we saw
previously, namely, $\nicefrac{\Gamma}{R_{1}}$ (figure
\ref{fig:dihedral_pair}(a)). At first sight one might wonder whether
theorem \ref{thm:mainthm} has any practical use, as the isospectral
quantum graphs we obtained from it are isometric. This fear is
ungrounded. Notice that $\nicefrac{\Gamma}{R}$ was constructed
according to a specific matrix representation of $R$, and we still
have not checked how a change of basis (which changes the matrix
representation) affects $\nicefrac{\Gamma}{R}$. There is indeed a
dependency on the choice of basis, as we now show. Another matrix
representation of $R$ is \begin{equation} \left\{ \begin{array}{ll}
\tau\sigma^{2}\mapsto\frac{1}{2}\left(\begin{array}{cc}
-1 & -\sqrt{3}\\
-\sqrt{3} & 1\end{array}\right), &
\tau\sigma^{3}\mapsto\frac{1}{2}\left(\begin{array}{cc}
\sqrt{3} & -1\\
-1 & -\sqrt{3}\end{array}\right)\end{array}\right\}
.\label{eq:repform2}\end{equation}
 We present only the matrices of the elements $\tau\sigma^{2},\tau\sigma^{3}$
since we saw that they suffice to construct the quotient. Figure
\ref{fig:intuitive_quotient5} is analogous to figure
\ref{fig:intuitive_quotient4}, for the current case. The fundamental
domain that we choose is the same, and the difference will appear in
the boundary conditions.
\begin{center}
\begin{figure}[!h]

\begin{centering}
\begin{minipage}[c][1\totalheight][t]{0.55\columnwidth}%
\begin{center}
\includegraphics[width=0.85\columnwidth]{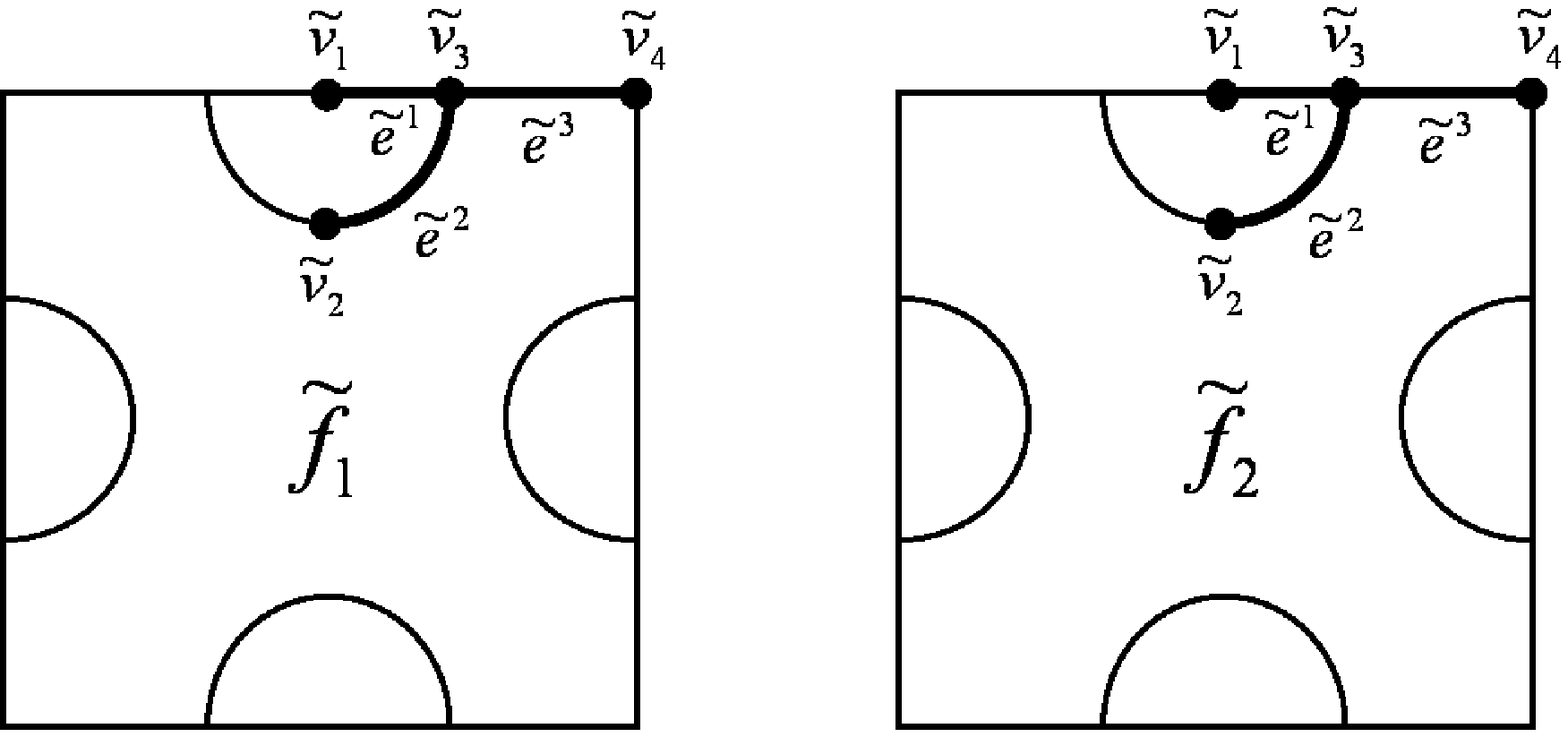}
\par\end{center}

\begin{center}
(a)
\par\end{center}%
\end{minipage}\hfill{}%
\begin{minipage}[c][1\totalheight][t]{0.25\columnwidth}%
\begin{center}
\includegraphics[width=0.9\columnwidth]{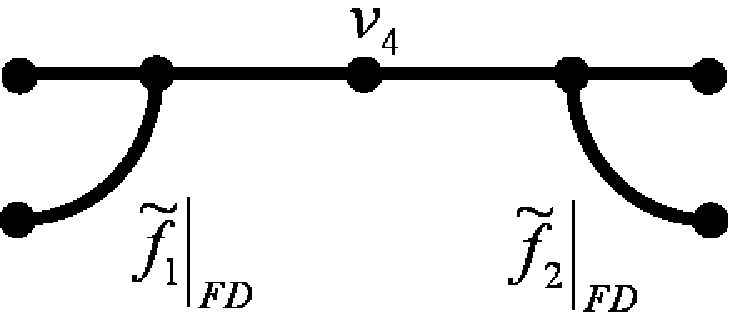}
\par\end{center}

\medskip{}

\begin{center}
(b)
\par\end{center}%
\end{minipage}\hfill{}%
\begin{minipage}[c][1\totalheight][t]{0.16\columnwidth}%
\begin{center}
\includegraphics[width=0.6\columnwidth]{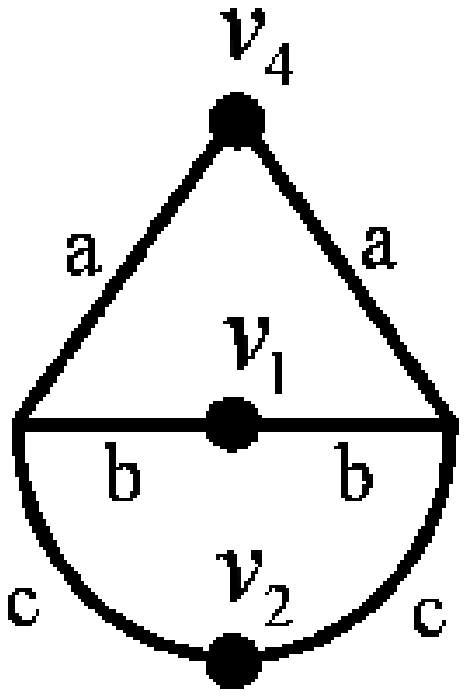}
\par\end{center}

\begin{center}
\medskip{}
 (c)
\par\end{center}%
\end{minipage}
\par\end{centering}

\caption{(a) Two copies of the graph $\Gamma$ which assist in the
visualization of the functions $\tilde{f}_{1},\,\tilde{f}_{2}$ on
$\Gamma$. The fundamental domain is marked in bold. (b) The first
stage in the gluing process. The encoding is done by restricting
$\tilde{f}_{1},\,\tilde{f}_{2}$ to the fundamental domain. (c) The
quotient graph $\nicefrac{\Gamma}{R}$ which is built using the
matrix representation (\ref{eq:repform2}). The boundary conditions
are described in (\ref{eq:v4_bc}), (\ref{eq:v12_bc}).}
\label{fig:intuitive_quotient5}
\end{figure}
\par\end{center}
We consider two functions
$\tilde{f}_{1},\tilde{f}_{2}\in\Phi_{\Gamma}(\lambda)$ (for some
$\lambda\in\mathbb{C}$) that transform according to the matrix
representation (\ref{eq:repform2}). The first column of the matrix
representing $\tau\sigma^{3}$ tells us that \begin{eqnarray}
\tau\sigma^{3}\tilde{f}_{1} & = & \nicefrac{\sqrt{3}}{2}\,\tilde{f}_{1}-\nicefrac{1}{2}\,\tilde{f}_{2}\label{eq:diagonal_reflection_values}\\
\tau\sigma^{3}\tilde{f}_{1}' & = &
\nicefrac{\sqrt{3}}{2}\,\tilde{f}_{1}'-\nicefrac{1}{2}\,\tilde{f}_{2}'\label{eq:diagonal_reflection_deriv}\end{eqnarray}
 Evaluating (\ref{eq:diagonal_reflection_values}),
 (\ref{eq:diagonal_reflection_deriv})
on $\tilde{v}_{4}$ and using the knowledge that $\tilde{f}_{1}$
obeys Neumann boundary conditions on $\tilde{v}_{4}$ (see equations
(\ref{eq:v4_temp_values}), (\ref{eq:v4_temp_deriv})) gives
\begin{eqnarray}
\left(1-\nicefrac{\sqrt{3}}{2}\right)\,\tilde{f}_{1}\at_{\tilde{e}^{3}}(\tilde{v}_{4})+\nicefrac{1}{2}\,\tilde{f}_{2}\at_{\tilde{e}^{3}}(\tilde{v}_{4}) & = & 0\,.\label{eq:v4_values}\\
\left(-1-\nicefrac{\sqrt{3}}{2}\right)\,\tilde{f}_{1}'\at_{\tilde{e}^{3}}(\tilde{v}_{4})+\nicefrac{1}{2}\,\tilde{f}_{2}'\at_{\tilde{e}^{3}}(\tilde{v}_{4})
& = & 0\,.\label{eq:v4_derivatives}\end{eqnarray}
 This indicates how we should start gluing the two subgraphs in order
to obtain the quotient. The first stage in this process, depicted in
figure \ref{fig:intuitive_quotient5}(b), is to identify the vertex
$\tilde{v}_{4}$ in the two copies and turn it into the vertex
$v_{4}$ of the quotient, with the boundary conditions that were
derived in (\ref{eq:v4_values}), (\ref{eq:v4_derivatives}):
\begin{eqnarray}
A_{v_{4}}=\left(\begin{array}{cc}1-\nicefrac{\sqrt{3}}{2} & \nicefrac{1}{2}\\
0 & 0\end{array}\right) & \quad,\quad & B_{v_{4}}=\left(\begin{array}{cc}0 & 0\\
-1-\nicefrac{\sqrt{3}}{2} &
\nicefrac{1}{2}\end{array}\right)\,.\label{eq:v4_bc}\end{eqnarray}
 After treating vertices $\tilde{v}_{1},\,\tilde{v}_{2}$ similarly we obtain the quotient $\nicefrac{\Gamma}{R}$ (figure
\ref{fig:intuitive_quotient5}(c)), whose remaining boundary
conditions are given by: \begin{eqnarray}
A_{v_{1}}=A_{v_{2}}=\left(\begin{array}{cc}\nicefrac{3}{2} & \nicefrac{\sqrt{3}}{2}\\
0 & 0\end{array}\right)\quad,\quad B_{v_{1}}=B_{v_{2}}=\left(\begin{array}{cc}0 & 0\\
-\nicefrac{1}{2} &
\nicefrac{\sqrt{3}}{2}\end{array}\right)\,.\label{eq:v12_bc}\end{eqnarray}

This last example demonstrates that for a multidimensional
representation the quotient graph depends on the explicit matrix
representation by which it is constructed. One choice of basis for
the matrix representation gave us a quotient which is identical to
one obtained previously (figure \ref{fig:dihedral_pair}(a)), while
another basis yielded a new quotient graph (figure
\ref{fig:intuitive_quotient5}(c)). As a matter of fact, all the
examples of quotient graphs with respect to the various
representations discussed so far (figures \ref{fig:dihedral_pair}(a)
and (b) and \ref{fig:intuitive_quotient3}(b)) can also be obtained
as quotients with respect to the representation $R$, by suitable
choices of bases for its matrix representation. Furthermore, there
are many other quantum graphs isospectral to these. For example, we
consider an arbitrary orthogonal matrix representation of $R$, which
is parameterized by :

\[
\left\{ \begin{array}{ccc} \tau\sigma^{2} & \mapsto &
\left(\begin{array}{cc}
\cos^{2}\theta-\sin^{2}\theta & -2\cos\theta\sin\theta\\
-2\cos\theta\sin\theta & -\cos^{2}\theta+\sin^{2}\theta\end{array}\right)\\
\\\tau\sigma^{3} & \mapsto & \left(\begin{array}{cc}
2\cos\theta\sin\theta & \cos^{2}\theta-\sin^{2}\theta\\
\cos^{2}\theta-\sin^{2}\theta &
-2\cos\theta\sin\theta\end{array}\right)\end{array}\right\} \]
 (e.g., (\ref{eq:repform2}) is obtained from $\theta=\nicefrac{\pi}{3}$).
Using the general construction method, which is described in the
next section, we obtain from this matrix representation the quotient
given in figure \ref{fig:intuitive_quotient5}(c), with the following
boundary conditions: \begin{eqnarray*} \fl
A_{v_{1}}=A_{v_{2}}=\left(\begin{array}{cc}2\sin^{2}\theta & \sin2\theta\\
\sin2\theta & 2-2\sin^{2}\theta\\
0 & 0\\
0 & 0\end{array}\right) & \qquad & A_{v_{4}}=\left(\begin{array}{cc}1-\sin2\theta & 2\sin^{2}\theta-1\\
2\sin^{2}\theta-1 & 1+\sin2\theta\\
0 & 0\\
0 & 0\end{array}\right)\\
\fl
B_{v_{1}}=B_{v_{2}}=\left(\begin{array}{cc}0 & 0\\
0 & 0\\
2-2\sin^{2}\theta & -\sin2\theta\\
-\sin2\theta & 2\sin^{2}\theta\end{array}\right) & \qquad & B_{v_{4}}=\left(\begin{array}{cc}0 & 0\\
0 & 0\\
1+\sin2\theta & 1-2\sin^{2}\theta\\
1-2\sin^{2}\theta & 1-\sin2\theta\end{array}\right).\end{eqnarray*}
 The matrices above are not square ones as is required from the definition of the boundary conditions.
However, since their role is to describe linear restrictions and due
to the fact that they are all of rank one, they can be reduced to
square matrices by deleting the appropriate rows\footnote{See also
the discussion that comes after
\eref{eq:final-boundary-conditions-1},
\eref{eq:final-boundary-conditions-2}.}. We thus get a continuous
family of isospectral graphs. We already met some of its members.
For example, $\theta=\nicefrac{3\pi}{4}$ gives the following
boundary conditions:
\begin{eqnarray}
A_{v_{1}}=A_{v_{2}}=\left(\begin{array}{cc}1 & -1\\
0 & 0\end{array}\right)\qquad & \qquad A_{v_{4}}=\left(\begin{array}{cc}2 & 0\\
0 & 0\end{array}\right)\\
B_{v_{1}}=B_{v_{2}}=\left(\begin{array}{cc}0 & 0\\
1 & 1\end{array}\right)\qquad & \qquad B_{v_{4}}=\left(\begin{array}{cc}0 & 0\\
0 & 2\end{array}\right)\end{eqnarray}
 When applying this to figure \ref{fig:intuitive_quotient5}(c), we
notice that $v_{4}$ does not remain a vertex of degree two, but
rather, splits into two vertices of degree one, one having Dirichlet
boundary condition and the other Neumann. The vertices $v_{1}$ and
$v_{2}$, however, stay connected and obtain Neumann boundary
conditions. The resulting quotient is thus the one that we have
already obtained as $\nicefrac{\Gamma}{R_{2}}$ (figure
\ref{fig:dihedral_pair}(b)). In a similar manner, the quotient
$\nicefrac{\Gamma}{R_{1}}$ (figure \ref{fig:dihedral_pair}(a)) is
obtained from the choice $\theta=0$. We conclude by pointing out
that the graph described in figure \ref{fig:intuitive_quotient5}(c)
is a good prototype for the isospectral family mentioned, yet it
also might be misleading, since some members of the family have
boundary conditions that tear apart the edges connected to some of
the vertices, and thus change the connectivity of the graph. One
should also pay attention to the fact that we have treated only
orthogonal representations of $R$. We may further extend the
isospectral family presented above by considering the broader case
of all matrix representations of $R$. In particular, the quotient
$\nicefrac{\Gamma}{R_{3}}$ (figure \ref{fig:intuitive_quotient3}(b))
is obtained from the unitary representation\[ \left\{
\begin{array}{cc} \sigma\mapsto\left(\begin{array}{cc}
i & 0\\
0 & -i\end{array}\right), & \tau\mapsto\left(\begin{array}{cc}
0 & -1\\
-1 & 0\end{array}\right)\end{array}\right\} \,.\]

\section{The rigorous construction of a quotient graph}

\label{sec:rigorous}

The rigorous formalism of the quotient graph construction is fully
described and proven in \cite{PB08}. Here we summarize this method
in accordance with the discussion and demonstrations presented so
far in this paper. Let $\Gamma$ be a quantum graph with a finite set
of vertices $V$ and a finite set $E$ of edges connecting the
vertices. Let $G$ be a finite group that acts on $\Gamma$, and $R$ a
$d$-dimensional representation of $G$. We assume for now that $G$
acts freely on the edges, i.e., $ge\neq e$ for $e\in E$, $id\neq
g\in G$, and leave the treatment of a non-free action on the edges
for later (section \ref{subsec:non_free_action}). We may choose an
ordered basis $B=\left(b_{j}\right)_{j=1}^{d}$ for $R$, with respect
to which we think of it as a matrix representation. We choose
representatives $\left\{ \tilde{e}^{i}\right\} _{i=1}^{I}$ for the
orbits $\nicefrac{E}{G}$, and likewise $\left\{
\tilde{v}_{k}\right\} _{k=1}^{K}$ for $\nicefrac{V}{G}$. We shall
assume, by adding {}``dummy'' vertices (vertices of valency two with
Neumann boundary conditions) if needed, that $G$ does not carry any
vertex in $V$ to one of its neighbors. This ensures that no edge is
transformed onto itself in the opposite direction, which would force
us to take half the edge as a representative. The quotient graph
$\nicefrac{\Gamma}{R}$ obtained from these choices is defined to
have $\left\{ v_{k}\right\} _{k=1}^{K}$ as its set of vertices, and
$\left\{ e_{j}^{i}\right\} _{j=1..d}^{i=1..I}$ for edges, where each
$e_{j}^{i}$ is of length $l_{\tilde{e}^{i}}$. The vertices $v_{k}$,
$v_{k'}$ of $\nicefrac{\Gamma}{R}$ are connected by the edge
$e_{j}^{i}$ if there exist $g,g'\in G$ such that $\tilde{e}^{i}$
connects $g\tilde{v}_{k}$ to $g'\tilde{v}_{k'}$ in $\Gamma$. In such
a case the vertices $v_{k}$, $v_{k'}$ are connected by all the edges
$\left\{ e_{j}^{i}\right\} _{j=1}^{d}$. Until now we have used only
the action of the group $G$ on $\Gamma$ in order to determine the
edges and vertices of the quotient graph and its connectivity. Now
we also need to use the information from $R$, $B$, and the boundary
conditions of $\Gamma$, in order to specify the boundary conditions
at each vertex $v_{k}$ in $\nicefrac{\Gamma}{R}$. Let the boundary
conditions at $\tilde{v}_{k}$ in $\Gamma$ be described by
$A_{\tilde{v}_{k}},\,B_{\tilde{v}_{k}}$. From here to the end of
this section we focus on the vertex $v_k$ and its predecessor
$\tilde{v}_k$, and keep in mind that our parameters depend on $k$
even when it is not reflected by the notations. The set of edges
incident to the vertex $\tilde{v}_{k}$ can be written as
$E_{\tilde{v}_{k}}=\left\{ g_{l}\tilde{e}^{\nu_{l}}\right\}
_{l=1}^{n}$, for some $\left\{ g_{l}\right\} _{l=1}^{n}$ in G. Note
that repetitions can occur among the $\nu_{l}$'s, and also among the
$g_{l}$'s, and that their total number is $n=d_{\tilde{v}_{k}}$. The
set of edges entering $v_{k}$ is $E_{v_{k}}=\left\{
e_{j}^{\mu_{l'}}\right\} _{{1\leq l'\leq m\atop 1\leq j\leq d}}$,
where $\left\{ \mu_{l'}\right\} _{l'=1}^{m}$ is defined to be the
set of distinct values among $\left\{ \nu_{l}\right\} _{l=1}^{n}$.
Obviously, $m\leq n$, and the relation between the sets $\left\{
\mu_{l'}\right\} _{l'=1}^{m}$, $\left\{ \nu_{l}\right\} _{l=1}^{n}$
is given by the $n\times m$ matrix
\begin{equation*}
\Theta'_{ll'}=\left\{\begin{array}{ll} 1 &\text{$\nu_{l}=\mu_{l'}$}\\
0 &\text{otherwise}\end{array}\right. .\end{equation*}
 The resulting matrices (see \cite{PB08}), describing the boundary
conditions at $v_{k}$, are: \begin{eqnarray}
\fl A_{v_{k}} & = & \left(A_{\tilde{v}_{k}}\otimes I_{d}\right)\cdot{\rm diag}\left(\left[\rho_{R}\left(g_{{1}}^{-1}\right)\right]_{B},\ldots,\left[\rho_{R}\left(g_{{n}}^{-1}\right)\right]_{B}\right)^{T}\cdot\left(\Theta'\otimes I_{d}\right)\label{eq:final-boundary-conditions-1}\\
\fl B_{v_{k}} & = & \left(B_{\tilde{v}_{k}}\otimes
I_{d}\right)\cdot{\rm
diag}\left(\left[\rho_{R}\left(g_{{1}}^{-1}\right)\right]_{B},\ldots,\left[\rho_{R}\left(g_{{n}}^{-1}\right)\right]_{B}\right)^{T}\cdot\left(\Theta'\otimes
I_{d}\right)\,.\label{eq:final-boundary-conditions-2}\end{eqnarray}
 Observe that the boundary conditions (\ref{eq:final-boundary-conditions-1}),
(\ref{eq:final-boundary-conditions-2}) encapsulate the information
of the boundary conditions on $\Gamma$ (given by
$A_{\tilde{v}_{k}},B_{\tilde{v}_{k}}$), the action of the group
($\Theta'$ and $\left\{ g_{i}\right\} $), the representation
($\rho_{R}$) and the basis that was chosen for the representation
($B$). However, they may fail to be square matrices as our
definition of a quantum graph calls for. We therefore need to
discuss the question of how many linearly independent restrictions
are dictated by those boundary conditions, i.e., find ${\rm
rank}\left(A_{v_{k}}\, B_{v_{k}}\right)$. Let us assume that the
original boundary conditions on $\tilde{v}_{k}$ were linearly
independent, i.e.
$\left(A_{\tilde{v}_{k}}\,B_{\tilde{v}_{k}}\right)$ is of maximal
rank, $d_{\tilde{v}_{k}}$. If the action of $G$ is free not only on
the edges but also on the vertices, then $\Theta'$ is a permutation
matrix and we get that $A_{v_{k}}$ and $B_{v_{k}}$ are square
matrices. Furthermore, ${\rm rank}\left(A_{v_{k}}\,
B_{v_{k}}\right)=d\cdot\mathrm{rank}\left(A_{\tilde{v}_{k}}\,B_{\tilde{v}_{k}}\right)=d\cdot
d_{\tilde{v}_{k}}=d_{v_{k}}$, which means that $\left(A_{v_{k}}\,
B_{v_{k}}\right)$ is also of maximal rank.

In the general case, it is shown in \cite{PB08} that if the matrix
$A_{\tilde{v}_{k}}\cdot B_{\tilde{v}_{k}}^{\dagger}$ is self-adjoint
then ${\rm rank}\left(A_{v_{k}}\, B_{v_{k}}\right)=d_{v_{k}}$.
 This means that we may eliminate rows from
$\left(A_{v_{k}}\, B_{v_{k}}\right)$, and remain with square
$A_{v_{k}},\, B_{v_{k}}$ which still describe the same boundary
conditions. Therefore, starting from a graph $\Gamma$ whose
Laplacian is self-adjoint, any $\nicefrac{\Gamma}{R}$ produced by
this construction method would be a valid quantum graph (whether the
Laplacian on $\nicefrac{\Gamma}{R}$ is also self-adjoint is
discussed in section \ref{subsec:self-adjoint}). This also
demonstrates the benefits of using quantum graphs to implement the
quotient construction. Although one may apply the above procedure to
other types of objects (see section \ref{sec:drums}), the resulting
quotients might not be objects of the same type as their
predecessors. The advantage of quantum graphs in this context is
that the self-adjointness condition guarantees that the quotient is
a quantum graph as well.

We now recall the intimate relation between the construction of
$\nicefrac{\Gamma}{R}$ and the encoding process, which for each
$\lambda\in\mathbb{C}$ takes functions
$\widetilde{f}_1,\ldots,\widetilde{f}_d\in\Phi_{\Gamma}(\lambda)$
that transform according to $R$, and maps them to a single function
$f\in\Phi_{\nicefrac{\Gamma}{R}}(\lambda)$. This encoding is
described by: \begin{eqnarray} f\at_{e_{j}^{i}} & \equiv &
\widetilde{f}_j\at_{\tilde{e}^{i}}.\label{eq:encoding}\end{eqnarray}
 This relation gives the decoding process as well $-$ starting from
$f\in\Phi_{\nicefrac{\Gamma}{R}}(\lambda)$, one reconstructs the
values of $\tilde{f}_1,\ldots,\tilde{f}_d\in\Phi_{\Gamma}(\lambda)$
on the edges' representatives $\left\{ {\tilde{e}^{i}}\right\} $.
The values on the rest of the edges are then determined by the
action of $G$ and the representation $R$.

We elaborate on the functions
$\tilde{f}_1,\ldots,\tilde{f}_d\in\Phi_{\Gamma}(\lambda)$ that play
a role in the decoding-encoding process. For two representations $R,
R'$ with corresponding carrier spaces $V, V'$, ${\rm
Hom}_{G}\left(V, V'\right)$ is the vector space of all linear
transformations $T:V\rightarrow V'$
 that respect the action of the
group $G$, i.e., $\forall g\in G, v\in V:\;
T(\rho_{R}(g)\,v)=\rho_{R'}(g)\,T(v)$. Such linear transformations
are called intertwiners of the representations $R,\,R'$. It is known
that $\left\langle
\chi_{R},\chi_{\Phi_{\Gamma}\left(\lambda\right)}\right\rangle
_{G}=\dim{\rm Hom}_{G}\left(R, \Phi_{\Gamma}\left(\lambda\right)
\right)$ (\cite{Curtis}, section 9C) and we recall that the quotient
graph's construction guarantees
$\sigma_{\nicefrac{\Gamma}{R}}(\lambda)=\left\langle
\chi_{R},\chi_{\Phi_{\Gamma}\left(\lambda\right)}\right\rangle
_{G}$. We therefore have that
$\Phi_{\nicefrac{\Gamma}{R}}(\lambda)\cong {\rm Hom}_{G}\left(R,
\Phi_{\Gamma}\left(\lambda\right) \right)$, which offers another
point of view on our quotient graphs. We demonstrate this by an
example. Let $G$ be a group and $R=S\oplus S\oplus S$, for some
irreducible representation $S$ of $G$, and let
$\lambda\in\mathbb{C}$ be such that $\Phi_{\Gamma}(\lambda)\cong
V_{1}^{S}\oplus V_{2}^{S}$, where $V_{1}^{S}, \,V_{2}^{S}$ are
carrier spaces of $S$. Since $\dim{\rm Hom}_{G}\left(R,
\Phi_{\Gamma}\left(\lambda\right) \right)=\left\langle
\chi_{R},\chi_{\Phi_{\Gamma}\left(\lambda\right)}\right\rangle
_{G}=\left\langle 3\chi_{S},2\chi_{S}\right\rangle _{G}=6$, we can
choose six linearly independent intertwiners from $V^{R}$ to
$\Phi_{\Gamma}\left(\lambda\right)$. One such choice of intertwiners
can be described as follows. We decompose a carrier space of $R$,
$V^R$, into three carrier spaces of $S$ and each intertwiner then
sends one of these three copies onto either
$V_1^S$ or $V_2^S$, and the other two copies to zero. \\
The quotient $\nicefrac{\Gamma}{R}$ is constructed with respect to a
certain basis $B$ of $R$. The image of this basis by each of the
above six intertwiners is a set of $d={\rm dim}\,R$ functions,
$\tilde{f}_1,\ldots,\tilde{f}_d\in\Phi_{\Gamma}(\lambda)$ which
transform according to $R$ \footnote{Note that in this example, this
set of $d$ functions is not a linearly independent set.}, and this
set can be encoded by a function
$f\in\Phi_{\nicefrac{\Gamma}{R}}(\lambda)$ according to
(\ref{eq:encoding}). We thus obtain six linearly independent
eigenfunctions of the eigenvalue $\lambda$ on the quotient
$\nicefrac{\Gamma}{R}$, which demonstrates the equivalence
$\Phi_{\nicefrac{\Gamma}{R}}(\lambda)\cong {\rm Hom}_{G}\left(R,
\Phi_{\Gamma}\left(\lambda\right) \right)$.

\section{Application of the method $-$ Further investigation}

\label{sec:further_investigation} Having revealed the key elements
of the construction method, we are ready to present its theoretical
implications, as well as various issues that may interest those who
wish to apply it.

\subsection{Sunada's isospectral theorem}

\label{subsec:sunada} In his famous paper \cite{Sunada}, Sunada
provides a method for producing isospectral Riemannian manifolds.
Phrasing his result somewhat loosely, given a manifold $M$ equipped
with a free action of a finite group $G$, and subgroups $H_{1},\,
H_{2}$ of $G$ which are \emph{almost conjugate}, the manifolds
$\nicefrac{M}{H_{1}}$ and $\nicefrac{M}{H_{2}}$ are isospectral.
$H_{1}$ and $H_{2}$ are said to be almost conjugate if $\,\,\forall
g\in G\quad|\{[g]\cap H_{1}\}|=|\{[g]\cap H_{2}\}|$, where $[g]$
denotes the conjugacy class of $g$ in G. As mentioned and proved in
\cite{Brooks-Sun}, $H_{1}$ and $H_{2}$ are almost conjugate iff
$\ind_{H_{1}}^{G}\mathbf{1}_{H_{1}}\cong\ind_{H_{2}}^{G}\mathbf{1}_{H_{2}}$,
where $\mathbf{1}_{H_{i}}$ is the trivial representation of $H_{i}$.
Comparing this to corollary \ref{cor:sunadapair}, we see that
Sunada's theorem is obtained as a particular case of the corollary
if $\nicefrac{M}{H_{i}}$ is isospectral to
$\nicefrac{M}{\mathbf{1}_{H_{i}}}$. This is indeed the case since
the space of functions on $\nicefrac{M}{H_{i}}$ is isomorphic to the
space of functions on $M$ which are invariant under the action of
$H_{i}$. But the latter is the trivial component of the space of
functions on $M$, which by definition is isomorphic to the space of
functions on $\nicefrac{M}{\mathbf{1}_{H_{i}}}$, so that \[
\forall\lambda\in\mathbb{C}\quad\Phi_{\nicefrac{M}{H_{i}}}(\lambda)\cong\Phi_{M}^{\mathbf{1}_{H_{i}}}(\lambda)\cong\Phi_{\nicefrac{M}{\mathbf{1}_{H_{i}}}}(\lambda),\]
as claimed.
 During the discussion above we have applied our isospectral theory
to manifolds, as opposed to quantum graphs, for which it was
developed in the previous sections. This should not bother us - the
method can be applied to other geometric objects, as will be
demonstrated in section \ref{sec:drums}. Finally, we note that the
equivalence of
$\ind_{H_{1}}^{G}\mathbf{1}_{H_{1}}\cong\ind_{H_{2}}^{G}\mathbf{1}_{H_{2}}$
and $H_{1},H_{2}$ being almost conjugate was already used by Pesce
\cite{Pesce} to give another proof for Sunada's theorem. Our
application of Frobenius Reciprocity for the proof of theorem
\ref{thm:mainthm} is similar to that of Pesce, whose proof is also
summarized comprehensibly in \cite{Brooks-Sun}.

\subsection{A non-free action on the edges}

\label{subsec:non_free_action} We now treat the case of a group
whose action on the edges is not free. We have already encountered
examples of non-free action on the vertices, and saw that this may
result in the vertex in the quotient graph having a degree smaller
than that of its predecessor. An action which is not free on the
edges results in some more interesting features. Denote by $\Gamma$
the quantum graph which is an equilateral tetrahedron equipped with
Neumann boundary conditions at all its vertices (figure
\ref{fig:tetrahedron}).
\begin{center}
\begin{figure}[!h]
\begin{centering}
\begin{minipage}[c][1\totalheight][t]{0.5\columnwidth}%
\begin{center}
\includegraphics[width=0.9\columnwidth]{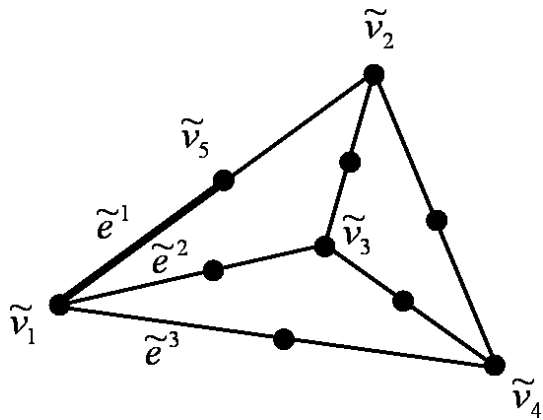}
\par\end{center}%
\end{minipage}
\par\end{centering}

\caption{An equilateral tetrahedron with Neumann boundary
conditions. It is equipped with an action of $S_{4}$ permuting its
vertices
$\tilde{v}_{1},\,\tilde{v}_{2},\,\tilde{v}_{3},\,\tilde{v}_{4}$.}

\label{fig:tetrahedron}
\end{figure}
\par\end{center}
Note that we have added {}``dummy'' vertices in the middle of the
edges, as described in the previous section. $S_{4}$ acts by
permuting the vertices
$\tilde{v}_{1},\,\tilde{v}_{2},\,\tilde{v}_{3},\,\tilde{v}_{4}$, and
this determines its action on the whole of $\Gamma$. Let $S$ denote
the sign representation of $S_{4}$: \begin{equation}
\forall\sigma\in
S_{4}\quad\rho_{S}\left(\sigma\right)=\left(\mathrm{sgn}\,\sigma\right)\,.\label{eq:sign_rep}\end{equation}
 We choose the edge $\tilde{e}^{1}$ and the vertices $\tilde{v}_{1},\,\tilde{v}_{5}$
as a fundamental domain for the action of $S_{4}$. A function
$\tilde{f}$ which transforms according to $S$ satisfies
$(3\,4)\,\tilde{f}=\rho_{S}\left[(3\,4)\right]\tilde{f}=-\tilde{f}$.
Evaluating this on the edge $\tilde{e}^{1}$ gives
$\tilde{f}\at_{\tilde{e}^{1}}=\tilde{f}\at_{(3\,4)\tilde{e}^{1}}=\left((3\,4)\tilde{f}\right)\at_{\tilde{e}^{1}}=-\tilde{f}\at_{\tilde{e}^{1}}$,
which means that $\tilde{f}$ vanishes on $\tilde{e}^{1}$ and
therefore also on the whole of $\Gamma$. We conclude that $\Gamma$'s
Laplacian has no eigenfunctions which transform according to $S$, so
that $\nicefrac{\Gamma}{S}$ is just the empty graph. Putting this
another way, the edge $\tilde{e}^{1}$ of $\Gamma$ was chosen as a
representative for the orbits of the edges under the action of the
group, $\nicefrac{E}{S_{4}}$. We therefore expected to have a single
edge $e_{1}^{1}$ in the quotient. However, this edge has disappeared
due to the action of the stabilizer of $\tilde{e}^{1}$.
Specifically, the stabilizer is $G_{\tilde{e}^{1}}=\left\{
id,(3\,4)\right\} $ and the restriction of the representation on it
is not trivial: $\rho_{S}\at_{G_{\tilde{e}^{1}}}\neq1$. This caused
the disappearance.

\begin{figure}[!h]

\begin{centering}
\hfill{}%
\begin{minipage}[c][1\totalheight]{0.58\columnwidth}%
\begin{center}
\includegraphics[scale=0.6]{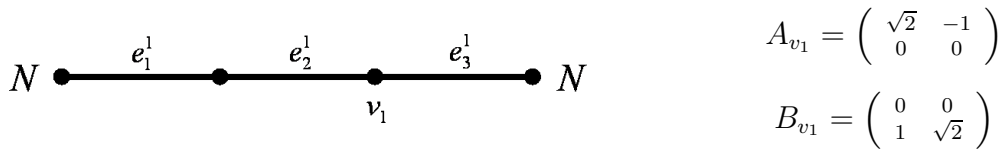}
\par\end{center}%
\end{minipage}\hfill{}%
\begin{minipage}[c][1\totalheight]{0.4\columnwidth}%
\begin{center}
$A_{v_{1}}=\scriptsize{\left(\begin{array}{cc}\sqrt{2} & -1\\
0 & 0\end{array}\right)}$
\par\end{center}

\begin{center}
$B_{v_{1}}=\scriptsize{\left(\begin{array}{cc}0 & 0\\
1 & \sqrt{2}\end{array}\right)}$
\par\end{center}%
\end{minipage}\hfill{}
\par\end{centering}

\caption{The quotient $\nicefrac{\Gamma}{S}$. The only non-Neumann
boundary condition is at $v_{1}$ and it is specified by the matrices
$A_{v_{1}},B_{v_{1}}$.}

\label{fig:tetrahedron_quotient}
\end{figure}

Moving to a more complex demonstration of this principle, we examine
the permutation representation of $S_{4}$, which we denote by $R$.
Some of the matrices of $R$ using its standard basis are:
\begin{equation} \left\{ \begin{array}{ll}
(1\,2)\mapsto\scriptsize{\left(\begin{array}{cccc}0 & 1 & 0 & 0\\
1 & 0 & 0 & 0\\
0 & 0 & 1 & 0\\
0 & 0 & 0 & 1\end{array}\right)}, & (3\,4)\mapsto\scriptsize{\left(\begin{array}{cccc}1 & 0 & 0 & 0\\
0 & 1 & 0 & 0\\
0 & 0 & 0 & 1\\
0 & 0 & 1 & 0\end{array}\right)}\\
\\(2\,3\,4)\mapsto \scriptsize{\left(\begin{array}{cccc}1 & 0 & 0 & 0\\
0 & 0 & 0 & 1\\
0 & 1 & 0 & 0\\
0 & 0 & 1 & 0\end{array}\right)}, & (2\,4\,3)\mapsto
\scriptsize{\left(\begin{array}{cccc}1 & 0 & 0 & 0\\
0 & 0 & 1 & 0\\
0 & 0 & 0 & 1\\
0 & 1 & 0 & 0\end{array}\right)}\end{array}\right\}
.\label{eq:s4_rep1}\end{equation} We choose the same fundamental
domain and consider the values of the functions
$\tilde{f}_1,\tilde{f}_2,\tilde{f}_3,\tilde{f}_4$ (which transform
according to (\ref{eq:s4_rep1})) on it. We now have four edges,
$e^1_1,e^1_2,e^1_3,e^1_4$, that are supposed to form the quotient
graph after we properly glue them and deduce the boundary
conditions. From the action of $(2\,3\,4)$ and $(2\,4\,3)$ on
$\tilde{f}_1$, the chosen matrix representation for $R$, and the
fact that $\tilde{v}_{1}$ has Neumann boundary conditions, it
follows that $\tilde{f}_1'\at_{\tilde{e}^{1}}(\tilde{v}_{1})=0$.
Likewise, from the action of $(1\,2)$ on $\tilde{f}_1$ we conclude
that
$\tilde{f}_1\at_{\tilde{e}^{1}}(\tilde{v}_{5})=\tilde{f}_2\at_{\tilde{e}^{1}}(\tilde{v}_{5})$
and
$\tilde{f}_1'\at_{\tilde{e}^{1}}(\tilde{v}_{5})+\tilde{f}_2'\at_{\tilde{e}^{1}}(\tilde{v}_{5})=0$
(where the derivatives are outgoing from $\tilde{v}_{5}$). This
allows us glue the edges $e_{1}^{1},\, e_{2}^{1}$ and deduce a
Neumann condition on the left side of $e_{1}^{1}$ (figure
\ref{fig:tetrahedron_quotient}). These considerations are similar to
those used before, and the new part comes when we observe that
$(3\,4)\tilde{f}_3=\tilde{f}_4$. This gives
$\tilde{f}_3\at_{\tilde{e}^{1}}=\tilde{f}_4\at_{\tilde{e}^{1}}$ and
means that we do not need both of the corresponding edges (i.e.,
$e_{3}^{1}$ and $e_{4}^{1}$) in the quotient graph \footnote{This is
because the quotient construction is motivated by encoding the
functions on $\Gamma$ which transform according to $R$, without
redundancies.}. We notice that the problem arises since the
representation restricted to the stabilizer
$G_{\tilde{e}^{1}}=\left\{ id,(3\,4)\right\} $ has a non trivial
component ($\rho_{R}\left[(3\,4)\right]$ has an eigenvalue not equal
to one). The solution for this situation is to change our basis in
the following way: \[
\begin{array}{ll}
\hat{f}_1=\tilde{f}_1, & \hat{f}_3=\frac{1}{\sqrt{2}}\left(\tilde{f}_3+\tilde{f}_4\right)\\
\hat{f}_2=\tilde{f}_2, &
\hat{f}_4=\frac{1}{\sqrt{2}}\left(\tilde{f}_3-\tilde{f}_4\right)\,,\end{array}\]
 This forces us to rewrite (\ref{eq:s4_rep1}) as \begin{equation}
\fl \left\{ \begin{array}{ll}
(1\,2)\mapsto\scriptsize{\left(\begin{array}{cccc}0 & 1 & 0 & 0\\
1 & 0 & 0 & 0\\
0 & 0 & 1 & 0\\
0 & 0 & 0 & 1\end{array}\right)}, & (3\,4)\mapsto\scriptsize{\left(\begin{array}{cccc}1 & 0 & 0 & 0\\
0 & 1 & 0 & 0\\
0 & 0 & 1 & 0\\
0 & 0 & 0 & -1\end{array}\right)}\\
\\(2\,3\,4)\mapsto\scriptsize{\left(\begin{array}{cccc}1 & 0 & 0 & 0\\
0 & 0 & \nicefrac{1}{\sqrt{2}} & -\nicefrac{1}{\sqrt{2}}\\
0 & \nicefrac{1}{\sqrt{2}} & \nicefrac{1}{2} & \nicefrac{1}{2}\\
0 & \nicefrac{1}{\sqrt{2}} & -\nicefrac{1}{2} &
-\nicefrac{1}{2}\end{array}\right)}, & (2\,4\,3)\mapsto
\scriptsize{\left(\begin{array}{cccc}1 & 0 & 0 & 0\\
0 & 0 & \nicefrac{1}{\sqrt{2}} & \nicefrac{1}{\sqrt{2}}\\
0 & \nicefrac{1}{\sqrt{2}} & \nicefrac{1}{2} & -\nicefrac{1}{2}\\
0 & -\nicefrac{1}{\sqrt{2}} & \nicefrac{1}{2} &
-\nicefrac{1}{2}\end{array}\right)}
\end{array}\right\}
.\label{eq:s4_rep2}\end{equation}
 Examining the action of $(3\,4)$ on $\hat{f}_4$, we get $\hat{f}_4\at_{\tilde{e}^{1}}\equiv0$
and this turns out to be the edge which disappears. The previous
considerations we made concerning $\tilde{f}_1\at_{\tilde{e}^{1}}$
and $\tilde{f}_2\at_{\tilde{e}^{1}}$ are still valid since
$\hat{f}_1=\tilde{f}_1,\,\hat{f}_2=\tilde{f}_2$. The last step in
the gluing process follows from the second column of the matrix
representations of $(2\,3\,4),\,(2\,4\,3)$, and the evaluation of
the corresponding relations on $\tilde{e}^{1}$:
\begin{eqnarray}
\hat{f}_2\at_{\tilde{e}^{3}}= & \left((2\,3\,4)\,\hat{f}_2\right)\at_{\tilde{e}^{1}} & =\nicefrac{1}{\sqrt{2}}\,\hat{f}_3\at_{\tilde{e}^{1}}+\nicefrac{1}{\sqrt{2}}\,\hat{f}_4\at_{\tilde{e}^{1}}\label{eq:v1_bc1}\\
\hat{f}_2\at_{\tilde{e}^{2}}= &
\left((2\,4\,3)\,\hat{f}_2\right)\at_{\tilde{e}^{1}} &
=\nicefrac{1}{\sqrt{2}}\,\hat{f}_3\at_{\tilde{e}^{1}}-\nicefrac{1}{\sqrt{2}}\,\hat{f}_4\at_{\tilde{e}^{1}}\label{eq:v1_bc2}\end{eqnarray}
 Recalling that $\hat{f}_4\at_{\tilde{e}^{1}}\equiv0$ and using
(\ref{eq:v1_bc1}), (\ref{eq:v1_bc2}) to obtain relations for the
derivatives as well, we finish the process and give the quotient
$\nicefrac{\Gamma}{S}$ that is shown in figure
\ref{fig:tetrahedron_quotient}.

We now return to the general construction procedure described in the
previous section, and discuss the adjustments that should be made in
order to apply it to the case of a non-free action on the edges.
After choosing the representatives $\left\{ \tilde{e}^{i}\right\}
_{i=1}^{I}$ for the orbits $\nicefrac{E}{G}$, we consider for each
$i$ the representation $\res_{G_{\tilde{e}^{i}}}^{G}S$ with its
carrier space $V^{i}$. We decompose $V^{i}$ into its trivial
component, $\left(V^{i}\right)^{\mathbf{1}_{G_{\tilde{e}^{i}}}}$,
and its complement, $\bigoplus\limits
_{S\ncong\mathbf{1}_{G_{\tilde{e}^{i}}}}\left(V^{i}\right)^{S}$. We
denote the dimension of the trivial component of $V^{i}$ by $d_{i}$
and choose a basis for it: $\left\{ b_{j}^{i}\right\}
_{j=1}^{d_{i}}$. We complete this to a basis for the whole of
$V^{i}$ by adding vectors from the complement of
$\left(V^{i}\right)^{\mathbf{1}_{G_{\tilde{e}^{i}}}}$, and we denote
this basis, $\left\{ b_{j}^{i}\right\} _{j=1}^{d}$, by $B^{i}$. Let
$\left\{ \widetilde{f}^i_j\at_{\tilde{e}^{i}}\right\} _{j=1}^{d}$ be
functions on the edge $\tilde{e}^{i}$ which transform according to
the representation $R$ as given by the basis $B^{i}$. As in the
preceding examples we have that $\forall j>d_{i},\;\;
\widetilde{f}^i_j\at_{\tilde{e}^{i}}=0$ \footnote{This can be
deduced from observation \eref{eq:sum_of_all_group_elements}.}.
Therefore, in the assembly of the quotient, we endow it with only
$d_{i}$ copies of the edge $\tilde{e}^{i}$ $-$ the copies which
correspond to the functions $\left\{
\widetilde{f}^i_j\at_{\tilde{e}^{i}}\right\} _{j=1}^{d_{i}}$.
Equations (\ref{eq:final-boundary-conditions-1}),
(\ref{eq:final-boundary-conditions-2}) are replaced in the general
case by the following: \begin{eqnarray}
A_{v_{k}} & = & \left(A_{\tilde{v}_{k}}\otimes I_{d}\right)\cdot{\rm diag}\left(\left[\rho_{R}\left(g_{{1}}^{-1}\right)\right]_{B^{\nu_{1}}}^{B},\ldots,\left[\rho_{R}\left(g_{{n}}^{-1}\right)\right]_{B^{\nu_{n}}}^{B}\right)^{T}\cdot\Theta\label{eq:final-boundary-conditions-3}\\
B_{v_{k}} & = & \left(B_{\tilde{v}_{k}}\otimes I_{d}\right)\cdot{\rm
diag}\left(\left[\rho_{R}\left(g_{{1}}^{-1}\right)\right]_{B^{\nu_{1}}}^{B},\ldots,\left[\rho_{R}\left(g_{{n}}^{-1}\right)\right]_{B^{\nu_{n}}}^{B}\right)^{T}\cdot\Theta\label{eq:final-boundary-conditions-4}\end{eqnarray}
 One obvious difference is the use of a separate basis $B^{i}$ for
each edge that enters the vertex. The other change is due to
$\Theta$ which is defined to be the $nd\times d_{v_{k}}$ matrix
obtained by removing from $\left(\Theta'\otimes I_{d}\right)$ the
columns $\left\{ \left(i-1\right)\cdot d+j\right\} _{{1\leq i\leq
m\atop d_{\mu_{i}}<j\leq d}}$; these are the columns which represent
the functions whose vanishing on the corresponding edge
($\tilde{e}^{i}$) of the graph $\Gamma$ causes the disappearance of
($d-d_{i}$) copies of that edge from the quotient. Note that
\eref{eq:final-boundary-conditions-1},
\eref{eq:final-boundary-conditions-2} are obtained as a special case
of \eref{eq:final-boundary-conditions-3},
\eref{eq:final-boundary-conditions-4} when the action is free on the
edges and a single choice of basis is made (i.e.,
$B^{1}=\ldots=B^{I}=B$). In addition, the
encoding$\setminus$decoding process in this case is then slightly
altered from the one described by \eref{eq:encoding}. For each
$\lambda \in \mathbb{C}$ and $1\leq i \leq I$, a set of functions
$\widetilde{f}^i_1,\ldots,\widetilde{f}^i_d\in\Phi_{\Gamma}(\lambda)$
that transform according to the basis $B^i$ of $R$ is mapped into a
single function $f\in\Phi_{\nicefrac{\Gamma}{R}}(\lambda)$ by:
\begin{eqnarray}
\forall 1\leq j \leq d_i, \;\; f\at_{e_{j}^{i}} & \equiv &
\widetilde{f}^i_j\at_{\tilde{e}^{i}}.\label{eq:encoding_nonfree}\end{eqnarray}
Note that the functions
$\left\{\widetilde{f}^i_j\at_{\tilde{e}^{i}}\right\}_{ 1\leq i \leq
I \atop{d_i< j \leq d}}$ are zero and therefore not encoded.
Finally, the discussion that comes after equations
(\ref{eq:final-boundary-conditions-1}),
(\ref{eq:final-boundary-conditions-2}) regarding the rank of
$A_{v_{k}}B_{v_{k}}$ is valid here as well: whenever the Laplacian
on $\Gamma$ is self-adjoint, the produced quotient obeys the
definition of a quantum graph.

\subsection{The self-adjointness of the Laplacian on the quotient graph}

\label{subsec:self-adjoint} A natural question to ask is whether the
Laplacian on the quotient graph produced by the above construction
is self-adjoint. The necessary and sufficient conditions for the
self-adjointness of a quantum graph were described in section
\ref{sec:graphs_intro}. Their examination in light of
(\ref{eq:final-boundary-conditions-3}),
(\ref{eq:final-boundary-conditions-4}) gives the following:

\begin{prop}
\label{prop:self-adj} Let $\nicefrac{\Gamma}{R}$ be a quotient
quantum graph constructed as explained in section
(\ref{subsec:non_free_action}).
\begin{enumerate}
\item If $\Gamma$'s Laplacian is self-adjoint, $G$ acts freely on $\Gamma$
(both on the edges and the vertices), and
$\left[\rho_{R}\left(g\right)\right]_{B}^{B^{i}}$ is unitary for all
$g\in G$, $1\leq i\leq I$, then the Laplacian on
$\nicefrac{\Gamma}{R}$ is self-adjoint.
\item If $\Gamma$ has Neumann boundary conditions and $\left[\rho_{R}\left(g\right)\right]_{B}^{B^{i}}$
is unitary for all $g,i$, then the Laplacian on
$\nicefrac{\Gamma}{R}$ is self-adjoint if and only if for every
vertex $\tilde{v}$ at least one of the following holds:

\begin{enumerate}
\item \textup{$\left\langle \chi_{R},\mathbf{1}\right\rangle _{G_{\tilde{v}}}=0$.}
\item All stabilizers $\left\{ G_{\tilde{e}}\right\} _{\tilde{e}\in E_{\tilde{v}}}$
are of equal order.
\end{enumerate}
\end{enumerate}
\end{prop}
\begin{rem*}
Since $G$ is assumed to be finite, one dimensional representations
are unitary in all bases.
\end{rem*}
\begin{proof}
We start by recalling the observations made in sections
\ref{sec:rigorous}, \ref{subsec:non_free_action}. When $\Gamma$ has
a self-adjoint Laplacian, our construction ensures that for every
vertex $v$ of the quotient we have ${\rm rank}\left(A_{v}\,
B_{v}\right)=d_{v}$. Therefore, one condition required for the
self-adjointness of the quotient is fulfilled, and we are left to
verify the other, $A_{v}\cdot B_{v}^{\dagger}=B_{v}\cdot
A_{v}^{\dagger}$.
\begin{enumerate}
\item Let $\tilde{v}$ be a vertex in $\Gamma$, with $E_{\tilde{v}}=\left\{ g_{i}\tilde{e}^{\nu_{i}}\right\} _{i=1}^{n}$,
and $v$ the corresponding vertex in $\nicefrac{\Gamma}{R}$. Since
the action is free both on the edges and on the vertices we have
$\Theta'=I_{n}$ (by appropriate indexing of the edges) and
$\Theta=\Theta'\otimes I_{d}=I_{d_{v}}$ (since the free action also
implies $d_{v}=d_{\tilde{v}}\cdot d=n\cdot d$). The unitarity of the
representation with respect to the bases $B^{i}$ and $B$ gives that
$D:={\rm
diag}\left(\left[\rho_{R}\left(g_{{1}}^{-1}\right)\right]^{B}_{B^{\nu_{1}}},\ldots,\left[\rho_{R}\left(g_{{n}}^{-1}\right)\right]^{B}_{B^{\nu_{n}}}\right)^{T}$
is a unitary matrix. Therefore, \begin{eqnarray*}
A_{v}\cdot B_{v}^{\dagger} & = & \left( A_{\tilde{v}}\otimes I_{d}\right)\cdot D\cdot I_{d_{v}}\cdot\left(\left( B_{\tilde{v}}\otimes I_{d}\right)\cdot D\cdot I_{d_{v}}\right)^{\dagger}\\
 & = & \left( A_{\tilde{v}}\otimes I_{d}\right)\cdot\left( B_{\tilde{v}}\otimes I_{d}\right)^{\dagger}=\left( A_{\tilde{v}}\cdot B_{\tilde{v}}^{\dagger}\right)\otimes I_{d}\\
 & = & \left( B_{\tilde{v}}\cdot A_{\tilde{v}}^{\dagger}\right)\otimes I_{d}=B_{v}\cdot A_{v}^{\dagger}\,\,.\end{eqnarray*}

\item Let $v$ be a vertex of $\nicefrac{\Gamma}{R}$, and $\tilde{v}$
its predecessor in $\Gamma$. Recall (section \ref{sec:rigorous})
that $E_{\tilde{v}}=\left\{ g_{i}\tilde{e}^{\nu_{i}}\right\}
_{i=1}^{n}$, and that $\left\{ \nu_{i}\right\} _{i=1}^{n}$ attain in
total $m$ distinct values, $\left\{ \mu_{i}\right\} _{i=1}^{m}$. We
assume (by reordering, if necessary) that $\left\{ \nu_{i}\right\}
_{i=1}^{m}$ are different: this means that $\left\{
g_{i}\tilde{e}^{\nu_{i}}\right\} _{i=1}^{m}$ are representatives for
the different orbits of the edges in $E_{\tilde{v}}.$ For each
$1\leq i\leq m$ we choose a set of representatives $\left\{
t_{i}^{j}\right\} _{j=1}^{r_{i}}$
($r_{i}=\left[G_{\tilde{v}}:G_{g_{i}\tilde{e}^{\nu_{i}}}\right]$)
for the left cosets of $G_{g_{i}\tilde{e}^{\nu_{i}}}$ in
$G_{\tilde{v}}$, so that when writing $E_{\tilde{v}}=\left\{
t_{i}^{j}g_{i}\tilde{e}^{\nu_{i}}\right\} _{{i=1..m\atop
j=1..r_{i}}}$ we obtain each edge in $E_{\tilde{v}}$ exactly once.
This is the ordering
 with which we represent the boundary
conditions at $\tilde{v}$.

Denoting
\begin{eqnarray*}
\mathfrak{g}_{i,j}=\left(\left[\rho_R\left(t_{i}^{j}g_{i}\right)^{-1}\right]_{B^{\nu_{i}}}^{B}\right)^{T}\,,\qquad\theta_{i}=\left(\begin{array}{c|c}
I_{d_{\nu^{i}}} & \vphantom{I_{d_{\nu^{i}}}}0\\
\hline \vphantom{I_{d_{\nu^{i}}}}0 &
\vphantom{I_{d_{\nu^{i}}}}0\end{array}\right)\,\in\, M_{d\times
d}\left(\mathbb{C}\right)\:, \\
\mathfrak{G}=\mathrm{diag}\left(\mathfrak{g}_{1,1},\ldots,\mathfrak{g}_{1,r_{1}},\mathfrak{g}_{2,1},\ldots\ldots,\mathfrak{g}_{m,r_{m}}\right),
\end{eqnarray*} we have
\begin{equation*}
\fl
\mathfrak{G}\Theta\Theta^{\dagger}\mathfrak{G}^{\dagger}=\scriptsize\left(\begin{array}{ccccccc}
\mathfrak{g}_{1,1}\theta_{1}\mathfrak{g}_{1,1}^{\dagger} & \cdots & \mathfrak{g}_{1,1}\theta_{1}\mathfrak{g}_{1,r_{1}}^{\dagger} & & & & \\
\vdots & \ddots & \vdots & & & &\\
\mathfrak{g}_{1,r_{1}}\theta_{1}\mathfrak{g}_{1,1}^{\dagger} & \cdots & \mathfrak{g}_{1,r_{1}}\theta_{1}\mathfrak{g}_{1,r_{1}}^{\dagger} & & & &\\
 &  &  & \ddots & & &\\
 &  &  &  & \mathfrak{g}_{m,1}\theta_{m}\mathfrak{g}_{m,1}^{\dagger} & \cdots & \mathfrak{g}_{m,1}\theta_{m}\mathfrak{g}_{m,r_{m}}^{\dagger}\\
 &  &  &  & \vdots & \ddots & \vdots\\
 &  &  &  & \mathfrak{g}_{m,r_{m}}\theta_{m}\mathfrak{g}_{m,1}^{\dagger} & \cdots & \mathfrak{g}_{m,r_{m}}\theta_{m}\mathfrak{g}_{m,r_{m}}^{\dagger}\end{array}\right)\:.
 \end{equation*}
The Neumann boundary conditions at $\tilde{v}$ can be expressed by
\[
A_{\tilde{v}}=\scriptsize{\left(\begin{array}{cccc}1 & -1\\
\vdots &  & \ddots\\
1 &  &  & -1\\
0 & \cdots & 0 & 0\end{array}\right)}\qquad B_{\tilde{v}}=\scriptsize{\left(\begin{array}{cccc}0 & 0 & \cdots &  0\\
\vdots & \vdots & \ddots &  \vdots\\
0 & 0 & \cdots &  0\\
1 & 1 & \cdots &  1\end{array}\right)}\,\] and using
(\ref{eq:final-boundary-conditions-3}),
(\ref{eq:final-boundary-conditions-4}) we have
$B_{v}A_{v}^{\dagger}=\left( B_{\tilde{v}}\otimes
I_{d}\right)\mathfrak{G}\Theta\Theta^{\dagger}\mathfrak{G}^{\dagger}\left(
A_{\tilde{v}}\otimes I_{d}\right)^{\dagger}$ and can write
$B_{v}A_{v}^{\dagger}$ in $d\times d$ blocks:
\[ \fl
B_{v}A_{v}^{\dagger}=\scriptsize\left(\begin{array}{cccccccc}\left[0\right] & \cdots & \left[0\right] & \left[0\right] & \cdots & \cdots & \left[0\right] & \left[0\right]\\
\vdots &  & \vdots & \vdots &  &  & \vdots & \vdots\\
\left[0\right] & \cdots & \left[0\right] & \left[0\right] & \cdots & \cdots & \left[0\right] & \left[0\right]\\
\left[M_{1,1}-M_{1,2}\right] & \cdots &
\left[M_{1,1}-M_{1,r_{1}}\right] & \left[M_{1,1}-M_{2,1}\right] &
\cdots & \cdots & \left[M_{1,1}-M_{m,r_{m}}\right] &
\left[0\right]\end{array}\right),\]
where for $1\leq i\leq m$, \, $1\leq j\leq r_{i}$ we denote
\begin{equation*}
M_{i,j}=\sum_{k=1}^{r_{i}}\mathfrak{g}_{i,k}\theta_{i}\mathfrak{g}_{i,j}^{\dagger}.
\end{equation*}
Due to the fact that $B_{v}A_{v}^{\dagger}$ is strictly
lower-triangular, it can only be self-adjoint if it is zero, i.e.,
if $M_{i,j}$ is the same for all $i,j$, and we wish to analyze when
this occurs.
 First, we note that\begin{equation}
\theta_{i}=\frac{1}{\left|G_{\tilde{e}^{\nu_{i}}}\right|}\sum\limits
_{g\in
G_{\tilde{e}^{\nu_{i}}}}\left[\rho_{R}\left(g\right)\right]_{B^{\nu_{i}}}^{B^{\nu_{i}}}\:,\label{eq:Delta_i_significance}\end{equation}
by the following observations:\\
\emph{(a) }For every irreducible representation $\rho_{S}$ of a
group $G$ we have
\begin{equation}
\frac{1}{\left|G\right|}\sum\limits _{g\in
G}\rho_{S}\left(g\right)=\left\{\begin{array}{ccc}
I & & S\cong\mathbf{1}_{G}\\
0 & & S\ncong\mathbf{1}_{G}\end{array}\right.
\label{eq:sum_of_all_group_elements}
\end{equation}
 \emph{(b) } As explained in section \ref{subsec:non_free_action}, we chose
$B^{\nu_{i}}=\left\{ b_{j}^{\nu_{i}}\right\} _{j=1}^{d}$ such that
$\left\{ b_{j}^{\nu_{i}}\right\} _{j=1}^{d_{\nu_{i}}}$ is a basis
for the trivial component of $\res^{G}_{G_{\tilde{e}^{\nu_{i}}}}R$,
and $\left\{ b_{j}^{\nu_{i}}\right\} _{j=d_{\nu_{i}}+1}^{d}$ is a
basis for $\bigoplus\limits
_{S\ncong\mathbf{1}_{G}}\left(\mathrm{Res}_{G_{\tilde{e}^{\nu_{i}}}}^{G}R\right)^{S}$.
\\
\\
By our unitarity assumption,
$\left(\left[\rho_R\left(t_{i}^{j}g_{i}\right)^{-1}\right]_{B^{\nu_{i}}}^{B}\right)^{\dagger}=\left[\rho_R\left(t_{i}^{j}g_{i}\right)\right]_{B}^{B^{\nu_{i}}}$,
so that \begin{eqnarray*} \overline{M_{i,j}} & = &
\sum_{k=1}^{r_{i}}\left[\rho_{R}\left(t_{i}^{k}g_{i}\right)\right]_{B}^{B^{\nu_{i}}}\theta_{i}\left[\rho_{R}\left(t_{i}^{j}g_{i}\right)^{-1}\right]_{B^{\nu_{i}}}^{B}\:.\end{eqnarray*}
Using (\ref{eq:Delta_i_significance}) we find\begin{eqnarray*}
\overline{M_{i,j}} & = & \frac{1}{\left|G_{\tilde{e}^{\nu_{i}}}\right|}\sum_{k=1}^{r_{i}}\left[\rho_{R}\left(t_{i}^{k}g_{i}\right)\right]_{B}^{B^{\nu_{i}}}\sum\limits _{g\in G_{\tilde{e}^{\nu_{i}}}}\left[\rho_{R}\left(g\right)\right]_{B^{\nu_{i}}}^{B^{\nu_{i}}}\left[\rho_{R}\left(t_{i}^{j}g_{i}\right)^{-1}\right]_{B^{\nu_{i}}}^{B}\\
 & = & \frac{1}{\left|G_{\tilde{e}^{\nu_{i}}}\right|}\sum_{k=1}^{r_{i}}\sum\limits _{g\in G_{\tilde{e}^{\nu_{i}}}}\left[\rho_{R}\left(t_{i}^{k}g_{i}g\left(t_{i}^{j}g_{i}\right)^{-1}\right)\right]_{B}^{B}\end{eqnarray*}
and since
$g_{i}G_{\tilde{e}^{\nu_{i}}}g_{i}^{-1}=G_{g_{i}\tilde{e}^{\nu_{i}}}$,\begin{eqnarray*}
\overline{M_{i,j}} & = & \frac{1}{\left|G_{\tilde{e}^{\nu_{i}}}\right|}\sum_{k=1}^{r_{i}}\sum\limits _{g\in G_{g_{i}\tilde{e}^{\nu_{i}}}}\left[\rho_{R}\left(t_{i}^{k}g\left(t_{i}^{j}\right)^{-1}\right)\right]_{B}^{B}\\
 & = & \frac{1}{\left|G_{\tilde{e}^{\nu_{i}}}\right|}\sum\limits _{g\in G_{\tilde{v}}}\left[\rho_{R}\left(g\left(t_{i}^{j}\right)^{-1}\right)\right]_{B}^{B}=\frac{1}{\left|G_{\tilde{e}^{\nu_{i}}}\right|}\sum\limits _{g\in G_{\tilde{v}}}\left[\rho_{R}\left(g\right)\right]_{B}^{B}\:,\end{eqnarray*}
as for every $i$ we have the disjoint union
$G_{\tilde{v}}=\bigcup_{k=1}^{r_{i}}t_{i}^{k}G_{g_{i}\tilde{e}^{\mu_{i}}}$,
and $t_{i}^{j}\in G_{\tilde{v}}$ for all $i$ and $j$. We therefore
have $M_{i,j}=M_{i',j'}$ iff either
$\left|G_{\tilde{e}^{\nu_{i}}}\right|=\left|G_{\tilde{e}^{\nu_{i'}}}\right|$
or $\sum_{g\in G_{\tilde{v}}}\rho_{R}\left(g\right)=0$. Observation
\eref{eq:sum_of_all_group_elements} shows that the latter happens if
and only if $\res^{G}_{G_{\tilde{v}}}R$ has no trivial component,
i.e., $\left\langle \chi_{R},\mathbf{1}\right\rangle
_{G_{\tilde{v}}}=0$.
\end{enumerate}
\end{proof}

\begin{cor}
\label{cor:SA-quotient-of-free-action}If $\Gamma$'s Laplacian is
self adjoint, and $G$ acts freely on $\Gamma$, then a quotient with
a self-adjoint Laplacian can be constructed for any representation
of $G$.
\end{cor}
\begin{proof}
This follows from the first part of the proposition; as the action
is free, we have no restrictions on the bases $\left\{B^i
\right\}_{i=1}^{I}$ used in the construction. By Weyl's unitary
trick, any finite dimensional representation of a finite group is
unitarisable (see for example \cite{Hall}) and therefore we can
choose a single basis $B$ with respect to which $R$ is a unitary
matrix representation, and take $B^{1}=\ldots=B^{I}=B$.
\end{proof}

\subsection{Transplantation}

\label{subsec:transplantation} The transplantation technique is one
of the simple and illuminating proofs for isospectrality. For two
isospectral objects, the transplantation is a linear map which
bijects each eigenspace of the first object's Laplacian onto the
corresponding eigenspace of the second. The existence of a
transplantation proves that the spectra of the two objects coincide,
including multiplicity. One usually encounters the transplantation
technique when dealing with isospectral objects consisting of
building blocks. Several copies of a certain fundamental building
block are pasted together along their boundaries to form the
complete object. The second isospectral object is composed of the
same building blocks, pasted in a different manner. The
transplantation then presents the restriction of an eigenfunction to
a building block of one object as a linear combination of such
restrictions to several building blocks of the other object. The
linearity of the eigenvalue problem ensures that the transplanted
function is indeed an eigenfunction of the second shape, as long as
it obeys the boundary conditions. The idea of transplantation is
further discussed and demonstrated in \cite{Buser,
Berard,Brooks-Sun,BuserConway,Jakobson,Levitin, Chapman,Okada}.

In order to discuss the notion of transplantation as it arises from
our method, it is crucial to examine the exact procedure by which
the isospectral quantum graphs are constructed. The main tools for
producing isospectral graphs are theorem \ref{thm:mainthm} and
corollary \ref{cor:sunadapair}. However, we have already noticed
that $\nicefrac{\Gamma}{R}$ is not a single graph, but rather a
continuum of isospectral graphs, each constructed by a certain
choice of basis for $R$. Therefore, isospectrality emerges also from
a change of basis for the representation. Furthermore, even when one
works with a single basis, $\nicefrac{\Gamma}{R}$ might depend on
the choice of representatives for the orbits $\nicefrac{E}{G}$,
$\nicefrac{V}{G}$. An example of producing isospectral graphs using
this freedom is shown in section \ref{subsec:freedom_of_repres}.
This variety of degrees of freedom (representations, bases,
representatives) motivates us to address each of them separately and
one can combine them appropriately in order to construct the
over-all transplantation.

\begin{prop}
\label{prop:transplantation} Let $\Gamma$ be a quantum graph
equipped with an action of a group $G$, and let $R$ be a
representation of $G$. Then any two $\nicefrac{\Gamma}{R}$ graphs
constructed according to the recipe given in sections
\ref{sec:rigorous}, \ref{subsec:non_free_action} are transplantable.
\end{prop}
\begin{proof}

We deal with the following degrees of freedom:
\begin{enumerate}
\item \emph{The freedom to choose a basis.} \\
 Let $\left\{ \tilde{e}^{i}\right\} _{i=1}^{I}$ be representatives
for the orbits $\nicefrac{E}{G}$ of $\Gamma$ (the choices of
representatives for the vertices do not affect the transplantation).
For every $1\leq i\leq I$, let ${\normalcolor B^{i}}$, ${\mathfrak{
B}^{i}}$ be two bases for $R$ chosen to correspond to the edge
$\tilde{e}^{i}$ (following the prescription given in the end of
section \ref{subsec:non_free_action}). Let $\Gamma_{1}$ be the
quotient obtained as $\nicefrac{\Gamma}{R}$ when we choose the bases
$\{B^{i}\}_{i=1}^{I}$ for $R$ and $\Gamma_{2}$ the quotient
$\nicefrac{\Gamma}{R}$ obtained by the bases
$\{\mathfrak{B}^{i}\}_{i=1}^{I}$. Our motivation is the following:
for every $\lambda\in\mathbb{C}$, $\Phi_{\Gamma_1}(\lambda)$ and
$\Phi_{\Gamma_2}(\lambda)$ serve to encode the same space (if $R$ is
irreducible then this is the $R$-isotypic component of
$\Phi_\Gamma\left(\lambda\right)$. In the general case, it the
corresponding space of intertwiners). Therefore, a bijection between
them can be obtained by composing the ``$\Gamma_1$-decoding'' with
the ``$\Gamma_2$-encoding''. However, since functions on the two
quotients encode functions on $\Gamma$ which transform according to
different bases, a suitable change of basis is required between the
decoding and encoding. Starting with
$f\in\Phi_{\Gamma_{1}}(\lambda)$, and denoting the corresponding
function in $\Phi_{\Gamma_{2}}(\lambda)$ by $\varphi$, we obtain

\begin{eqnarray*}
f\at_{e_{j}^{i}} & = & \widetilde{f}^i_j\at_{\tilde{e}^{i}}\\
 & = & \sum_{k=1}^{d}\left[\left[I\right]_{\mathfrak{B}^i}^{B^i}\right]_{k,j}\widetilde{\varphi}^i_k\at_{\tilde{e}^{i}}\\
 & = & \sum_{k=1}^{d}\left[\left[I\right]_{\mathfrak{B}^i}^{B^i}\right]_{k,j}\varphi\at_{e_{k}^{i}}\:,\end{eqnarray*}
where $\left[I\right]_{\mathfrak{B}^i}^{B^i}$ is the change of basis
matrix and $\left\{ \widetilde{f}^i_j\right\} _{j=1}^{d},\left\{
\widetilde{\varphi}^i_k\right\} _{k=1}^{d}$ are the functions in
$\Phi_{\Gamma}\left(\lambda\right)$ described by
\eref{eq:encoding_nonfree}. By the construction, $\left\{
\widetilde{f}^i_j\right\} _{j=1}^{d}$ transform according to $B^i$,
and $\left\{ \widetilde{\varphi}^i_k\right\} _{k=1}^{d}$ according
to $\mathfrak{B}^{i}$.
 We now note that the bijection we have constructed is indeed
a transplantation, as it represents the restrictions of one function
to each edge as a linear combination of restrictions of the second
function to edges of the same length. We also note that the choice
of the global basis $B$ (see (\ref{eq:final-boundary-conditions-3}),
(\ref{eq:final-boundary-conditions-4})) does not affect the
transplantation.
\item \emph{The freedom to choose representatives.} \\
Let $\left\{ \tilde{e}^{i}\right\} _{i=1}^{I}$ , $\left\{
\tilde{\varepsilon}^{i}\right\} _{i=1}^{I}$ be two sets of
representatives for the orbits $\nicefrac{E}{G}$. We describe the
connection between them by choosing $g^i$ such that
$g^i\tilde{e}^{i}=\tilde{\varepsilon}^{i}$ for $1\leq i\leq I$. For
each $i$, let ${\normalcolor B^{i}=\left\{ b^i_j\right\}_{j=1}^{d}}$
be a basis for $R$ chosen to correspond to the edge $\tilde{e}^{i}$.
A natural choice for a basis ${\mathfrak{B}^{i}=\left\{
\mathfrak{b}^i_j\right\}_{j=1}^{d}}$ which corresponds to the edge
$\tilde{\varepsilon}^{i}$ would then be ${\mathfrak{b}^i_j=g^i
b^i_j}$. Let $\Gamma_{1}$ be the quotient obtained as
$\nicefrac{\Gamma}{R}$ by the choice of the representatives $\left\{
\tilde{e}^{i}\right\} _{i=1}^{I}$ and the bases
$\{B^{i}\}_{i=1}^{I}$, and $\Gamma_{2}$ the quotient
$\nicefrac{\Gamma}{R}$ obtained from $\left\{
\tilde{\varepsilon}^{i}\right\} _{i=1}^{I}$ and
$\{\mathfrak{B}^{i}\}_{i=1}^{I}$. Working with these bases for
$\Gamma_2$ does not limit generality, as other choices can be
handled by the first part of the proposition. The merit of
$\{\mathfrak{B}^{i}\}_{i=1}^{I}$ is that it gives an extremely
simple transplantation between $f\in\Phi_{\Gamma_{1}}(\lambda)$ and
the corresponding $\varphi\in\Phi_{\Gamma_{2}}(\lambda)$:
\[
f\at_{e_{j}^{i}}=\widetilde{f}^i_j\at_{\tilde{e}^{i}}=g^i\widetilde{f}^i_j\at_{g^i\tilde{e}^{i}}=\widetilde{\varphi}^i_j\at_{\tilde{\varepsilon}^{i}}=\varphi\at_{\varepsilon_{j}^{i}}\;.\]
 We have used (\ref{eq:encoding_nonfree}) on the first and last equalities.
The second follows from the action of $G$ on
$\Phi_{\Gamma}\left(\lambda\right)$, and the third from the choice
of bases for the matrix representation (as before, $\left\{
\widetilde{f}^i_j\right\} _{j=1}^{d}$, $\left\{
\widetilde{\varphi}^i_j\right\} _{j=1}^{d}$ are functions in
$\Phi_{\Gamma}\left(\lambda\right)$ that transform according to the
matrix representations of $R$ as given by the bases
$B^{i},\mathfrak{B}^{i}$).
\end{enumerate}
\end{proof}
\begin{prop}
\label{prop:ind_transplantation}Let $\Gamma$ be a quantum graph
equipped with an action of a group $G$, $H$ a subgroup of $G$, and
$R$ a representation of $H$. Then there exists a transplantation
between every two graphs $\nicefrac{\Gamma}{R}$ and
$\nicefrac{\Gamma}{\ind_{H}^{G}R}$ which are constructed by the
recipe given in sections \ref{sec:rigorous},
\ref{subsec:non_free_action}.
\end{prop}
\begin{proof}
The outline of the proof is similar to that of proposition
\ref{prop:transplantation}. We start by describing a convenient
choice of representatives and bases for the construction of each of
the quotients $\nicefrac{\Gamma}{R}$ and
$\nicefrac{\Gamma}{\ind_{H}^{G}R}$. For every $\lambda\in\mathbb{C}$
we obtain a connection between two sets of functions in
$\Phi_{\Gamma}(\lambda)$, which transform according to those two
sets of bases. This connection yields the transplantation between
the functions in $\Phi_{\nicefrac{\Gamma}{R}}(\lambda)$ and
$\Phi_{\nicefrac{\Gamma}{\ind_{H}^{G}R}}(\lambda)$.
 Let $\left\{ \tilde{\varepsilon}^{i}\right\} _{i=1}^{I}$ be
the representatives for the orbits $\nicefrac{E}{G}$ of $\Gamma$
used in the construction of $\nicefrac{\Gamma}{\ind_H^G R}$. We deal
separately with each representative $\tilde{\varepsilon}^{i}$ and
address the simpler case of
$G_{\tilde{\varepsilon}^{i}}=\left\{id\right\}$ first. Let
$B^{i}=\left\{b^i_j\right\}_{j=1}^d$ be a basis for the
representation $R$. We choose representatives for the left cosets of
$H$ in $G$: $\left\{ t_k\right\}_{k=1}^{[G:H]}$. A possible basis
for $\ind_H^GR$ \footnote{See \ref{appendix}} is
$\mathfrak{B}^i=\left\{t_k b^i_j\right\}_{1\leq k\leq [G:H]\atop
1\leq j\leq d}$. Let $\lambda\in\mathbb{C}$. For
$\left\{\widetilde{f}^i_j\right\}_{1\leq j\leq d}$, a set of
functions in $\Phi_{\Gamma}(\lambda)$ which transform according to
$B^{i}$, we consider $\left\{\widetilde{\varphi}^i_{(k,j)}=t_k
\widetilde{f}^i_j\right\}_{1\leq k\leq [G:H]\atop 1\leq j\leq d}$, a
set of functions in $\Phi_{\Gamma}(\lambda)$ which transforms
according to $\mathfrak{B}_i$. We handle the case in which
$\nicefrac{\Gamma}{R}$ is constructed by choosing the
representatives for the orbits
$\nicefrac{G\tilde{\varepsilon}^i}{H}$ of $\Gamma$ to be
$\left\{\tilde{e}^{(i,k)}=t_k^{-1}\tilde{\varepsilon}^i\right\}_{k=1}^{[G:H]}$,
for which we obtain the transplantation:

\begin{eqnarray}
\varphi\at_{\varepsilon^i_{(k,j)}}&=&
\widetilde{\varphi}^i_{(k,j)}\at_{\tilde{\varepsilon}^i}=
t_k\widetilde{f}^i_j\at_{\tilde{\varepsilon}^i} =
\widetilde{f}^i_j\at_{t_k^{-1}\tilde{\varepsilon}^i} =
\widetilde{f}^i_j\at_{\tilde{e}^{(i,k)}} = f\at_{e^{(i,k)}_j}.
\end{eqnarray}

We now treat the case of a non-trivial $G_{\tilde{\varepsilon}^i}$.
Obviously, the construction of each quotient depends on the various
choices for bases and the edge representatives. We proceed in the
following order: starting with any representatives for
$\nicefrac{E}{G}$, $\left\{\tilde{\varepsilon}_i\right\}$, we pick
representatives for $\nicefrac{E}{H}$,
$\left\{\tilde{e}_{(i,k)}\right\}$. We then assume we are given any
bases for $R$, $\left\{B^{(i,k)} \right\}$, which fit
$\left\{\tilde{e}_{(i,k)}\right\}$ and use them to produce bases
$\left\{\mathfrak{B}^i\right\}$, which fit
$\left\{\tilde{\varepsilon}_i\right\}$. The instrument which enables
us to advance in this manner is the double coset structure.
\\
 We start by
choosing representatives for the $(G_{\tilde{\varepsilon}^i}, H)$
double cosets in $G$, $\left\{t_k\right\}_{k=1}^{l}$. We then denote
$\left\{t_{(k,1)}\right\}_{k=1}^{l}=\left\{t_k\right\}_{k=1}^{l}$
and complete this to a set of representatives of the left cosets of
$H$ in $G$, $\left\{ t_{(k,m)}\right\}_{1\leq k\leq l\atop 1\leq
m\leq n_k}$ ($\sum_{k=1}^{l}n_k~=~[G:H]$), obeying the condition
\begin{equation*}
\forall 1\leq k\leq l, \;\;\; \forall \, 1\leq m_1, m_2 \leq n_k,
\;\;\; \exists \, g\in G_{\tilde{\varepsilon}^i} \;\;\; s.t.\;\;
t_{(k,m_1)}=g\,t_{(k,m_2)}.
\end{equation*}
Note that the choice of $\left\{ t_{(k,m)}\right\}$ obviously
depends on $\tilde{\varepsilon}^i$, but we omit the $i$ index here
to simplify the notation. Returning to the graph, the condition
above means that
\begin{equation} \forall \, 1\leq k\leq l, \;\;\; \forall \, 1\leq m_1, m_2 \leq n_k,
\;\;\;
t_{(k,m_1)}^{-1}\tilde{\varepsilon}^i=t_{(k,m_2)}^{-1}\tilde{\varepsilon}^i.
\label{eq:edges_equality} \end{equation} It is now possible to
describe the representatives for the orbits $\nicefrac{E}{H}$, used
in $\nicefrac{\Gamma}{R}$'s construction. For each $1\leq i \leq I$
we can choose the representatives for the orbits
$\nicefrac{G\tilde{\varepsilon}^i}{H}$ as $\left\{
t_{(k,1)}^{-1}\tilde{\varepsilon}^i\right\}_{1\leq k \leq
l}~=~\left\{ t_{k}^{-1}\tilde{\varepsilon}^i\right\}_{1\leq k \leq
l}$. We denote these representatives by
$\tilde{e}^{(i,k)}=t_{k}^{-1}\tilde{\varepsilon}^i$. The union of
all these, $\left\{\tilde{e}^{(i,k)}\right\}_{1\leq i \leq I
\atop{1\leq k \leq l}}$, forms our choice of representatives for the
orbits $\nicefrac{E}{H}$.

We now assume that for each $k$ we are given
$B^{(i,k)}=\left\{b^{(i,k)}_j\right\}_{j=1}^{d}$, a basis for $R$
which corresponds to the edge $\tilde{e}^{(i,k)}$. Namely,
$\left\{b^{(i,k)}_j\right\}_{j=1}^{d_k}$ is a basis for the trivial
component of $\res^{H}_{H_{\tilde{e}^{(i,k)}}}R$, and
$\left\{b^{(i,k)}_j\right\}_{j=d_k+1}^{d}$ is a basis for
$\bigoplus\limits
_{S\ncong\mathbf{1}}\left(\res^{H}_{H_{\tilde{e}^{(i,k)}}}R\right)^{S}$
(note that $d_k$ depends on $i$, but we do not mention this to
simplify the notation).
 A possible basis for $\ind_H^GR$ is
$\left\{t_{(k,m)}b^{(i,k)}_j \right\}{{\atop {1\leq k\leq l}}\atop
{1\leq m \leq n_k \atop 1\leq j\leq d}}$, but the construction of
$\nicefrac{\Gamma}{\ind^{G}_{H}R}$ requires choosing a basis which
corresponds to the edge $\widetilde{\varepsilon}^i$, in the sense
mentioned above. By lemma \ref{lem:triv_basis} which follows this
proof, $\left\{
\frac{1}{n_k}\sum\limits_{m=1}^{n_k}t_{(k,m)}b^{(i,k)}_j\right\}_{1\leq
k\leq l\atop 1\leq j\leq d_k}$ is a basis for the trivial component
of $\res^{G}_{G_{\tilde{\varepsilon}^i}}\ind^{G}_{H}R$. We can
complete it to a basis of $\ind_H^GR$ by adding any bases for the
non-trivial irreducible components of
$\res^{G}_{G_{\tilde{\varepsilon}^i}}\ind^{G}_{H}R$. The exact form
of this completion does not affect the quotient
$\nicefrac{\Gamma}{\ind_H^G R}$ (nor the transplantation) since the
functions that correspond to these basis elements vanish on the edge
$\tilde{\varepsilon}^i$. We denote this basis of $\ind_H^GR$ by
$\mathfrak{B}^i$, and this finishes the description of the bases and
the representatives by which the two quotients are constructed. We
summarize them and the encoding$\setminus$decoding relations that
they imply:
\begin{itemize}
\item \Large$ \nicefrac{\Gamma}{R}$: \normalsize We have
$\left\{\tilde{e}^{(i,k)}=
t_{k}^{-1}\tilde{\varepsilon}^i\right\}_{1\leq k \leq l}$ as
representatives for the orbits
$\nicefrac{G\tilde{\varepsilon}^i}{H}$. The resulting edges of the
quotient are $\left\{e^{(i,k)}_{j}\right\}_{1\leq k \leq l \atop
{1\leq j \leq d_k}}$, where for each $k$, the edges
$\left\{e^{(i,k)}_{j}\right\}_{1\leq j \leq d_k}$ correspond to the
basis $\left\{b^{(i,k)}_j\right\}_{1\leq j \leq d_k}$. Let $f$ be a
function on $\nicefrac{\Gamma}{R}$. For each $k$, we can decode its
restrictions to the edges $\left\{e^{(i,k)}_{j}\right\}_{1\leq j
\leq d_k}$ into the set of functions
$\left\{\widetilde{f}^{(i,k)}_j\right\}_{1\leq j \leq d}$, whose
restrictions to $H\tilde{e}^{(i,k)}$ transform according to the
representation $R$ given by the basis $B^{(i,k)}$. The decoding is
given by $\widetilde{f}^{(i,k)}_j\at_{\tilde{e}^{(i,k)}}=\left\{
\begin{array}{lc} f\at_{e^{(i,k)}_j} & 1\leq j \leq d_k \\
0 & d_k< j \leq d \end{array} \right.$.

\item \Large$ \nicefrac{\Gamma}{\ind^{G}_{H}R}$: \normalsize Obviously, we have $\tilde{\varepsilon}^i$ as
as a representative for the orbit
$\nicefrac{G\tilde{\varepsilon}^i}{G}$. The resulting edges of the
quotient are $\left\{\varepsilon^{i}_{(k,j)}\right\}_{1\leq k \leq l
\atop {1\leq j \leq d_k}}$, and they correspond to the basis
$\left\{
\frac{1}{n_k}\sum\limits_{m=1}^{n_k}t_{(k,m)}b^{(i,k)}_j\right\}_{1\leq
k\leq l\atop 1\leq j\leq d_k}$. For $\varphi$, a function on
$\nicefrac{\Gamma}{\ind_H^G R}$, we decode its restrictions into the
edges $\left\{\varepsilon^{i}_{(k,j)}\right\}_{1\leq k \leq l \atop
{1\leq j \leq d_k}}$ by the set of functions
$\left\{\widetilde{\varphi}^{i}_{(k,j)}\right\}_{1\leq k \leq l
\atop 1\leq j \leq d}$ whose restrictions to
$G\tilde{\varepsilon}^{i}$ transform according to the basis
$\mathfrak{B}^i$. The decoding is given by
$\widetilde{\varphi}^i_{(k,j)}\at_{\tilde{\varepsilon}^{i}}=\left\{
\begin{array}{lc} \varphi\at_{\varepsilon^{i}_{(k,j)}} & 1\leq j \leq d_k \\
0 & d_k< j \leq d \end{array} \right.$.
\end{itemize}
We recall that we wish to find an isomorphism between
$\Phi_{\nicefrac{\Gamma}{R}}(\lambda)$ and
$\Phi_{\nicefrac{\Gamma}{\ind_H^GR}}(\lambda)$ for every
$\lambda\in\mathbb{C}$. In order to do that we compose the
``$\nicefrac{\Gamma}{R}$-decoding'' with the
``$\nicefrac{\Gamma}{\ind_H^GR}$-encoding'', but in the middle we
need to establish a bijection between the sets of functions whose
restrictions to $G\tilde{\varepsilon}^{i}$ transform according to
$\bigcup_{k=1}^{l}B^{(i,k)}$, and those whose restrictions to
$G\tilde{\varepsilon}^{i}$ transform according to $\mathfrak{B}^i$.
The bijection suggested by the bases we have chosen is:
\begin{equation}
\widetilde{\varphi}^{i}_{(k,j)}=\frac{1}{n_k}\sum_{m=1}^{n_k}t_{(k,m)}\widetilde{f}^{(i,k)}_{j}.
\label{eq:induced_functions1}
\end{equation}
By \eref{eq:edges_equality}, $\forall 1\leq m \leq n_k ,\;\;
{t_{(k,m)}^{-1}\tilde{\varepsilon}^i}=
{t_{k}^{-1}\tilde{\varepsilon}^i}=\tilde{e}^{(i,k)}$, 
so that \eref{eq:induced_functions1} simplifies to
\begin{equation}
\widetilde{\varphi}^{i}_{(k,j)}\at_{\tilde{\varepsilon}^i}
=\frac{1}{n_k}\sum_{m=1}^{n_k}t_{(k,m)}\widetilde{f}^{(i,k)}_{j}\at_{\tilde{\varepsilon}^i}
=\frac{1}{n_k}\sum_{m=1}^{n_k}\widetilde{f}^{(i,k)}_{j}\at_{t_{(k,m)}^{-1}\tilde{\varepsilon}^i}=
\widetilde{f}^{(i,k)}_{j}\at_{\tilde{e}^{(i,k)}}
\label{eq:induced_functions2} .\end{equation} Having set the stage,
the transplantation is simply given by:
\begin{equation}
\varphi\at_{{\varepsilon}^{i}_{(k,j)}}
=\widetilde{\varphi}^{i}_{(k,j)}\at_{\tilde{\varepsilon}^i}
=\widetilde{f}^{(i,k)}_{j}\at_{\tilde{e}^{(i,k)}}
=f\at_{e^{(i,k)}_j}.
\end{equation}
\end{proof}

\begin{lem} \label{lem:triv_basis}
Following the notations of the proof above, $\left\{
\frac{1}{n_k}\sum\limits_{m=1}^{n_k}t_{(k,m)}b^{i}_{(k,j)}\right\}_{1\leq
k\leq l\atop 1\leq j\leq d_k}$ is a basis for the trivial component
of $\res^{G}_{G_{\tilde{\varepsilon}^i}}\ind^{G}_{H}R$.
\end{lem}
\begin{proof}
Let $1\leq k \leq l$ and $1\leq m \leq n_k$.
\begin{eqnarray}
\nonumber \fl G_{\tilde{\varepsilon}^i}\,t_{(k,m)}&=&
\bigcup_{m'=1}^{n_k}\left\{G_{\tilde{\varepsilon}^i}\,t_{(k,m)}
\bigcap t_{(k,m')}H \right\} = \bigcup_{m'=1}^{n_k}
t_{(k,m')}\left\{t_{(k,m')}^{-1}\,G_{\tilde{\varepsilon}^i}\,t_{(k,m)}
\bigcap H \right\} \\
\nonumber &=&
\bigcup_{m'=1}^{n_k}t_{(k,m')}\left\{t_{(k,m)}^{-1}\,G_{\tilde{\varepsilon}^i}\,t_{(k,m)}
\bigcap H \right\} =
\bigcup_{m'=1}^{n_k}t_{(k,m')}\left\{G_{t_{(k,m)}^{-1}\tilde{\varepsilon}^i}
\bigcap H \right\} \\
 &=& \bigcup_{m'=1}^{n_k}t_{(k,m')}
H_{t_{(k,m)}^{-1}\tilde{\varepsilon}^i} =
\bigcup_{m'=1}^{n_k}t_{(k,m')} H_{\tilde{e}^{(i,k)}}.
\label{eq:G_e_coset}
\end{eqnarray}
The first equality is due to
$G_{\tilde{\varepsilon}^i}\,t_{(k,m)}H=\bigcup_{m'=1}^{n_k}t_{(k,m')}H$
and the third one is by \eref{eq:edges_equality}. In particular we
observe that for every $k$,
$|G_{\tilde{\varepsilon}^i}|=n_k|H_{\tilde{e}^{(i,k)}}|$. For
$b^{(i,k)}_{j}$ with $1\leq j \leq d_k$, we have by
\eref{eq:G_e_coset}
\begin{equation}
\fl \frac{1}{|G_{\tilde{\varepsilon}^i}|}\sum_{g\in
G_{\tilde{\varepsilon}^i}}g\,t_{(k,m)}\,b^{(i,k)}_{j} =
\frac{1}{|G_{\tilde{\varepsilon}^i}|}
\sum_{m'=1}^{n_k}t_{(k,m')}\sum_{h\in H_{\tilde{e}^{(i,k)}}}h\,
b^{(i,k)}_{j} = \frac{1}{n_k}\sum_{m'=1}^{n_k}t_{(k,m')}\,
b^{(i,k)}_{j}, \label{eq:invariant_vector1}
\end{equation}
where the last equality is since $H_{\tilde{e}^{(i,k)}}$ acts
trivially on $b^{(i,k)}_{j}$. Obviously, the l.h.s of
\eref{eq:invariant_vector1} is invariant under the action of
$G_{\tilde{\varepsilon}^i}$. Therefore, the set $\left\{
\frac{1}{n_k}\sum\limits_{m'=1}^{n_k}t_{(k,m')}b^{(i,k)}_{j}\right\}_{1\leq
k\leq l\atop 1\leq j\leq d_k}$ lies in the trivial component of
$\res^{G}_{G_{\tilde{\varepsilon}^i}}\ind^{G}_{H}R$. In addition, it
is linearly independent since $\left\{t_{(k,m)}b^{(i,k)}_j
\right\}{{\atop {1\leq k\leq l}}\atop {1\leq m \leq n_k \atop 1\leq
j\leq d}}$ is a basis for $\ind^{G}_{H}R$. To finish the proof we
show that its size, $\sum_{k=1}^{l}d_{k}$, equals $\left\langle
\res_{G_{\tilde{\varepsilon}^{i}}}^{G}\ind_{H}^{G}R,\mathbf{1}\right\rangle
_{G_{\tilde{\varepsilon}^{i}}}$.

If $R$ is a representation of $H\leq G$, then for any $g\in G$ we
also have $R$ as a representation of $gHg^{-1}$, with $ghg^{-1}\in
gHg^{-1}$ acting as $h\in H$. Mackey's Decomposition Theorem
(\cite{Curtis}, section 10B) uses this to relate induction and
restriction in the following way:
\begin{eqnarray*}
\fl \res_{G_{\tilde{\varepsilon}^{i}}}^{G}\ind_{H}^{G}R & \cong &
\bigoplus_{k=1}^{l}\ind_{t_{k}H_{\tilde{e}^{(i,k)}}t_{k}^{-1}}^{G_{\tilde{\varepsilon}^{i}}}\res_{t_{k}H_{\tilde{e}^{(i,k)}}t_{k}^{-1}}^{t_{k}Ht_{k}^{-1}}R,\end{eqnarray*}
which shows that
\begin{eqnarray*}
\fl \left\langle \res_{G_{\tilde{\varepsilon}^{i}}}^{G}\ind_{H}^{G}R,\mathbf{1}\right\rangle _{G_{\tilde{\varepsilon}^{i}}}  &=&  \sum_{k=1}^{l}\left\langle \ind_{^{t_{k}}H_{\tilde{e}^{(i,k)}}t_{k}^{-1}}^{G_{\tilde{\varepsilon}^{i}}}\res_{t_{k}H_{\tilde{e}^{(i,k)}}t_{k}^{-1}}^{t_{k}Ht_{k}^{-1}}R,\mathbf{1}\right\rangle _{G_{\tilde{\varepsilon}^{i}}}\\
  &=& \sum_{k=1}^{l}\left\langle \res_{t_{k}H_{\tilde{e}^{(i,k)}}t_{k}^{-1}}^{t_{k}Ht_{k}^{-1}}R,\mathbf{1}\right\rangle _{t_{k}H_{\tilde{e}^{(i,k)}}t_{k}^{-1}}
 \\ &=& \sum_{k=1}^{l}\left\langle \res_{H_{\tilde{e}^{(i,k)}}}^{H}R,\mathbf{1}\right\rangle _{H_{\tilde{e}^{(i,k)}}}=\sum_{k=1}^{l}d_{k}.\end{eqnarray*}

\end{proof}

\begin{cor}
\label{cor:transplantation} If $G$ acts on $\Gamma$ and
$H_{1},H_{2}$ are subgroups of $G$ with corresponding
representations $R_{1},R_{2}$ such that
$\ind_{H_{1}}^{G}R_{1}\cong\ind_{H_{2}}^{G}R_{2}$, then
$\nicefrac{\Gamma}{R_{1}}$ and $\nicefrac{\Gamma}{R_{2}}$ which are
constructed by the recipe described are transplantable.
\end{cor}

We end this section by demonstrating the process of finding the
transplantation for our basic isospectral example (figure
\ref{fig:dihedral_pair}). The representations (\ref{eq:r1_rep}),
(\ref{eq:r2_rep}) that were used to form those quotients are one
dimensional and we denote their bases by $B=\left\{b\right\}$,
$\mathfrak{B}=\left\{\mathfrak{b}\right\}$. We choose the
representatives $\left\{ e,\sigma\right\} $ to form the induction
$\ind_{H_{1}}^{D_{4}}R_{1}$ whose basis we choose to be $\left\{ eb,
\,\sigma b\right\} $.

The matrix representation corresponding to this basis is
\begin{equation} \fl \left\{
\begin{array}{cccc}
e\mapsto\left(\scriptsize{\begin{array}{cc}1 & 0\\
0 & 1\end{array}}\right), & \sigma\mapsto\left(\scriptsize{\begin{array}{cc}0 & 1\\
-1 & 0\end{array}}\right), & \sigma^{2}\mapsto\left(\scriptsize{\begin{array}{cc}-1 & 0\\
0 & -1\end{array}}\right), & \sigma^{3}\mapsto\left(\scriptsize{\begin{array}{cc}0 & -1\\
1 & 0\end{array}}\right),\\
\\\tau\mapsto\left(\scriptsize{\begin{array}{cc}-1 & 0\\
0 & 1\end{array}}\right), & \tau\sigma\mapsto\left(\scriptsize{\begin{array}{cc}0 & -1\\
-1 & 0\end{array}}\right), & \tau\sigma^{2}\mapsto\left(\scriptsize{\begin{array}{cc}1 & 0\\
0 & -1\end{array}}\right), & \tau\sigma^{3}\mapsto\left(\scriptsize{\begin{array}{cc}0 & 1\\
1 & 0\end{array}}\right)\end{array}\right\}
.\label{eq:induced_rep1}\end{equation}
 We choose the same representatives $\left\{ e,\sigma\right\} $ and
get $\ind_{H_{2}}^{D_{4}}R_{2}$ with the basis $\left\{
e\mathfrak{b},\,\sigma\mathfrak{b}\right\} $ and the following
matrix representation
\begin{equation} \fl \left\{
\begin{array}{cccc}
e\mapsto\left(\scriptsize{\begin{array}{cc}1 & 0\\
0 & 1\end{array}}\right), & \sigma\mapsto\left(\scriptsize{\begin{array}{cc}0 & 1\\
-1 & 0\end{array}}\right), & \sigma^{2}\mapsto\left(\scriptsize{\begin{array}{cc}-1 & 0\\
0 & -1\end{array}}\right), & \sigma^{3}\mapsto\left(\scriptsize{\begin{array}{cc}0 & -1\\
1 & 0\end{array}}\right),\\
\\\tau\mapsto\left(\scriptsize{\begin{array}{cc}0 & 1\\
1 & 0\end{array}}\right), & \tau\sigma\mapsto\left(\scriptsize{\begin{array}{cc}-1 & 0\\
0 & 1\end{array}}\right), & \tau\sigma^{2}\mapsto\left(\scriptsize{\begin{array}{cc}0 & -1\\
-1 & 0\end{array}}\right), & \tau\sigma^{3}\mapsto\left(\scriptsize{\begin{array}{cc}1 & 0\\
0 & -1\end{array}}\right)\end{array}\right\}
.\label{eq:induced_rep2}\end{equation}
 The relation between these two bases is \begin{eqnarray*}
e b & = & \nicefrac{1}{\sqrt{2}}\left(e\mathfrak{b}-\sigma\mathfrak{b}\right)\\
\sigma b & = &
\nicefrac{1}{\sqrt{2}}\left(e\mathfrak{b}+\sigma\mathfrak{b}\right)\:.\end{eqnarray*}

\begin{figure}[!h]
\begin{centering}
\hfill{}%
\begin{minipage}[c][1\totalheight]{0.3\columnwidth}%
\begin{center}
\includegraphics[scale=0.55]{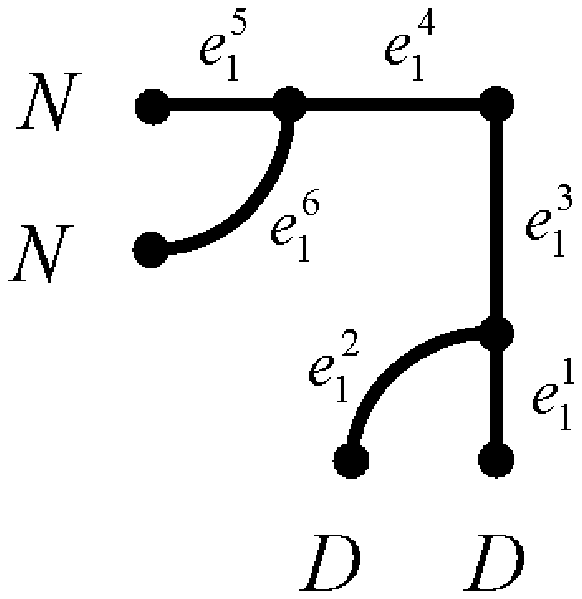}
\par\end{center}

\begin{center}
(a)
\par\end{center}%
\end{minipage}\hfill{}%
\begin{minipage}[c][1\totalheight]{0.3\columnwidth}%
\begin{center}
\includegraphics[scale=0.55]{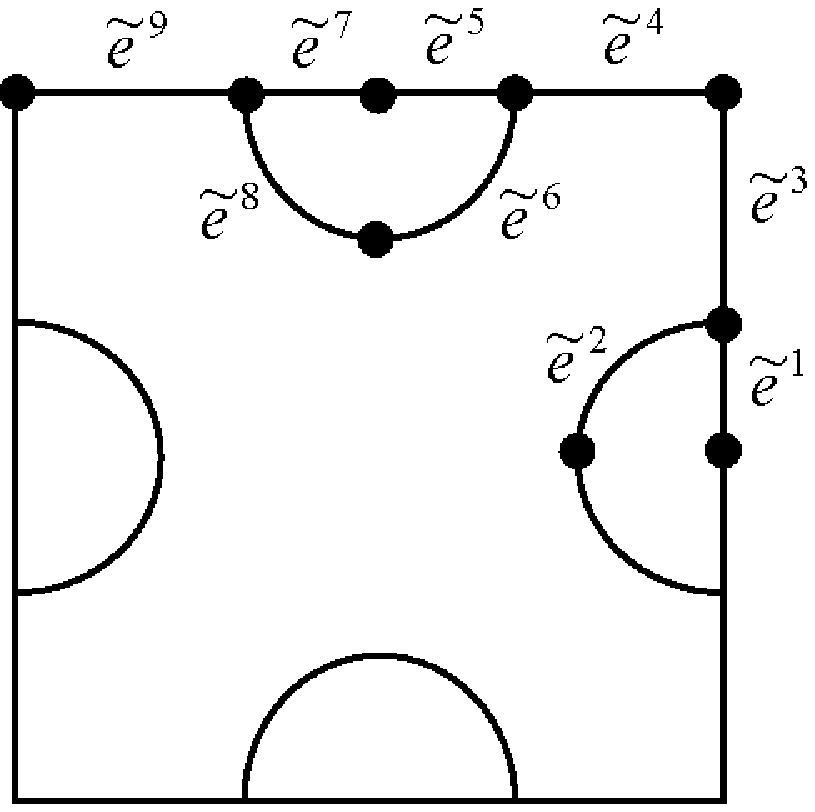}
\par\end{center}

\begin{center}
(b)
\par\end{center}%
\end{minipage}\hfill{}\hfill{}%
\begin{minipage}[c][1\totalheight]{0.3\columnwidth}%
\begin{center}
\includegraphics[scale=0.55]{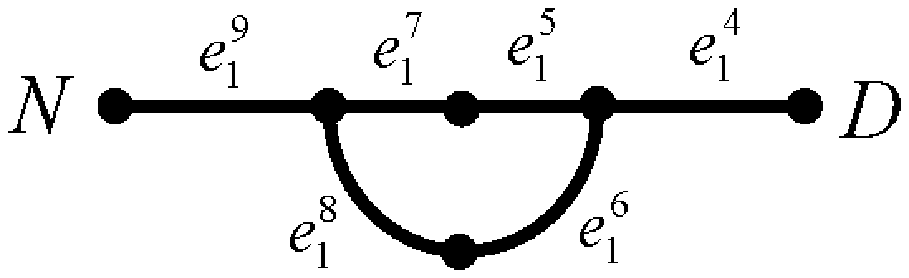}
\par\end{center}

\bigskip{}

\begin{center}
(c)
\par\end{center}%
\end{minipage}\hfill{}
\par\end{centering}

\caption{(a) The graph $\nicefrac{\Gamma}{R_{1}}$ with its edges
labeled. (b) The graph $\Gamma$ whose relevant edges are denoted.
(c) The graph $\nicefrac{\Gamma}{R_{2}}$.}

\label{fig:dihedral_transplantation}
\end{figure}

Let $\lambda\in\mathbb{C}$. Let
$f\in\Phi_{\nicefrac{\Gamma}{R_{1}}}(\lambda)$,
$\varphi\in\Phi_{\nicefrac{\Gamma}{R_{2}}}(\lambda)$. Decoding these
functions we get
$\widetilde{f}_{1}\in\Phi_{\Gamma}^{R_{1}}(\lambda)$,
$\widetilde{\varphi}_{1}\in\Phi_{\Gamma}^{R_{2}}(\lambda)$. Note
that in the general construction we also have a superscript index
denoting the number of the edge on which the function is defined.
However, in this example we have a free action on the edges so that
we can choose a single basis ($B$ for $\nicefrac{\Gamma}{R_{1}}$ and
$\mathfrak{B}$ for $\nicefrac{\Gamma}{R_{2}}$) to work with. Using
the general notation we therefore have $\forall 1\leq i \leq 6, \;\;
\widetilde{f}_{1}=\widetilde{f}^{i}_{1}$ and similarly for
$\widetilde{\varphi}_{1}$. Forming the inductions we have $\left\{
e\widetilde{f}_{1},\,\sigma\widetilde{f}_{1}\right\}
\subset\Phi_{\Gamma}^{\ind_{H_{1}}^{D_{4}}R_{1}}(\lambda), \;\;
\left\{
e\widetilde{\varphi}_{1},\,\sigma\widetilde{\varphi}_{1}\right\}
\subset\Phi_{\Gamma}^{\ind_{H_{2}}^{D_{4}}R_{2}}(\lambda)$. These
functions correspond to the bases $\left\{ eb, \,\sigma b\right\},
\, \left\{ e\mathfrak{b},\,\sigma\mathfrak{b}\right\} $ and
transform accordingly. We now describe the transplantation by
expressing $f$ in terms of $\varphi$ (the edges' labeling is given
in figure \ref{fig:dihedral_transplantation}).\begin{eqnarray*}
f\at_{e_{1}^{1}}  &= &
\widetilde{f}_{1}\at_{\tilde{e}^{1}}=e\widetilde{f}_{1}\at_{\tilde{e}^{1}}=
\nicefrac{1}{\sqrt{2}}\left(e\widetilde{\varphi}_{1}\at_{\tilde{e}^{1}}-\sigma\widetilde{\varphi}_{1}\at_{\tilde{e}^{1}}\right)
\\ &=&
\nicefrac{1}{\sqrt{2}}\left(\widetilde{\varphi}_{1}\at_{\tilde{e}^{1}}-\widetilde{\varphi}_{1}\at_{\sigma^{-1}\tilde{e}^{1}}\right)
 =
\nicefrac{1}{\sqrt{2}}\left(\widetilde{\varphi}_{1}\at_{\left(\tau\sigma^{3}\right)^{-1}\tilde{e}^{5}}-\widetilde{\varphi}_{1}\at_{\sigma^{-2}\tilde{e}^{7}}\right)
\\ &=&
\nicefrac{1}{\sqrt{2}}\left(\rho_{R_{2}}\left(\tau\sigma^{3}\right)\widetilde{\varphi}_{1}\at_{\tilde{e}^{5}}-\rho_{R_{2}}\left(\sigma^{2}\right)\widetilde{\varphi}_{1}\at_{\tilde{e}^{7}}\right)=\nicefrac{1}{\sqrt{2}}\left(-\widetilde{\varphi}_{1}\at_{\tilde{e}^{5}}+\widetilde{\varphi}_{1}\at_{\tilde{e}^{7}}\right)
\\ &=&
\nicefrac{1}{\sqrt{2}}\left(-\varphi\at_{e_{1}^{5}}+\varphi\at_{e_{1}^{7}}\right)\end{eqnarray*}
 \begin{eqnarray*}
f_{1}\at_{e_{1}^{5}} & = &
\widetilde{f}_{1}\at_{\tilde{e}^{5}}=e\widetilde{f}_{1}\at_{\tilde{e}^{5}}
  =  \nicefrac{1}{\sqrt{2}}\left(e\widetilde{\varphi}_{1}\at_{\tilde{e}^{5}}-\sigma\widetilde{\varphi}_{1}\at_{\tilde{e}^{5}}\right)\\
 & = & \nicefrac{1}{\sqrt{2}}\left(\widetilde{\varphi}_{1}\at_{\tilde{e}^{5}}-\widetilde{\varphi}_{1}\at_{\sigma^{-1}\tilde{e}^{5}}\right)=\nicefrac{1}{\sqrt{2}}\left(\widetilde{\varphi}_{1}\at_{\tilde{e}^{5}}-\widetilde{\varphi}_{1}\at_{\left(\tau\sigma^{3}\right)^{-1}\tilde{e}^{7}}\right)\\
 & = & \nicefrac{1}{\sqrt{2}}\left(\widetilde{\varphi}_{1}\at_{\tilde{e}^{5}}-\rho_{R_{2}}\left(\tau\sigma^{3}\right)\widetilde{\varphi}_{1}\at_{\tilde{e}^{7}}\right)=\nicefrac{1}{\sqrt{2}}\left(\widetilde{\varphi}_{1}\at_{\tilde{e}^{5}}+\widetilde{\varphi}_{1}\at_{\tilde{e}^{7}}\right)\\
 & = & \nicefrac{1}{\sqrt{2}}\left(\varphi\at_{e_{1}^{5}}+\varphi\at_{e_{1}^{7}}\right).\end{eqnarray*}
 The transplantation on all the other edges is obtained similarly
and the result is \begin{eqnarray*}
f\at_{e_{1}^{2}} & = & \nicefrac{1}{\sqrt{2}}\left(-\varphi\at_{e_{1}^{6}}+\varphi\at_{e_{1}^{8}}\right)\\
f\at_{e_{1}^{3}} & = & \nicefrac{1}{\sqrt{2}}\left(-\varphi\at_{e_{1}^{4}}+\varphi\at_{e_{1}^{9}}\right)\\
f\at_{e_{1}^{4}} & = & \nicefrac{1}{\sqrt{2}}\left(\varphi\at_{e_{1}^{4}}+\varphi\at_{e_{1}^{9}}\right)\\
f\at_{e_{1}^{6}} & = &
\nicefrac{1}{\sqrt{2}}\left(\varphi\at_{e_{1}^{6}}+\varphi\at_{e_{1}^{8}}\right).\end{eqnarray*}

\section{A gallery of isospectral graphs}  \label{sec:gallery_of_graphs}

We present three interesting examples of isospectral graphs. The
first example shows that once the algebraic conditions stated in
theorem \ref{thm:mainthm} or in corollary \ref{cor:sunadapair} are
satisfied, we can choose any graph that obeys the group symmetries
and obtain isospectral quotients. The second example shows a
symmetry group larger then $D_4$, as well as the effect of using
different choices of representatives while constructing the
quotient. The third example demonstrates the special case of a group
that acts freely on a graph which enables us to get quotients whose
boundary conditions are all of the Neumann type.

\subsection{$G=D_{4}$ $-$ Another Choice for $\Gamma$}

\begin{center}
\begin{figure}[!h]
\begin{centering}
\includegraphics[scale=0.35]{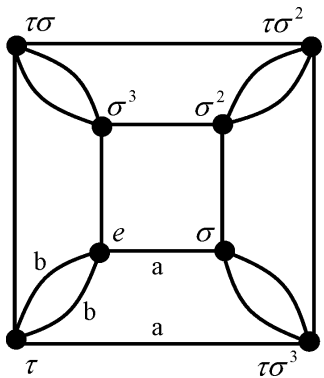}
\par\end{centering}
\caption{The Cayley graph of $D_{4}$ with the generating set
$\{\sigma,\tau\}$. The lengths of some edges of this quantum graph
are marked.} \label{fig:Cayley_Full}
\end{figure}
\end{center}

\begin{figure}[!h]
\begin{centering}
\begin{minipage}[c][1\totalheight]{0.1\columnwidth}%
\begin{center}
\includegraphics[scale=0.35]{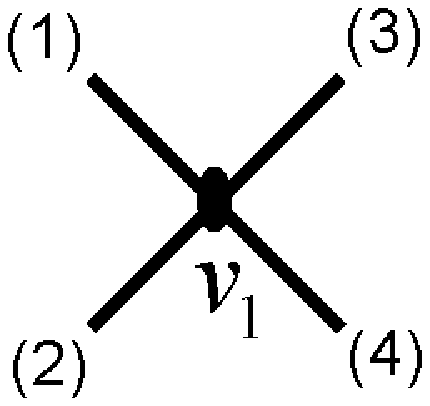}
\par\end{center}
\begin{center}
\includegraphics[scale=0.35]{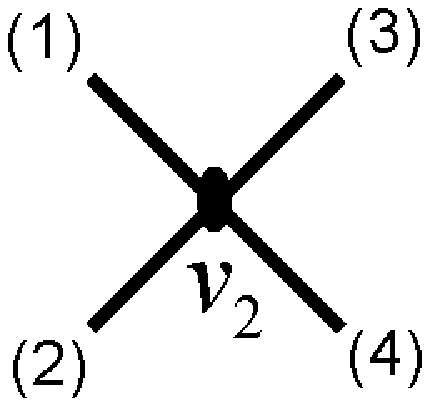}
\par\end{center}%
\end{minipage}%
\begin{minipage}[c][1\totalheight]{0.05\columnwidth}%
\begin{center}
\par\end{center}%
\end{minipage}%
\begin{minipage}[c][1\totalheight]{0.44\columnwidth}%
\begin{center}
\includegraphics[width=0.95\columnwidth]{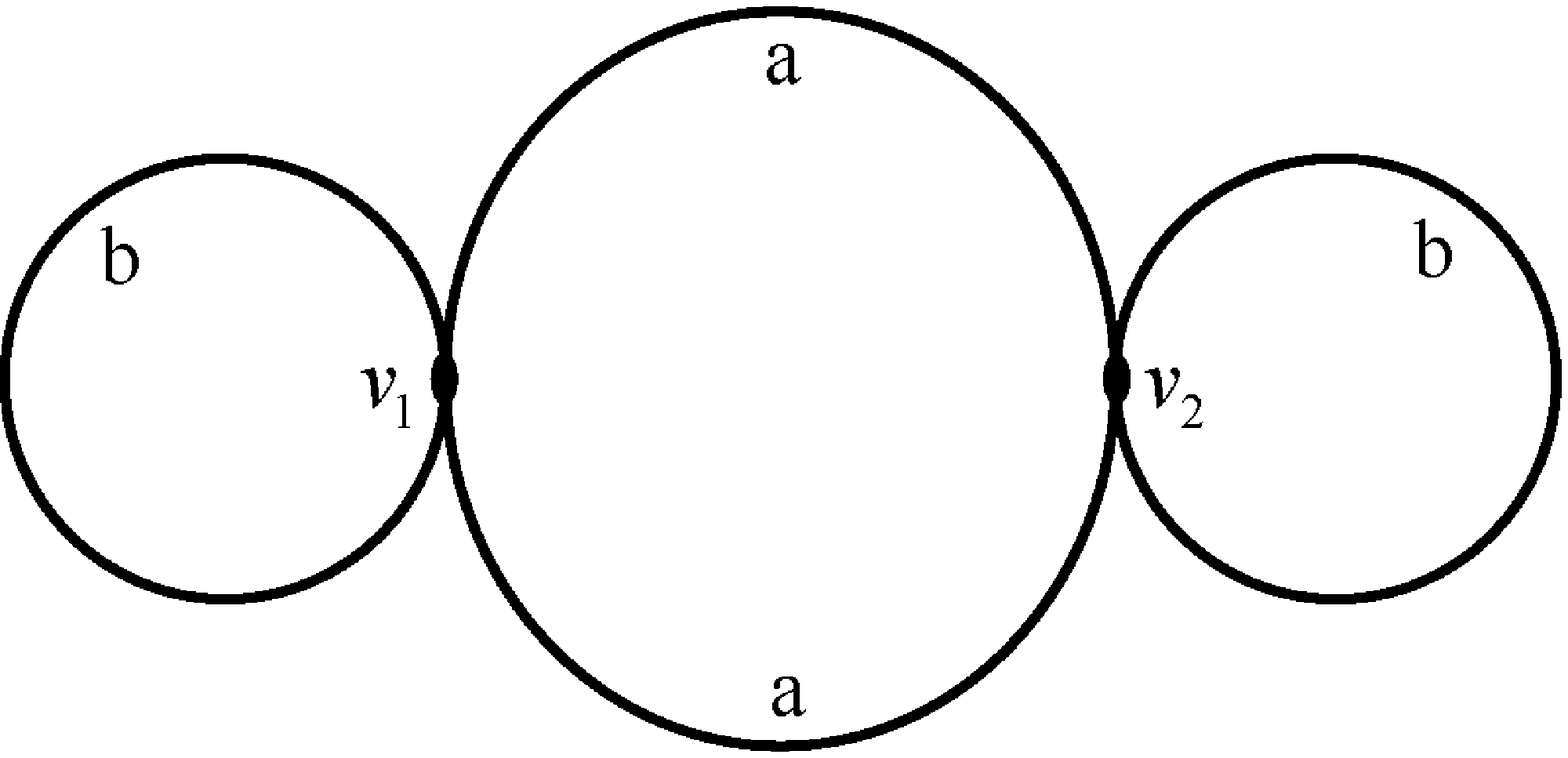}
\par\end{center}%
\end{minipage}%
\begin{minipage}[c][1\totalheight]{0.05\columnwidth}%
\begin{center}
\par\end{center}%
\end{minipage}%
\begin{minipage}[c][1\totalheight]{0.28\columnwidth}%
\begin{eqnarray*}
\fl A_{v_{1}}=A_{v_{2}}=\left(\scriptsize{\begin{array}{cccc} 1 & 1 & 0 & 0\\
1 & 0 & 1 & 0\\
1 & 0 & 0 & 1\\
0 & 0 & 0 & 0\end{array}}\right)\\
\\\fl B_{v_{1}}=B_{v_{2}}=\left(\scriptsize{\begin{array}{cccc} 0 & 0 & 0 & 0\\
0 & 0 & 0 & 0\\
0 & 0 & 0 & 0\\
1 & -1 & -1 & -1\end{array}}\right)\end{eqnarray*}
\end{minipage}
\par\end{centering}

\bigskip{}
\begin{centering}
(a)
\par\end{centering}

\bigskip{}
\begin{centering}
\begin{minipage}[c][1\totalheight]{0.1\columnwidth}%
\begin{center}
\includegraphics[scale=0.35]{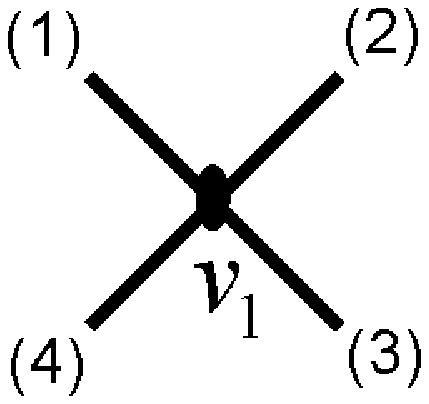}
\par\end{center}

\begin{center}
\includegraphics[scale=0.35]{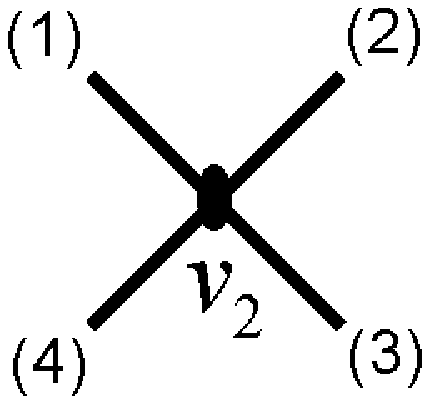}
\par\end{center}%
\end{minipage}%
\begin{minipage}[c][1\totalheight]{0.02\columnwidth}%
\begin{center}
\par\end{center}%
\end{minipage}%
\begin{minipage}[c][1\totalheight]{0.35\columnwidth}%
\begin{center}
\includegraphics[scale=0.35]{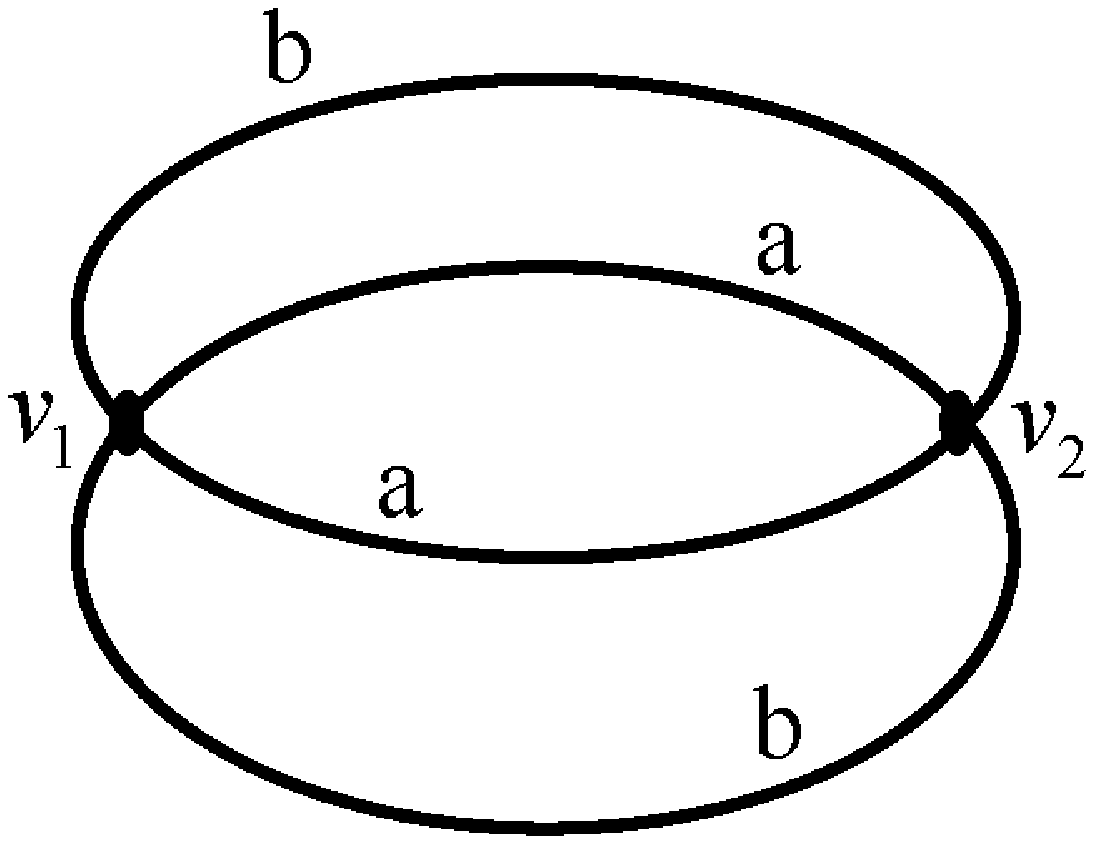}
\par\end{center}%
\end{minipage}%
\begin{minipage}[c][1\totalheight]{0.02\columnwidth}%
\begin{center}
\par\end{center}%
\end{minipage}%
\begin{minipage}[c][1\totalheight]{0.44\columnwidth}%
$\begin{array}{cc}
{\scriptstyle A_{v_{1}}=\left(\scriptsize{\begin{array}{cccc}1 & 1 & 0 & 0\\
1 & 0 & 1 & 0\\
1 & 0 & 0 & 1\\
0 & 0 & 0 & 0\end{array}}\right)} & {\scriptstyle B_{v_{1}}=\left(\scriptsize{\begin{array}{cccc}0 & 0 & 0 & 0\\
0 & 0 & 0 & 0\\
0 & 0 & 0 & 0\\
1 & -1 & -1 & -1\end{array}}\right)}\\
\\{\scriptstyle A_{v_{2}}=\left(\scriptsize{\begin{array}{cccc}1 & 1 & 0 & 0\\
1 & 0 & -1 & 0\\
1 & 0 & 0 & 1\\
0 & 0 & 0 & 0\end{array}}\right)} & {\scriptstyle B_{v_{2}}=\left(\scriptsize{\begin{array}{cccc}0 & 0 & 0 & 0\\
0 & 0 & 0 & 0\\
0 & 0 & 0 & 0\\
-1 & 1 & -1 & 1\end{array}}\right)}\end{array}$%
\end{minipage}
\par\end{centering}
\bigskip{}
\begin{centering}
(b)
\par\end{centering}
\bigskip{}
\begin{centering}
\begin{minipage}[c][1\totalheight]{0.1\columnwidth}%
\begin{center}
\includegraphics[scale=0.35]{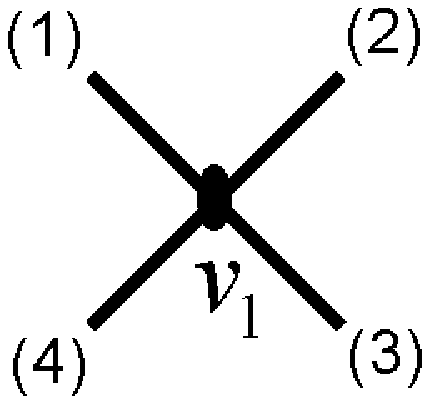}
\par\end{center}
\begin{center}
\includegraphics[scale=0.35]{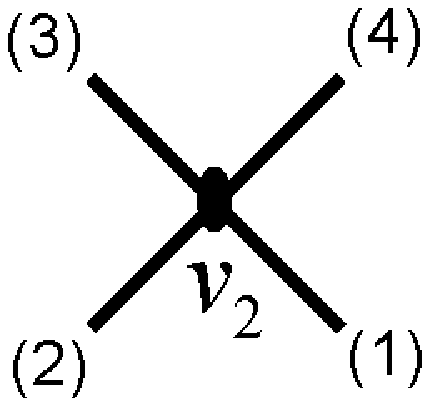}
\par\end{center}%
\end{minipage}%
\begin{minipage}[c][1\totalheight]{0.05\columnwidth}%
\begin{center}
\par\end{center}%
\end{minipage}%
\begin{minipage}[c][1\totalheight]{0.45\columnwidth}%
\begin{center}
\includegraphics[width=0.95\columnwidth]{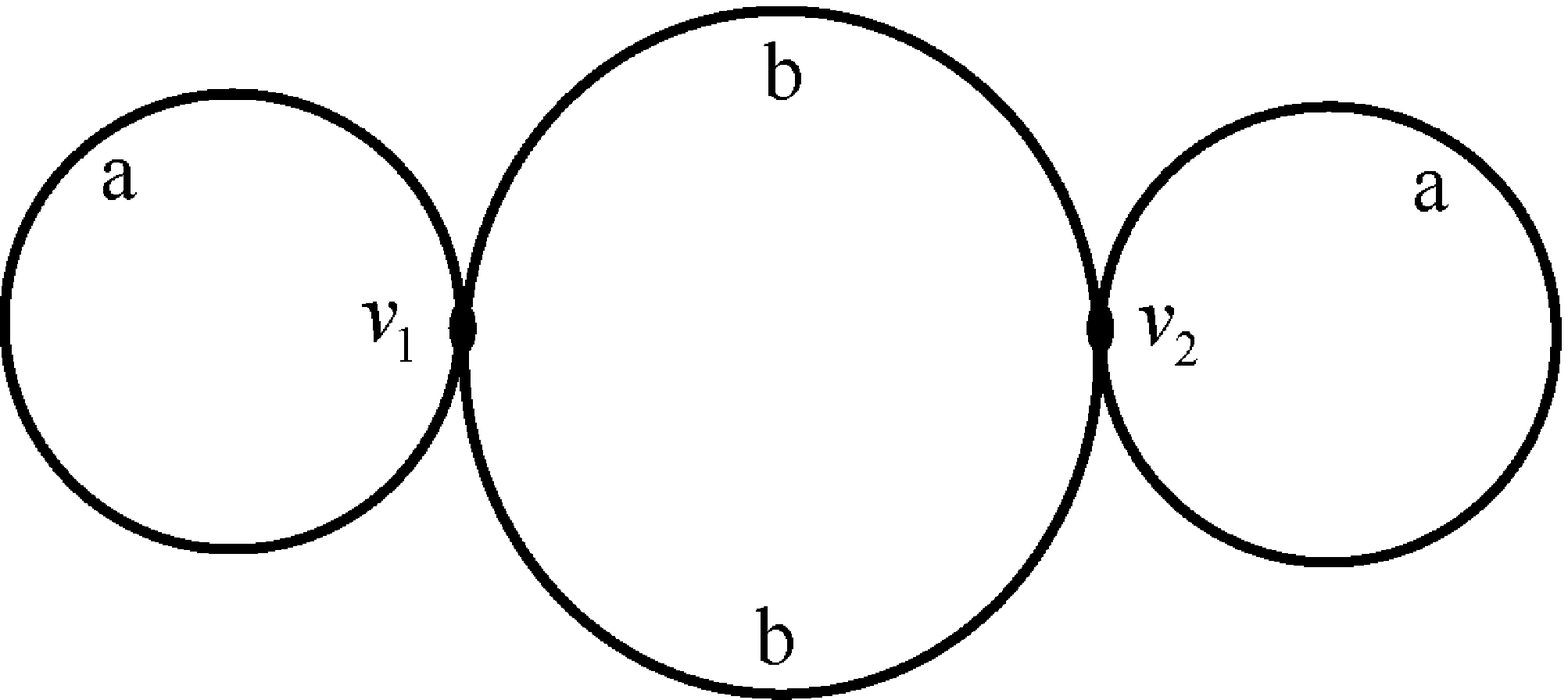}
\par\end{center}%
\end{minipage}%
\begin{minipage}[c][1\totalheight]{0.05\columnwidth}%
\begin{center}
\par\end{center}%
\end{minipage}%
\begin{minipage}[c][1\totalheight]{0.33\columnwidth}%
\begin{eqnarray*}
\fl A_{v_{1}}=A_{v_{2}}=\left(\scriptsize{\begin{array}{cccc}i & 1 & 0 & 0\\
i & 0 & 1 & 0\\
i & 0 & 0 & 1\\
0 & 0 & 0 & 0\end{array}}\right)\\
\\\fl B_{v_{1}}=B_{v_{2}}=\left(\scriptsize{\begin{array}{cccc}0 & 0 & 0 & 0\\
0 & 0 & 0 & 0\\
0 & 0 & 0 & 0\\
-i & 1 & 1 & 1\end{array}}\right)\end{eqnarray*}
\end{minipage}
\par\end{centering}

\bigskip{}
\begin{centering}
(c)
\par\end{centering}
\bigskip{}
\caption{The three quotient graphs obtained from the Cayley graph.
(a) $\nicefrac{\Gamma}{R_{1}}$, (b)$\nicefrac{\Gamma}{R_{2}}$, (c)
$\nicefrac{\Gamma}{R_{3}}$. The matrices describe the boundary
conditions at the various vertices. The diagrams on the left relate
the edges to the rows of the corresponding matrices, numbered 1-4.}
\label{fig:CayleyGraphs}
\end{figure}

We present another example using the group $D_{4}$, to emphasize
that it is possible to obtain isospectral quotient graphs from any
graph that obeys the symmetry of $D_{4}$. We now denote the Cayley
graph of $D_{4}$ with respect to the generating set
$S=\{\sigma,\tau\}$ by $\Gamma$ (figure \ref{fig:Cayley_Full}).
Since a Cayley graph can be constructed for any discrete group, this
example can easily be generalized to create other isospectral sets.
The set of edges of $\Gamma$ is then $E=\{(g,gs)\}_{g\in D_{4}, s\in
S}$, and the action of $D_{4}$ on the edges of $\Gamma$ is by
$\forall h\in D_{4}$, $h(g,gs)=(hg,hgs)$, where $s\in S$. Taking the
same subgroups of $D_{4}$ and representations as in sections
\ref{sec:basic_example} and \ref{sec:extending_example}, we obtain
three isospectral quotient graphs $\nicefrac{\Gamma}{R_{1}}$,
$\nicefrac{\Gamma}{R_{2}}$, $\nicefrac{\Gamma}{R_{3}}$, shown in
figure \ref{fig:CayleyGraphs}. The representatives chosen to
construct the graphs are:
\[
\begin{array}{llll}
\nicefrac{V}{H_{1}}=\{e,\sigma\} &  &  & \nicefrac{E}{H_{1}}=\{(e,\sigma),\,(\sigma^{3},e),\,(e,\tau),\,(\sigma,\tau\sigma^{3})\}\\
\nicefrac{V}{H_{2}}=\{e,\sigma\} &  &  & \nicefrac{E}{H_{2}}=\{(e,\sigma),\,(\sigma^{3},e),(e,\tau),\,(\sigma,\tau\sigma^{3})\}\\
\nicefrac{V}{H_{3}}=\{e,\tau\} &  &  &
\nicefrac{E}{H_{3}}=\{(e,\sigma),\,(\tau\sigma^{3},\tau),\,(\tau,e),(e,\tau)\}\end{array}\]

\subsection{$G=O_{h}$ $-$ Freedom to Choose Representatives}
\label{subsec:freedom_of_repres}

\begin{figure}[!h]
\begin{centering}
\begin{minipage}[c][1\totalheight]{0.49\columnwidth}%
\begin{center}
\includegraphics[width=0.85\columnwidth]{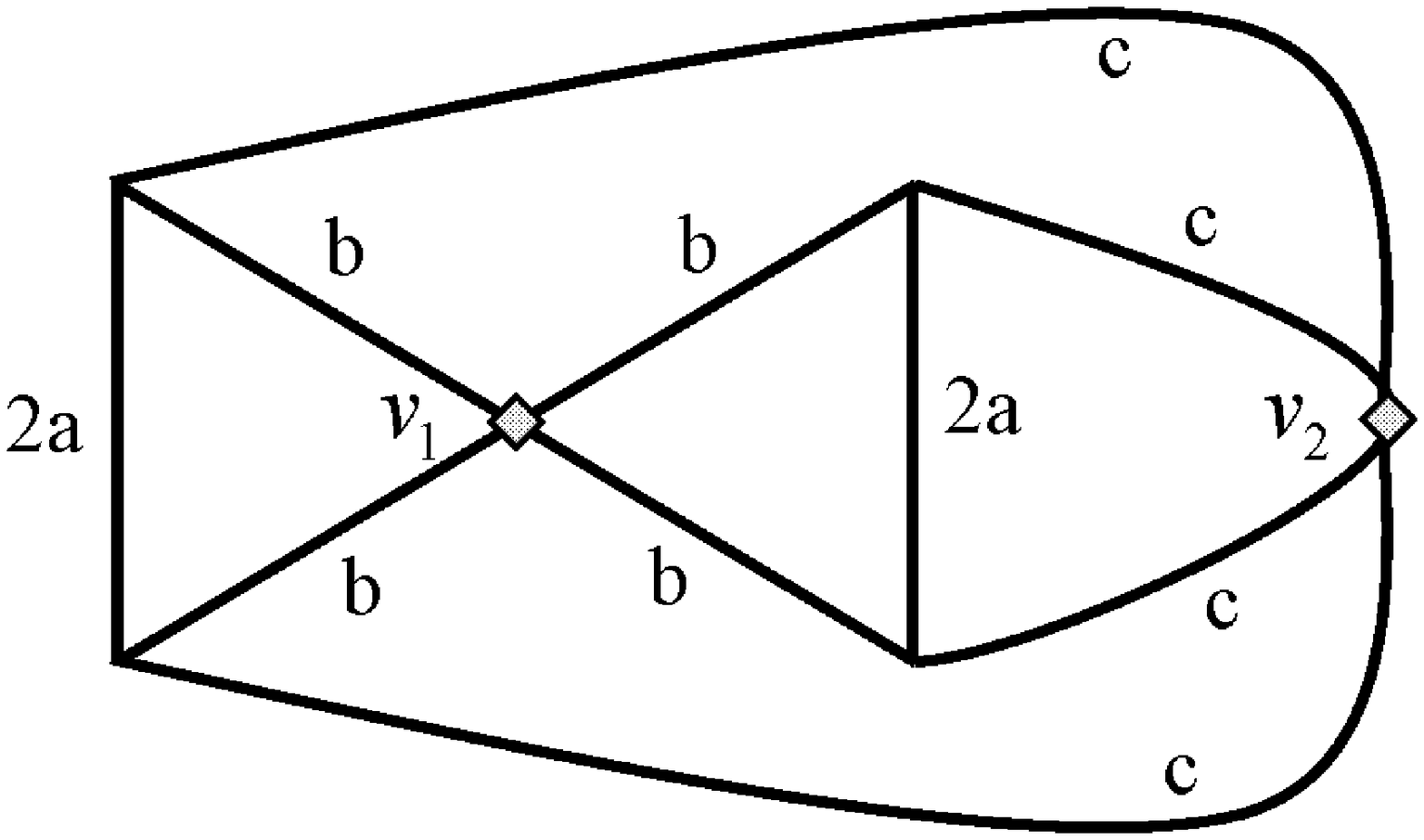}
\par\end{center}%
\end{minipage}%
\begin{minipage}[c][1\totalheight]{0.5\columnwidth}%
\begin{center}
$\begin{array}{cc}
{\scriptstyle A_{v_{1}}=\left(\scriptsize{\begin{array}{cccc}1 & c & 0 & -s\\
0 & s & -1 & c\\
0 & 0 & 0 & 0\\
0 & 0 & 0 & 0\end{array}}\right)} & {\scriptstyle B_{v_{1}}=\left(\scriptsize{\begin{array}{cccc}0 & 0 & 0 & 0\\
0 & 0 & 0 & 0\\
1 & -c & 0 & s\\
0 & -s & -1 & -c\end{array}}\right)}\\
\\{\scriptstyle A_{v_{2}}=\left(\scriptsize{\begin{array}{cccc}1 & c & 0 & s\\
0 & s & 1 & -c\\
0 & 0 & 0 & 0\\
0 & 0 & 0 & 0\end{array}}\right)} & {\scriptstyle B_{v_{2}}=\left(\scriptsize{\begin{array}{cccc}0 & 0 & 0 & 0\\
0 & 0 & 0 & 0\\
1 & -c & 0 & -s\\
0 & -s & 1 & c\end{array}}\right)}\end{array}$\\

\par\end{center}

\begin{center}
where $c=cos(\pi/3)$ and $s=sin(\pi/3)$
\par\end{center}%
\end{minipage}\bigskip{}

\par\end{centering}

\begin{centering}
(a)
\par\end{centering}

\bigskip{}

\begin{centering}
\begin{minipage}[c][1\totalheight]{0.49\columnwidth}%
\begin{center}
\includegraphics[width=0.95\columnwidth]{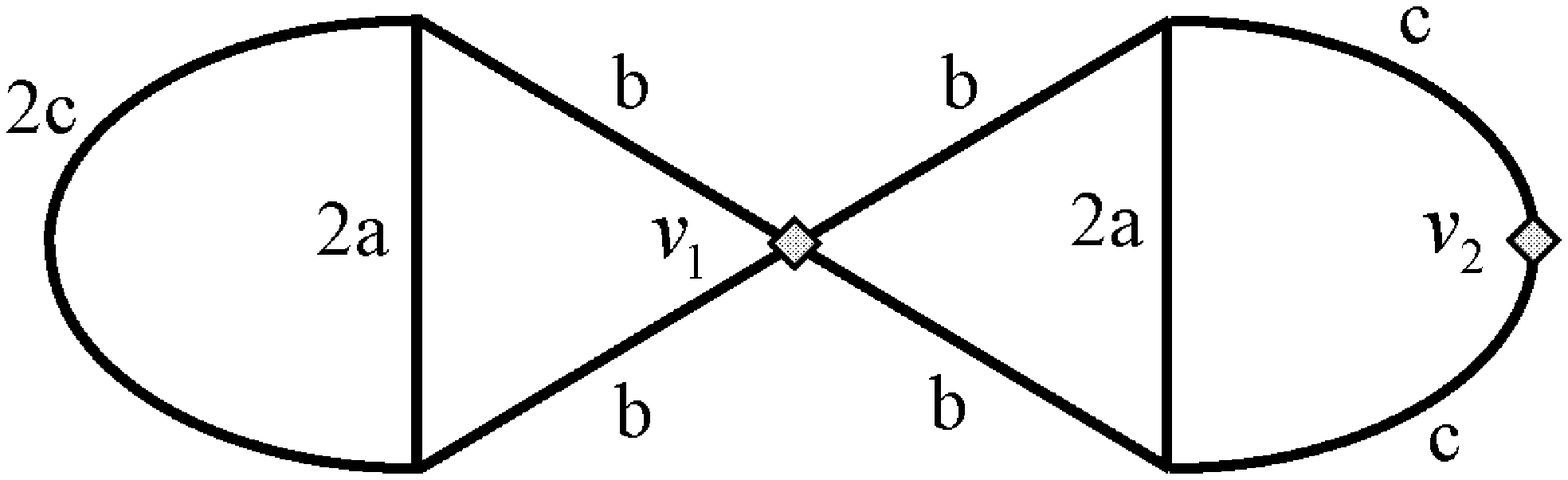}
\par\end{center}%
\end{minipage}%
\begin{minipage}[c][1\totalheight]{0.49\columnwidth}%
\begin{center}
$\begin{array}{cc}
{\scriptstyle A_{v_{1}}=\left(\scriptsize{\begin{array}{cccc}1 & c & 0 & -s\\
0 & s & -1 & c\\
0 & 0 & 0 & 0\\
0 & 0 & 0 & 0\end{array}}\right)} & {\scriptstyle B_{v_{1}}=\left(\scriptsize{\begin{array}{cccc}0 & 0 & 0 & 0\\
0 & 0 & 0 & 0\\
1 & -c & 0 & s\\
0 & -s & -1 & -c\end{array}}\right)}\\
\\{\scriptstyle A_{v_{2}}=}\left(\scriptsize{\begin{array}{cccc}1 & 1\\
0 & 0\end{array}}\right) & {\scriptstyle B_{v_{2}}=}\left(\scriptsize{\begin{array}{cccc}0 & 0\\
1 & -1\end{array}}\right)\end{array}$
\par\end{center}%
\end{minipage}
\par\end{centering}

\bigskip{}

\begin{centering}
(b)
\par\end{centering}

\bigskip{}

\begin{centering}
\begin{minipage}[c][1\totalheight]{0.5\columnwidth}%
\begin{center}
\includegraphics[width=0.9\columnwidth]{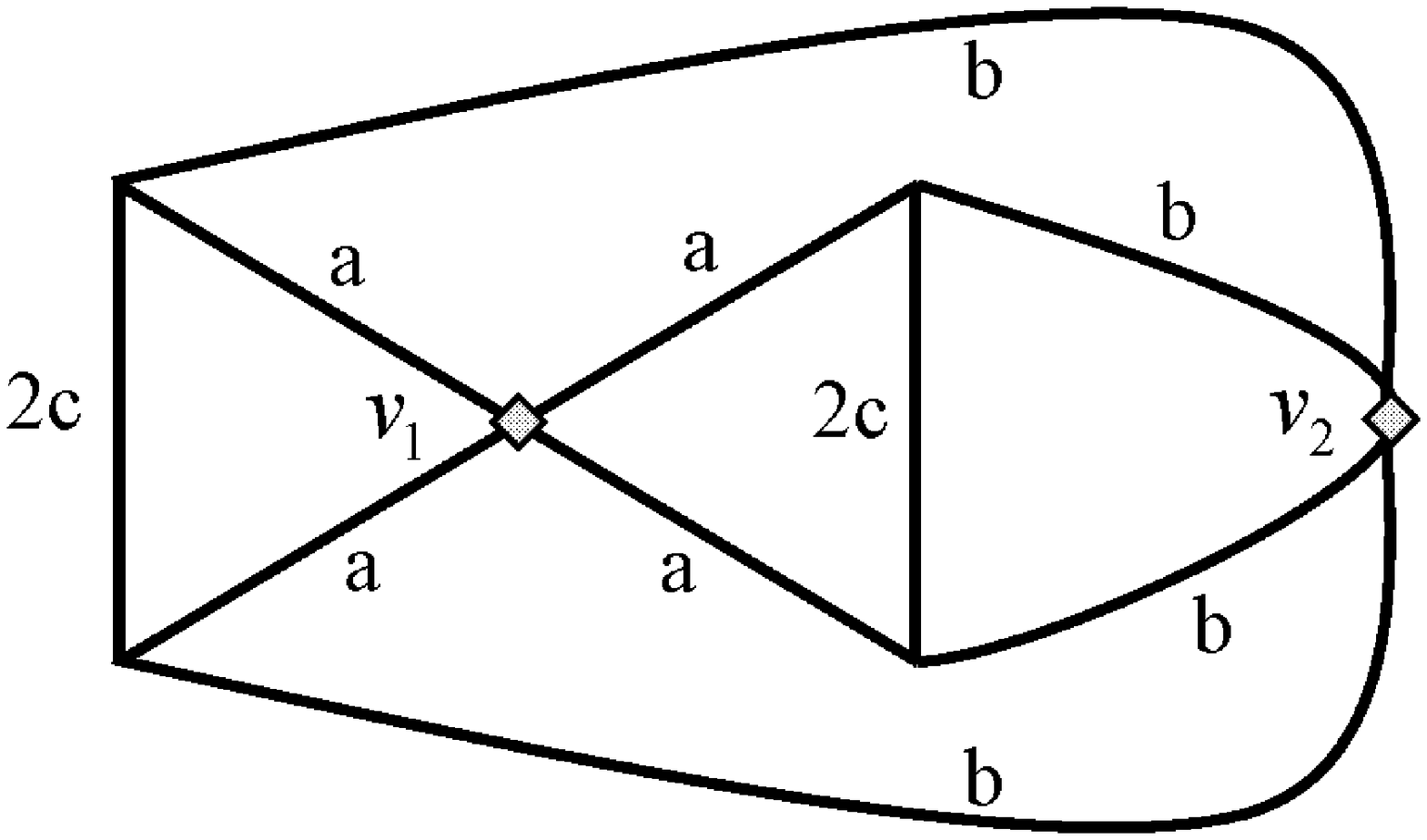}
\par\end{center}%
\end{minipage}%
\begin{minipage}[c][1\totalheight]{0.49\columnwidth}%
\begin{center}
$\begin{array}{cc}
{\scriptstyle A_{v_{1}}=\left(\scriptsize{\begin{array}{cccc}1 & c & 0 & s\\
0 & s & 1 & -c\\
0 & 0 & 0 & 0\\
0 & 0 & 0 & 0\end{array}}\right)} & {\scriptstyle B_{v_{1}}=\left(\scriptsize{\begin{array}{cccc}0 & 0 & 0 & 0\\
0 & 0 & 0 & 0\\
1 & -c & 0 & -s\\
0 & -s & 1 & c\end{array}}\right)}\\
\\{\scriptstyle A_{v_{2}}=\left(\scriptsize{\begin{array}{cccc}1 & c & 0 & -s\\
0 & s & 1 & c\\
0 & 0 & 0 & 0\\
0 & 0 & 0 & 0\end{array}}\right)} & {\scriptstyle B_{v_{2}}=\left(\scriptsize{\begin{array}{cccc}0 & 0 & 0 & 0\\
0 & 0 & 0 & 0\\
1 & -c & 0 & s\\
0 & -s & 1 & -c\end{array}}\right)}\end{array}$
\par\end{center}%
\end{minipage}
\par\end{centering}

\bigskip{}

\begin{centering}
(c)
\par\end{centering}

\bigskip{}

\begin{centering}
\begin{minipage}[c][1\totalheight]{0.5\columnwidth}%
\begin{center}
\includegraphics[width=0.65\columnwidth]{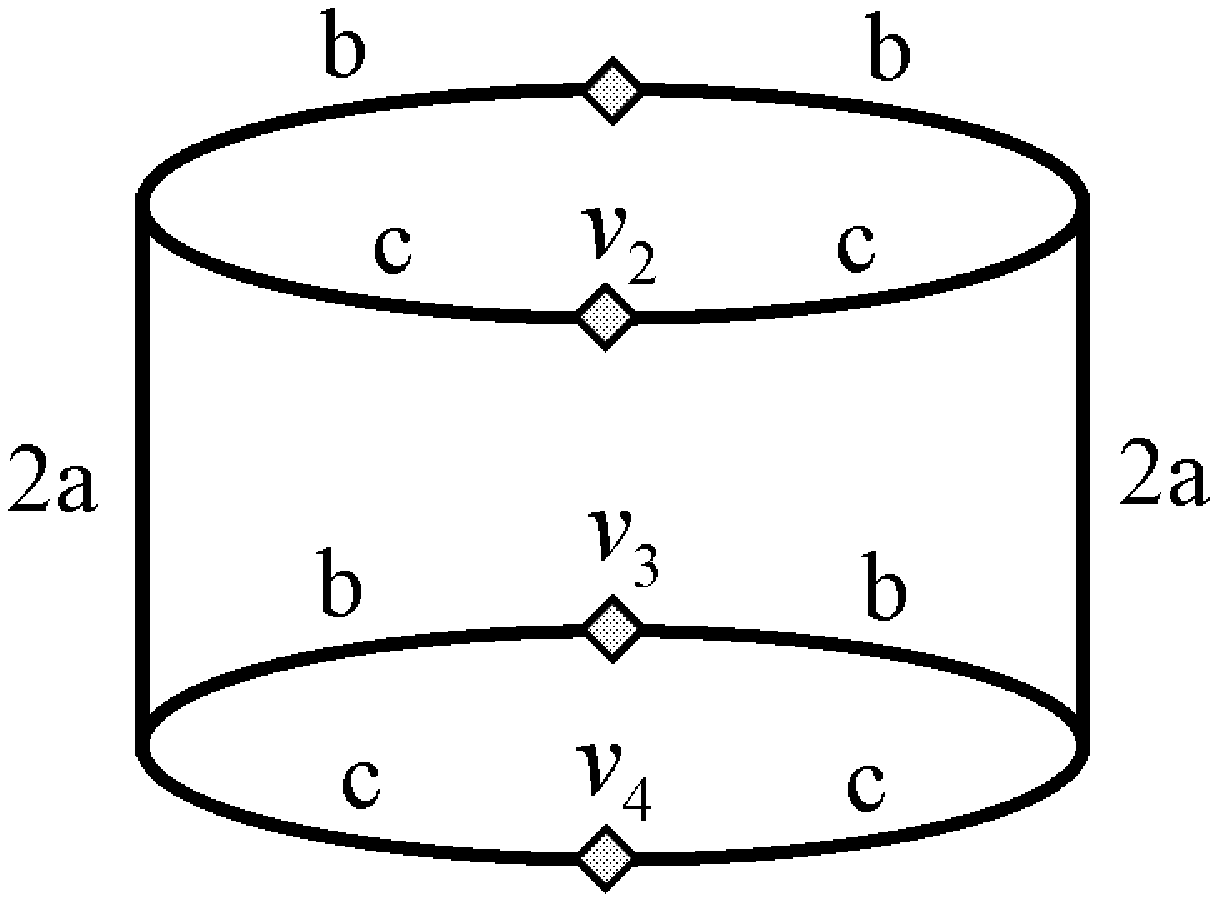}
\par\end{center}%
\end{minipage}%
\begin{minipage}[c][1\totalheight]{0.49\columnwidth}%
\begin{center}
$\begin{array}{cc}
{\scriptstyle A_{v_{1}}=A_{v_{3}}=\left(\scriptsize{\begin{array}{cc}1+c & -s\\
0 & 0\end{array}}\right)} & {\scriptstyle A_{v_{2}}=\left(\scriptsize{\begin{array}{cc}1+c & s\\
0 & 0\end{array}}\right)}\\
\\{\scriptstyle B_{v_{1}}=B_{v_{3}}=\left(\scriptsize{\begin{array}{cc}0 & 0\\
1-c & s\end{array}}\right)} & {\scriptstyle B_{v_{2}}=\left(\scriptsize{\begin{array}{cc}0 & 0\\
1-c & -s\end{array}}\right)}\end{array}$
\par\end{center}%
\end{minipage}
\par\end{centering}

\bigskip{}

\begin{centering}
(d)
\par\end{centering}

\bigskip{}

\caption{The isospectral graphs obtained from the cube. (a)-(c) are
all $\nicefrac{\Gamma}{R_{1}}$, formed by choosing different
representatives for $\nicefrac{E}{O}$ and (d) is the graph
$\nicefrac{\Gamma}{R_{2}}$. Unless stated otherwise, Neumann
boundary conditions are assumed. The boundary conditions of the
marked vertices are given by the matrices on the right.}

\label{fig:CubicGraphs}
\end{figure}

We now consider a larger group, and treat $O_{h}$, the group of
symmetries of the cube. The group $O_{h}$ contains forty-eight
elements: the identity, twenty-three rotations, and twenty-four
reflections. A fundamental domain for the action of $O_{h}$ on the
cube is a tetrahedron. Each of the forty-eight tetrahedra that
together form the cube has one of its faces along the external
boundary of the cube, and its other three faces inside the cube. We
begin by describing a graph, denoted by $\Gamma$, that obeys the
symmetries of $O_{h}$. We consider star graphs with three edges,
such that each tetrahedron contains one such star graph, and the
edges of the star graph are connected to the centre of the interior
faces of the tetrahedron (one edge goes to each face). This star
graph is constructed to have three edges of different lengths,
labeled $a,b$, and $c$. Any two star graphs in neighbouring
tetrahedra are then connected at the center of their common face.
$\Gamma$ is the union of these star graphs. It is a three-regular,
bipartite graph, with forty-eight vertices, and seventy-two edges of
lengths $2a,2b$ and $2c$, each with equal multiplicity. We will use
the subgroups $O,T_{d}\leq O_{h}$, along with appropriate
representations, to form the isospectral quotient graphs.

We first consider the subgroup $O\leq O_{h}$, known as the
octahedral group, which contains the identity and the twenty-three
rotation elements. We take representatives for $\nicefrac{E}{O}$ and
$\nicefrac{V}{O}$. Since $O$ is a subgroup of index two, one
possiblitiy is to choose the edges and vertices contained in two
tetrahedra as representatives. $O$ has two 1-dimensional, one
2-dimensional, and two 3-dimensional irreducible representations
\cite{CubicAlgebra}. We work with the two dimensional
representation, using the basis given in \cite{CubicAlgebra} and we
denote it as $R_{1}$. The quotient graph $\nicefrac{\Gamma}{R_{1}}$
is shown in figure \ref{fig:CubicGraphs}(a). As noted in sections
\ref{sec:rigorous}, \ref{subsec:transplantation}, it is possible to
make different choices for the representatives. We present three
different quotient graphs for $\nicefrac{\Gamma}{R_{1}}$,
corresponding to three different choices of representatives, in
figure \ref{fig:CubicGraphs}(a)-(c).

We now examine the subgroup $T_{d}\leq O_{h}$. The vertices of the
cube consist of two sets, each of which forms the vertices of an
equilateral tetrahedron. $T_{d}$ contains all the elements of
$O_{h}$ whose action does not mix between the two sets. It should be
noted that $O\cong T_{d}\cong S_4$. In particular, $T_{d}$ and $O$
have the same irreducible representations. We will take the matrix
representation that was used for $O$, and use it for $T_{d}$,
denoting it by $R_{2}$. The quotient graph
$\nicefrac{\Gamma}{R_{2}}$
is shown in figure \ref{fig:CubicGraphs}(d).\\
 The isospectrality of the quotient graphs formed from $O$ and $T_{d}$
follows from the fact that
$\ind_{O}^{O_{h}}R_{1}\cong\ind_{T_{d}}^{O_{h}}R_{2}$.

\subsection{$G=D_{3}$ $-$ A free action on $\Gamma$}
\label{subsec:holy_triangle}

\begin{figure}[!h]

\begin{minipage}[b][1\totalheight]{0.48\columnwidth}
\begin{center}
\includegraphics[width=0.95\textwidth]{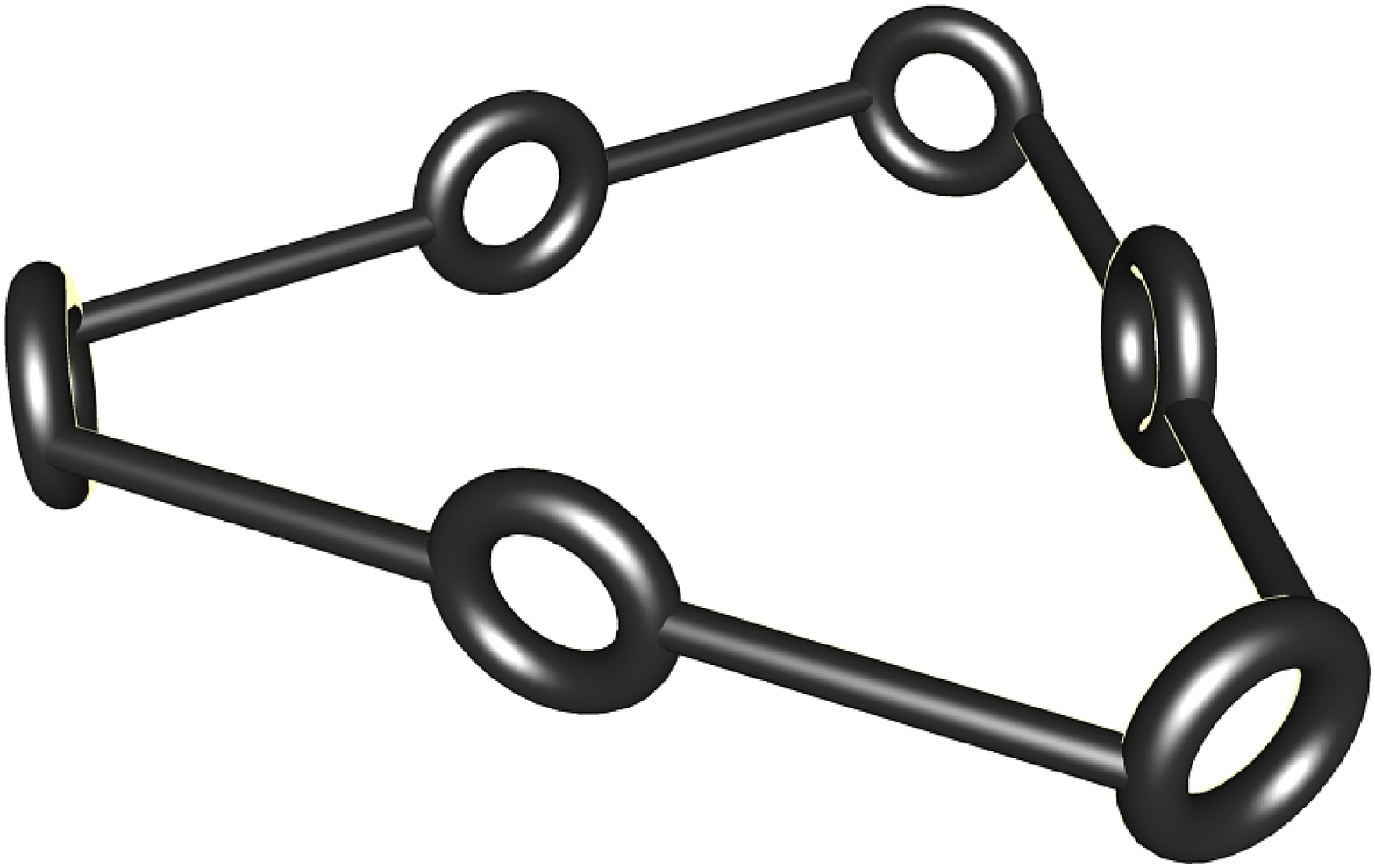}
\end{center}
\vspace{0pt}
\begin{center}
(a)  \;\;
\end{center}
\end{minipage} \hfill{}
\begin{minipage}[b][1\totalheight]{0.48\columnwidth}
\begin{center}
\includegraphics[scale=0.35]{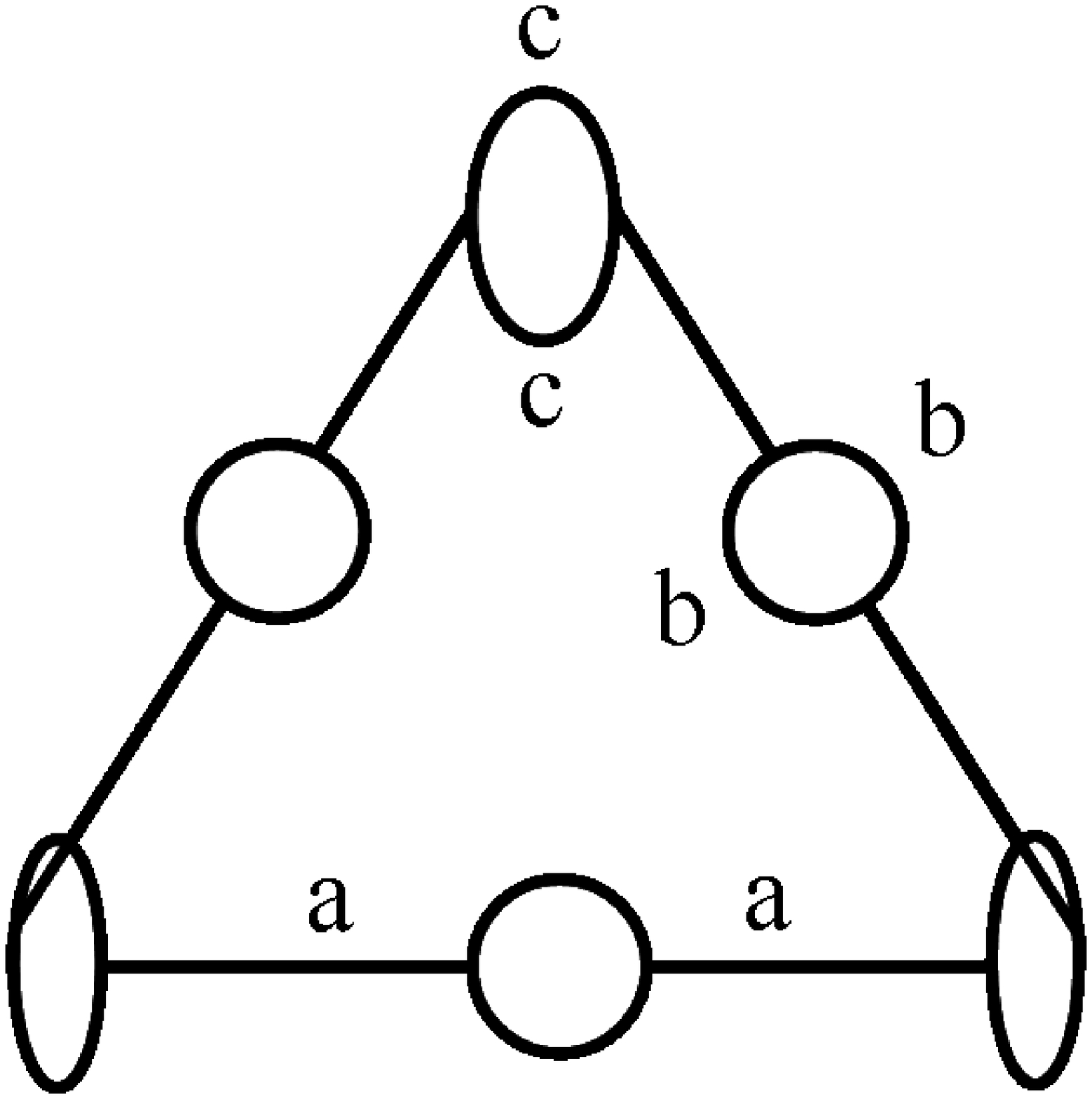}
\end{center}
\vspace{-8pt}
\begin{center}
(b)  \;\;
\end{center}
\end{minipage}

 \caption{Two visualizations of the graph $\Gamma$ which is symmetric
under $D_{3}$. (a) A three dimensional visualization. The loops are
perpendicular to the plane of the triangle. $\sigma$ is a rotation
of $\nicefrac{2\pi}{3}$ about an axis perpendicular to the plane of
the triangle, and $\tau$ is a rotation of $\pi$ about a height of
the triangle. (b) A two dimensional visualization whose edges'
lengths are marked.}

\label{fig:HolyTriangle_Full}
\end{figure}

We now consider an example that demonstrates the possibility of
constructing quotient graphs that have only Neumann boundary
conditions. We show that this can be achieved even when dividing by
multidimensional representations. We consider the graph $\Gamma$,
shown in figure \ref{fig:HolyTriangle_Full}, which is symmetric
under the action of $D_{3}\cong S_{3}$. The group acts freely on
both the vertices and edges of $\Gamma$. This is ensured by having
the circles perpendicular to the plane of the triangle. We have
$D_{3}=\{e,\sigma,\sigma^{2},\tau,\tau\sigma,\tau\sigma^{2}\}$,
where $\sigma$ is a rotation of $\nicefrac{2\pi}{3}$ about the axis
perpendicular to the plane of the triangle, and $\tau$ is a rotation
of $\pi$ about the height of the triangle.
%

 Rather than working
with representations of various subgroups, we use various bases of a
representation of the entire group to create the quotient graphs.
$D_{3}$ has three irreducible representations: the trivial
representation $S_{1}$, the sign representation $S_{2}$, and the
standard representation $S_{3}$, which is of dimension two. We take
the eight-dimensional representation $R~=~3S_{1}+S_{2}+2S_{3}$. We
begin by choosing a basis such that the matrix representation of $R$
is block diagonalized as follows: \[ \left(\begin{array}{c|c|c}
S_{1}+S_{2}+2S_{3} & \multicolumn{2}{c}{}\\
\cline{1-2} & S_{1}\\
\cline{2-3}\multicolumn{2}{c|}{} & S_{1}\end{array}\right).\]
 This choice of basis is not unique. We therefore specify it, by identifying
the top block as the regular representation of $D_{3}$, and taking
the standard basis for this representation. Due to the fact that the
matrix representation is block diagonalized, the quotient graph
created consists of three disjoint graphs. Each graph corresponds to
one block, and these are shown in figure
\ref{fig:HolyTriangleGraphs1}.
\begin{center}
\begin{figure}[!h]

\begin{centering}
\begin{minipage}[c][1\totalheight]{0.5\columnwidth}%
\begin{center}
\includegraphics[scale=0.28]{Holy_triangle_2d}
\par\end{center}

\medskip{}

\begin{center}
(a)
\par\end{center}%
\end{minipage}%
\begin{minipage}[c][1\totalheight]{0.49\columnwidth}%
\begin{center}
\includegraphics[scale=0.4]{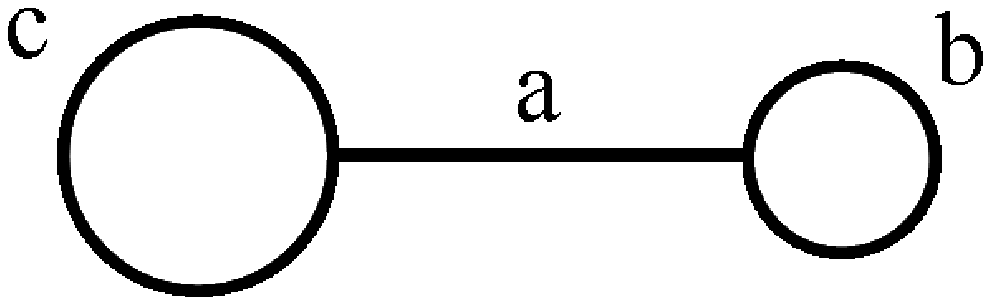}
\par\end{center}

\begin{center}
(b)
\par\end{center}

\medskip{}

\begin{center}
\includegraphics[scale=0.4]{Holy_quotient_1}
\par\end{center}

\begin{center}
(c)
\par\end{center}%
\end{minipage}
\par\end{centering}

\caption{The first quotient graph
$\nicefrac{\Gamma}{S_{1}+S_{2}+2S_{3}}\cup\nicefrac{\Gamma}{S_{1}}\cup\nicefrac{\Gamma}{S_{1}}$.
All vertices have Neumann boundary conditions. (a) The graph formed
from the first block of the representation. Note that in this case,
we have recovered the original graph. (b)-(c) Two copies, coming
from the two blocks containing the trivial representation.}

\label{fig:HolyTriangleGraphs1}
\end{figure}

\par\end{center}
Note that this representation consists of permutation matrices.
This, together with the fact that the action of $D_{3}$ on $\Gamma$
is free, ensures Neumann boundary conditions on the quotient graph.
We now choose a new basis, such that the matrix representation of
$R$ takes the form: \[ \left(\begin{array}{c|c|c}
S_{1}+S_{3} & \multicolumn{2}{c}{}\\
\cline{1-2} & S_{1}+S_{3}\\
\cline{2-3}\multicolumn{2}{c|}{} & S_{1}+S_{2}\end{array}\right).\]
 Again we must further specify the choice of basis. Each of the first two
blocks is the permutation representation of the symmetric group of
size three, and we choose its standard basis. For the third block,
we choose the basis in which: \begin{eqnarray*}
\sigma\mapsto\left(\begin{array}{cc}1 & 0\\
0 & 1\end{array}\right)  , \;\; \tau\mapsto\left(\begin{array}{cc}0 & 1\\
1 & 0\end{array}\right)\,.\end{eqnarray*}
 Using this basis, we again obtain a quotient graph consisting of three
disjoint graphs, corresponding to the three blocks of the matrix
representation, as shown in figure \ref{fig:HolyTriangleGraphs2}.
This quotient also has only Neumann boundary conditions for the same
reason. The two quotient graphs, namely
$\nicefrac{\Gamma}{S_{1}+S_{2}+2S_{3}}\cup\nicefrac{\Gamma}{S_{1}}\cup\nicefrac{\Gamma}{S_{1}}$
and
$\nicefrac{\Gamma}{S_{1}+S_{3}}\cup\nicefrac{\Gamma}{S_{1}+S_{3}}\cup\nicefrac{\Gamma}{S_{1}+S_{2}}$,
are isospectral.

\begin{figure}[!h]
\begin{centering}
\begin{minipage}[c][1\totalheight]{0.6\columnwidth}%
\begin{center}
\includegraphics[scale=0.4]{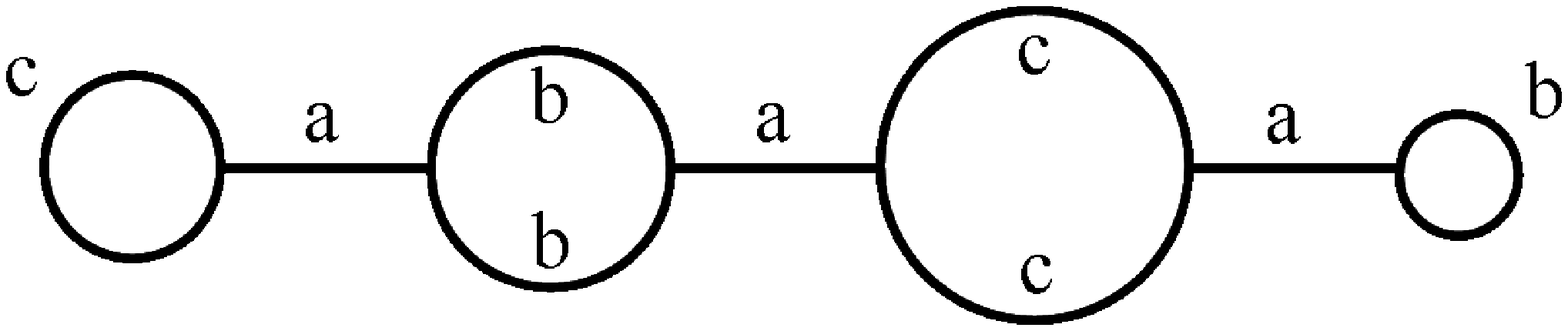}
\par\end{center}

\begin{center}
(a)
\par\end{center}

\medskip{}

\begin{center}
\includegraphics[scale=0.4]{Holy_quotient_3}
\par\end{center}

\begin{center}
(b)
\par\end{center}%
\end{minipage}%
\begin{minipage}[c][1\totalheight]{0.39\columnwidth}%
\begin{center}
\includegraphics[scale=0.4]{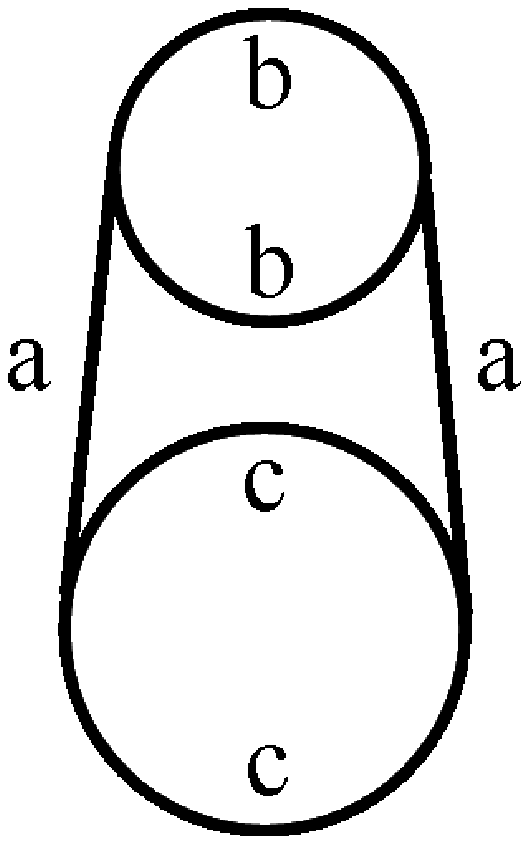}
\par\end{center}

\medskip{}

\begin{center}
(c)
\par\end{center}%
\end{minipage}
\par\end{centering}

\caption{The second quotient graph
$\nicefrac{\Gamma}{S_{1}+S_{3}}\cup\nicefrac{\Gamma}{S_{1}+S_{3}}\cup\nicefrac{\Gamma}{S_{1}+S_{2}}$.
All vertices have Neumann boundary conditions. (a)-(b) The graphs
formed from the first two blocks of the representation. (c) The
graph obtained from the third block.}
\label{fig:HolyTriangleGraphs2}
\end{figure}

\begin{rem*}
As a matter of fact, all the quotients obtained in this subsection
can also be obtained as quotients by the trivial representations of
$D_3$'s subgroups, as
\begin{eqnarray*}
\ind_{D_3}^{D_3}\mathbf{1} &\cong& S_1 \\
\ind^{D_3}_{\left<\sigma\right>}\mathbf{1} &\cong& S_1+S_2 \\
\ind^{D_3}_{\left<\tau\right>}\mathbf{1}&\cong&  S_1+S_3 \\
\ind^{D_3}_{\left\{id\right\}}\mathbf{1} &\cong& S_1+S_2+2S_3.
\end{eqnarray*}
\end{rem*}

\section{Drums and Manifolds}

\label{sec:drums}

We now apply theorem (\ref{thm:mainthm}) to other objects. In
particular, we reconstruct some existing examples using our method,
and comment on some new results.

\subsection{Jakobson, Levitin, et al.}

We begin by examining the isospectral domains presented by Jakobson
et al.\ and Levitin et al.~\cite{Jakobson,Levitin}. It is possible
to recover all the isospectral examples described in these papers as
quotients by representations with isomorphic inductions. We consider
first an interesting example, consisting of the four isospectral
domains shown in figure \ref{fig:isospectral_quartet} (See also
figure 7 in \cite{Levitin}).
\begin{center}
\begin{figure}[!h]
\begin{centering}
\begin{minipage}[c][1\totalheight]{0.24\columnwidth}%
\begin{center}
\includegraphics[width=0.9\textwidth]{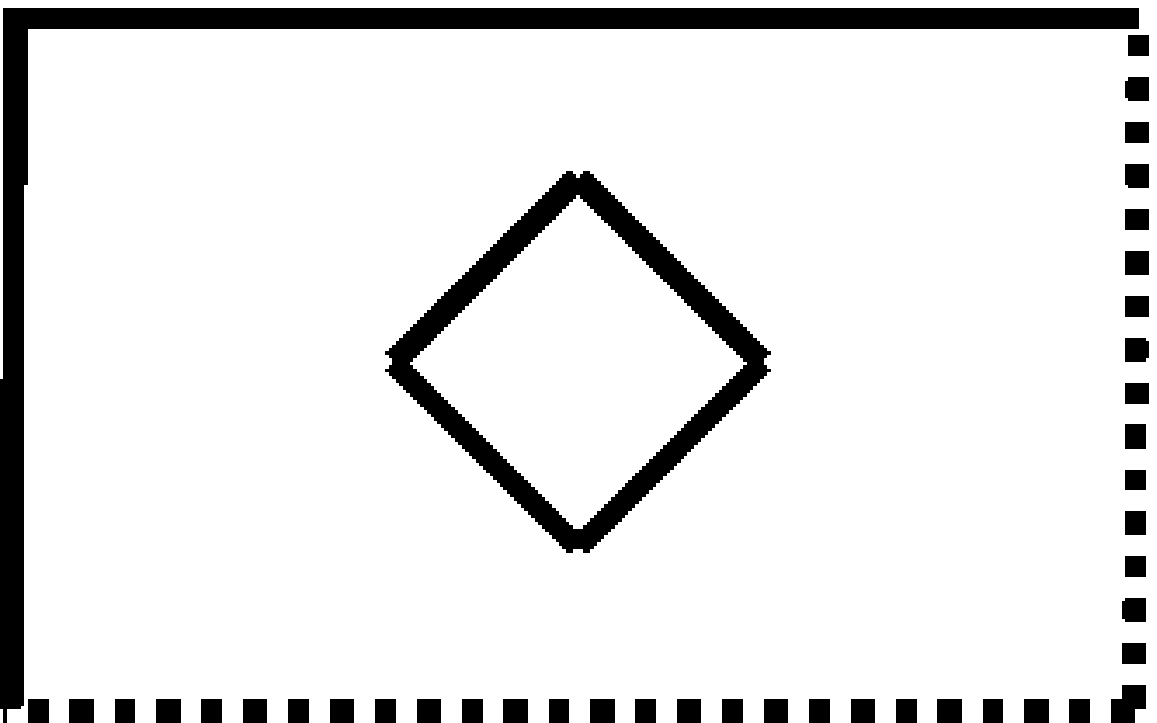}
\par\end{center}

\begin{center}
$\nicefrac{\mathbb{T}}{R_{1}\otimes R_{1}}$
\par\end{center}%
\end{minipage}%
\begin{minipage}[c][1\totalheight]{0.24\columnwidth}%
\begin{center}
\includegraphics[width=0.9\textwidth]{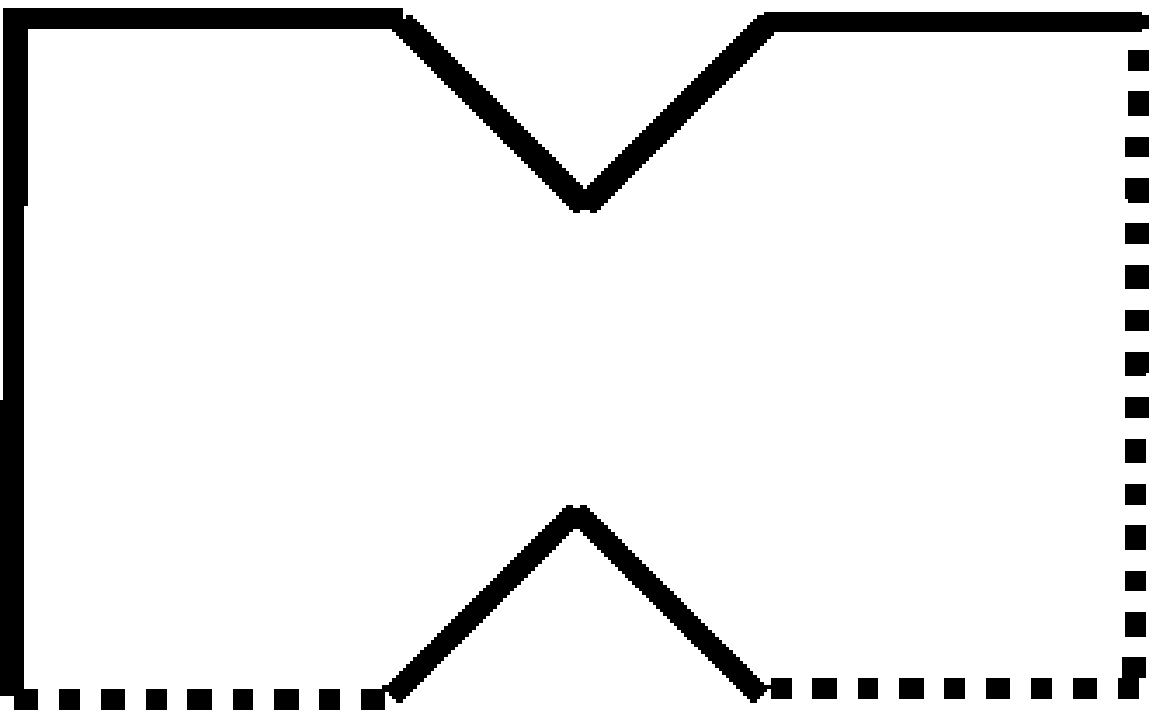}
\par\end{center}

\begin{center}
$\nicefrac{\mathbb{T}}{R_{1}\otimes R_{2}}$
\par\end{center}%
\end{minipage}%
\begin{minipage}[c][1\totalheight]{0.24\columnwidth}%
\begin{center}
\includegraphics[width=0.9\textwidth]{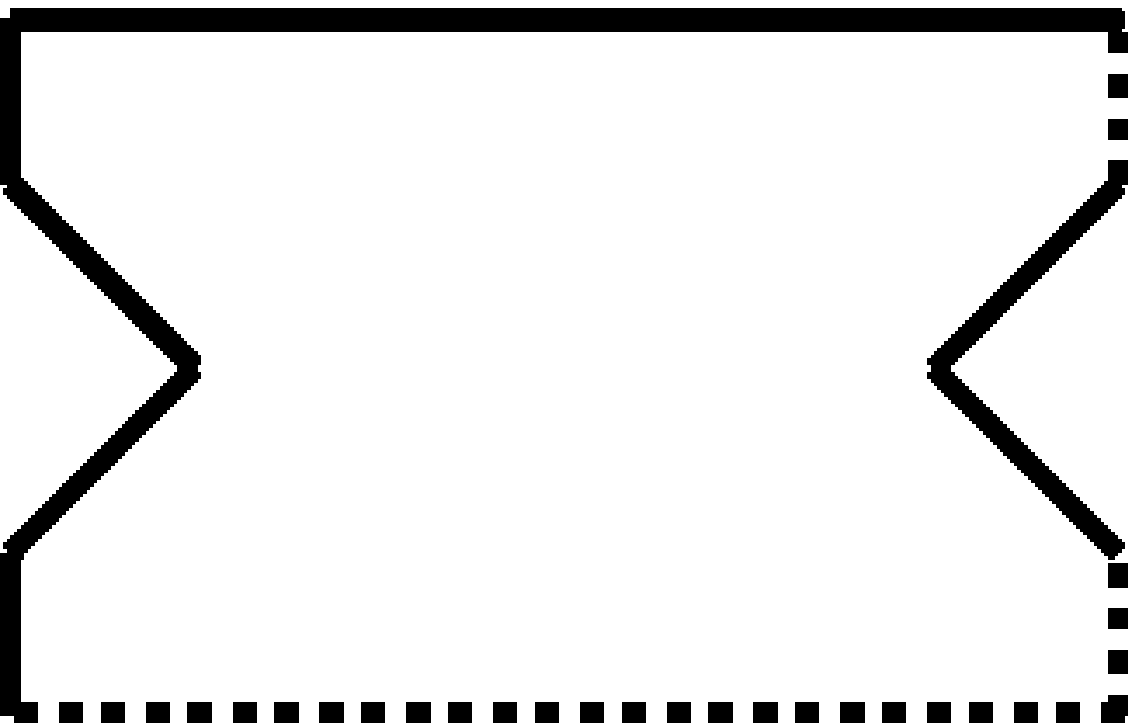}
\par\end{center}

\begin{center}
$\nicefrac{\mathbb{T}}{R_{2}\otimes R_{1}}$
\par\end{center}%
\end{minipage}%
\begin{minipage}[c][1\totalheight]{0.24\columnwidth}%
\begin{center}
\includegraphics[width=0.9\textwidth]{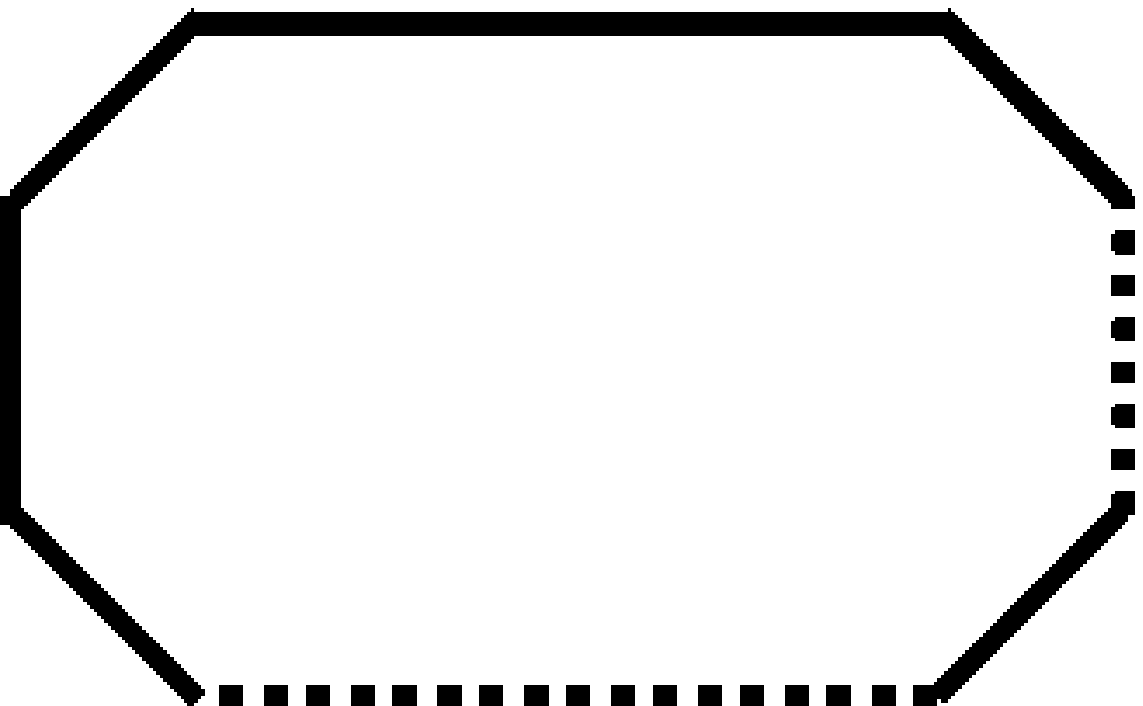}
\par\end{center}

\begin{center}
$\nicefrac{\mathbb{T}}{R_{2}\otimes R_{2}}$
\par\end{center}%
\end{minipage}
\par\end{centering}

\caption{The four isospectral domains presented in \cite{Levitin},
obtained as quotients of the torus $\mathbb{T}$ (figure
\ref{fig:quartet_torus}). solid lines indicate Dirichlet boundary
conditions, and dotted ones Neumann.}

\label{fig:isospectral_quartet}
\end{figure}

\par\end{center}
 We take a torus, $\mathbb{T}$, with
sixteen diamond-shaped holes, as shown in figure
\ref{fig:quartet_torus}. We can express the torus as $T_{a}\times
T_{b}$, where $T_{a},\, T_{b}$ are circles with circumferences $a$
and $b$ respectively. Taking a rigid action of $D_{4}$ on each of
the circles $T_{a}$ and $T_{b}$, we obtain an action of
$G=D_{4}\times D_{4}$ on the torus $T_{a}\times T_{b}$: the action
of the first $D_{4}$ in the direct product is by horizontal
translations and reflections, and that of the second $D_{4}$ is by
vertical ones. For example, the element $(\sigma,e)$ transforms
$\mathbb{T}$ rigidly onto itself so that the four columns of
diamonds are cyclically shifted by one, and the element $(\tau,e)$
swaps the first column with the fourth, and the second with the
third. Similarly, the action of elements of the form $(e,g)$ is by
transformations that permute the rows of diamonds.
\begin{center}
\begin{figure}[!h]
\begin{centering}
\includegraphics[scale=0.6]{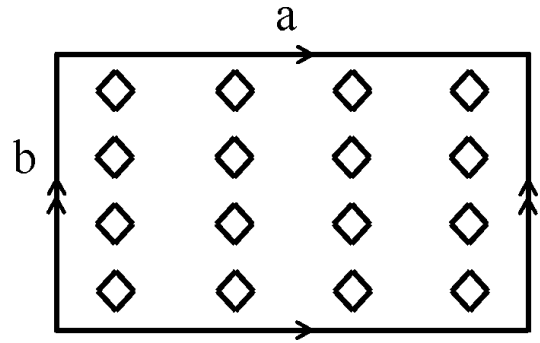}
\par\end{centering}

\caption{The torus $\mathbb{T}$, of which the four domains in figure
\ref{fig:isospectral_quartet} are quotients.}
\label{fig:quartet_torus}
\end{figure}

\par\end{center}
 Adopting the notations of
section~\ref{sec:basic_example}, we examine the subgroups $\left\{
H_{i}\times H_{j}\right\} _{i,j\in\left\{ 1,2\right\} }$ of
$D_{4}\times D_{4}$ and their corresponding representations $\left\{
R_{i}\otimes R_{j}\right\} _{i,j\in\left\{ 1,2\right\} }$. Since
$\ind_{H_{1}}^{D_{4}}R_{1}\cong\ind_{H_{2}}^{D_{4}}R_{2}$, all of
these representations have isomorphic inductions in $D_{4}\times
D_{4}$. Applying corollary \ref{cor:sunadapair}, we obtain that
$\left\{ \nicefrac{\mathbb{T}}{R_{i}\otimes R_{j}}\right\}
_{i,j\in\left\{ 1,2\right\} }$ are isospectral (figure
\ref{fig:isospectral_quartet}).
\begin{center}
\begin{figure}[!h]
\begin{centering}
\begin{minipage}[c][1\totalheight]{0.69\columnwidth}%
\begin{center}
\includegraphics[width=0.9\textwidth]{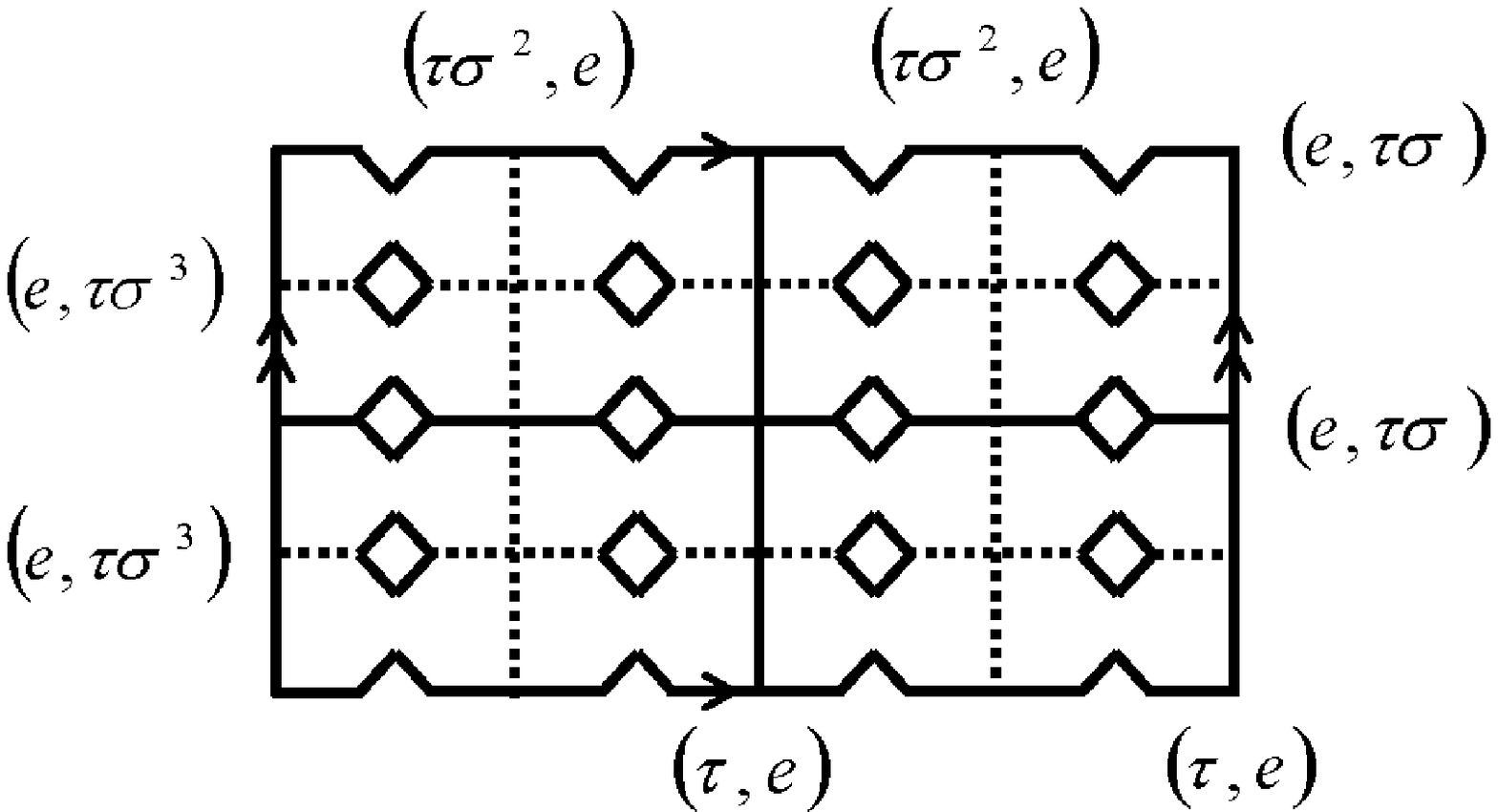}
\par\end{center}

\begin{center}
(a)
\par\end{center}%
\end{minipage}%
\begin{minipage}[c][1\totalheight]{0.3\columnwidth}%
\begin{center}
\includegraphics[width=0.7\columnwidth]{quartet2}
\par\end{center}

\medskip{}

\begin{center}
(b)
\par\end{center}%
\end{minipage}
\par\end{centering}

\caption{(a) The information we have on
$\tilde{f}\in\Phi_{\mathbb{T}}^{R_{1}\otimes R_{2}}(\lambda)$. The
function vanishes along the solid lines and its normal derivatives
at the dotted lines are zero. (b) The quotient planar domain
$\nicefrac{\mathbb{T}}{{R_{1}\otimes R_{2}}}$ which encodes this
information. Solid lines stand for Dirichlet boundary conditions and
dotted lines for Neumann.} \label{fig:quartet_torus2}
\end{figure}
\par\end{center}
To demonstrate how the technicalities work for these domains, we
construct
$\nicefrac{\mathbb{T}}{R_{1}\otimes R_{2}}$. 
From (\ref{eq:r1_rep}) and (\ref{eq:r2_rep}), we obtain information
on $\tilde{f}\in\Phi_{\mathbb{T}}^{R_{1}\otimes R_{2}}(\lambda)$: we
see that $\tilde{f}$ vanishes on the axes of reflection of
$(\tau,e)$, as it is anti-symmetric with respect to this reflection,
and similarly for $(e,\tau\sigma)$. Since $\tilde{f}$ is symmetric
with respect to the axes of $(\tau\sigma^{2},e)$ and
$(e,\tau\sigma^{3})$, its normal derivatives with respect to these
axes are zero. This is summarized in figure
\ref{fig:quartet_torus2}(a) and the resulting quotient is shown in
figure \ref{fig:quartet_torus2}(b). Notice that the two parts of
figure \ref{fig:quartet_torus2} serve the same purpose as those of
figure \ref{fig:intuitive_quotient1}, i.e., to demonstrate how the
information on a function belonging to a certain isotypic component
is encoded in the boundary conditions of the quotient. We end by
remarking that all the constructions demonstrated in section
\ref{sec:extending_example} can be applied analogously to
$\mathbb{T}$ to enlarge the isospectral quartet in figure
\ref{fig:isospectral_quartet}. However, the quotients obtained from
most choices of representations and bases will not be planar
domains, or even manifolds. For example, consider the representation
$R_3$ of $H_3$, given in \eref{eq:r3_rep}. The quotient
$\nicefrac{\mathbb{T}}{R_{1}\otimes R_{3}}$ (figure
\ref{fig:quartet_torus3}) is a cylinder with Dirichlet and Neumann
conditions at its boundaries, and a ``factor of $i$'' condition
along a section line normal to the boundaries (compare with the
quotient $\nicefrac{\Gamma}{R_{3}}$ introduced in section
\ref{sec:extending_example}). This quotient is a manifold with a
singularity. Other types of singularities may arise when considering
quotients with respect to multidimensional representations.
\begin{center}
\begin{figure}[!h]
\begin{centering}
\begin{minipage}[c][1\totalheight]{1.\columnwidth}%
\begin{center}
\includegraphics[width=0.3\columnwidth]{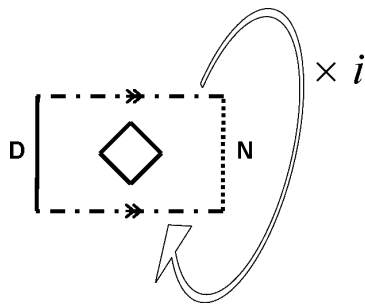}
\par\end{center}
\end{minipage}
\par\end{centering}
\caption{The quotient $\nicefrac{\mathbb{T}}{{R_{1}\otimes R_{3}}}$.
Solid lines stand for Dirichlet boundary conditions and dotted lines
for Neumann. There is a factor of i between the values and
derivatives at the top and bottom edges.} \label{fig:quartet_torus3}
\end{figure}
\par\end{center}

We now examine a more general example presented in \cite{Levitin}.
Consider the cylindrical drum $\mathbb{D}$, shown in
figure~\ref{fig:D2n}. In the figure, the left and the right edges
are identified. The three dots imply that the basic pattern which
appears in figure \ref{fig:D2n_FD} is repeated and $\mathbb{D}$
consists of $8n$ copies of it. $\mathbb{D}$ is symmetric under the
action of the dihedral group $D_{4n}$, where $\sigma$ rotates the
cylinder and $\tau$ is a reflection whose axis is shown in figure
\ref{fig:D2n}.

\begin{center}
\begin{figure}[!h]
\begin{centering}
\begin{minipage}[c][1\totalheight]{0.96\columnwidth}%
\begin{center}
\includegraphics[width=.95\textwidth]{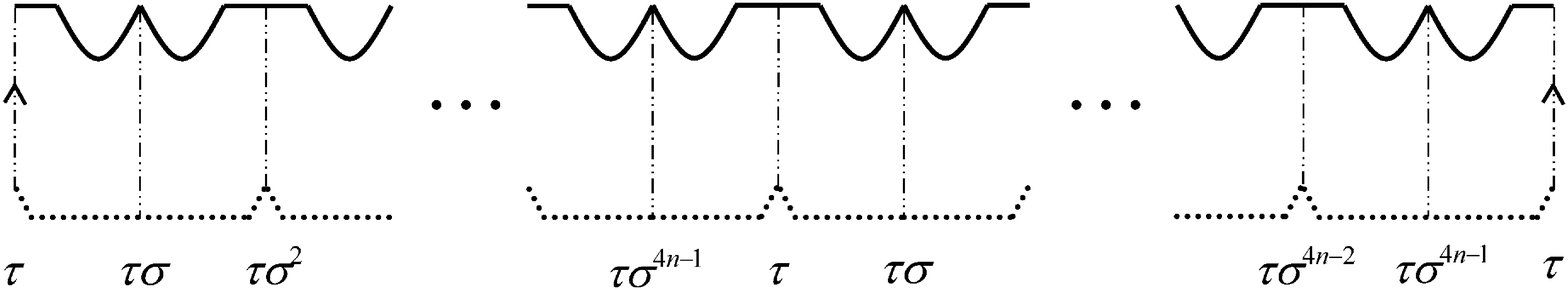}
\par\end{center}
\end{minipage}
\par\end{centering}
\caption{The cylindrical drum $\mathbb{D}$. The left and right edges
are identified. Solid lines stand for Dirichlet boundary conditions
and dotted ones for Neumann. The reflection elements in $D_{4n}$ and
their axes are indicated.} \label{fig:D2n}
\end{figure}
\par\end{center}

\begin{center}
\begin{figure}[!h]
\begin{centering}
\includegraphics[width=.07\textwidth]{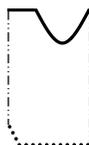}
\par\end{centering}
\caption{A fundamental domain of $\mathbb{D}$ under the action of
$D_{4n}$.} \label{fig:D2n_FD}
\end{figure}
\par\end{center} We consider the subgroups $H_{1},H_{2}\leq D_{4n}$,
where
\begin{eqnarray*}
H_{1} & = & \{e,\,\tau,\,\tau\sigma^{2n},\,\sigma^{2n}\}\\
H_{2} & = &
\{e,\,\tau\sigma,\,\tau\sigma^{2n+1},\,\sigma^{2n}\},\end{eqnarray*}
 equipped with the one-dimensional representations \begin{eqnarray*}
R_{1} & : & \left\{ \begin{array}{llll}
e\mapsto\left(1\right), & \tau\mapsto\left(-1\right), & \tau\sigma^{2n}\mapsto\left(1\right), & \sigma^{2n}\mapsto\left(-1\right)\end{array}\right\} \\
R_{2} & : & \left\{ \begin{array}{llll} e\mapsto\left(1\right), &
\tau\sigma\mapsto\left(1\right), &
\tau\sigma^{2n+1}\mapsto\left(-1\right), &
\sigma^{2n}\mapsto\left(-1\right)\end{array}\right\}
.\end{eqnarray*}
 We find that $\ind_{H_{1}}^{D_{4n}}R_{1}\cong\ind_{H_{2}}^{D_{4n}}R_{2}$,
and therefore we can form two isospectral quotients, each one a
quarter the width of the original drum, with Dirichlet boundary
conditions on one side, and Neumann on the other. In
figure~\ref{fig:D2n_quotients} the drum $\mathbb{D}$ and the
resulting quotients are shown for the case $n=2$. In conclusion, we
have provided an alternate proof for theorem~4.2 in~\cite{Levitin}.
In fact, this proof is valid for any number $n$, whereas the
original theorem in \cite{Levitin} treated only the case $n=2^{k}$.
The reader might wonder why did we limit our attention to $D_{4n}$.
The answer is that if we consider the more general $D_n$, then in
order to define $H_1$ and $H_2$, $n$ must be even. If $n$ is even,
but not a multiple of four, then the proof works, but in that case
$H_1$ and $H_2$ are also conjugate in $D_4$, so that the isospectral
domains thus obtained are also isometric.
\begin{center}
\begin{figure}
\begin{centering}
\begin{minipage}[c][1\totalheight]{0.96\columnwidth}%
\begin{center}
\includegraphics[width=0.95\textwidth]{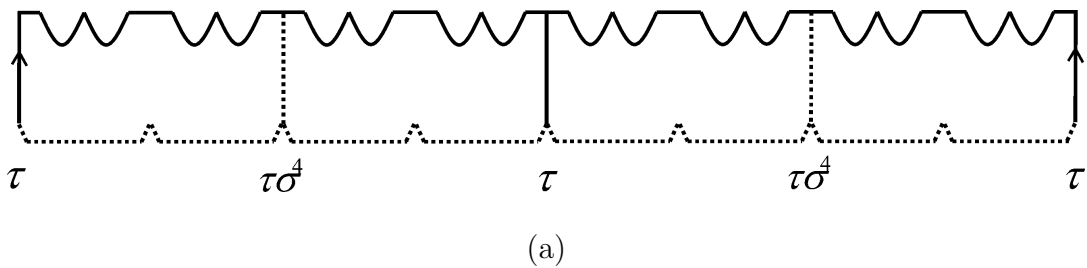}
\par\end{center}

\begin{center}
(a)
\par\end{center}%
\end{minipage}\\

\par\end{centering}

\vspace{10mm}

\begin{centering}
\begin{minipage}[c][1\totalheight]{0.48\columnwidth}%
\begin{center}
\includegraphics[width=0.6\textwidth]{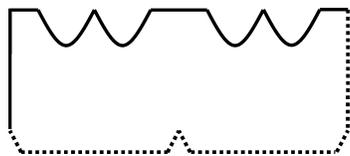}
\par\end{center}

\begin{center}
(b) - $\nicefrac{\mathbb{D}}{R_{1}}$
\par\end{center}%
\end{minipage}%
\begin{minipage}[c][1\totalheight]{0.48\columnwidth}%
\begin{center}
\includegraphics[width=0.6\textwidth]{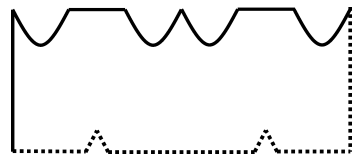}
\par\end{center}

\begin{center}
(c) - $\nicefrac{\mathbb{D}}{R_{2}}$
\par\end{center}%
\end{minipage}
\par\end{centering}

\caption{(a) The information we have on
$\tilde{f}\in\Phi_{\mathbb{D}}^{R_{1}}(\lambda)$. The function
vanishes along the solid lines, and its normal derivatives at the
dotted lines are zero. (b),(c) The quotient planar domains
$\nicefrac{\mathbb{D}}{R_{1}}$ and $\nicefrac{\mathbb{D}}{R_{2}}$.
Solid lines stand for Dirichlet boundary conditions and dotted ones
for Neumann. These domains illustrate the case $n=2$.}

\label{fig:D2n_quotients}
\end{figure}
\par\end{center}

\subsection{Gordon, Webb and Wolpert}
The famous domains of Gordon, Webb and Wolpert, originally presented
in \cite{Gordon1,Gordon2}, can be similarly constructed by our
method. Buser et al.\ \cite{BuserConway} have shown that they can be
constructed as quotients of the hyperbolic plane. This is done by
considering an epimorphism from the symmetry group of this plane
onto $\mathrm{PSL_{3}\left(\mathbb{F}_{2}\right)}$, and taking the
inverse images of two subgroups in
$\mathrm{PSL_{3}\left(\mathbb{F}_{2}\right)}$, each isomorphic to
$S_{4}$. In our formalism, the drums of Gordon et al.\ are obtained
(with Neumann boundary conditions) from the quotient of the
hyperbolic plane by the corresponding trivial representations of
$S_{4}$. Using the sign representation of $S_{4}$ instead we obtain
the same drums with different boundary conditions (figure
\ref{fig:new_Gordon}). The conditions of corollary
\ref{cor:sunadapair} are satisfied in this case as well, so that
this pair of drums is isospectral. A more detailed explanation is
given in \cite{PB08}.
\begin{center}
\begin{figure}[!h]
\hfill{}%
\begin{minipage}[t][1\totalheight]{0.4\columnwidth}%
\begin{center}
\includegraphics[width=0.5\textwidth]{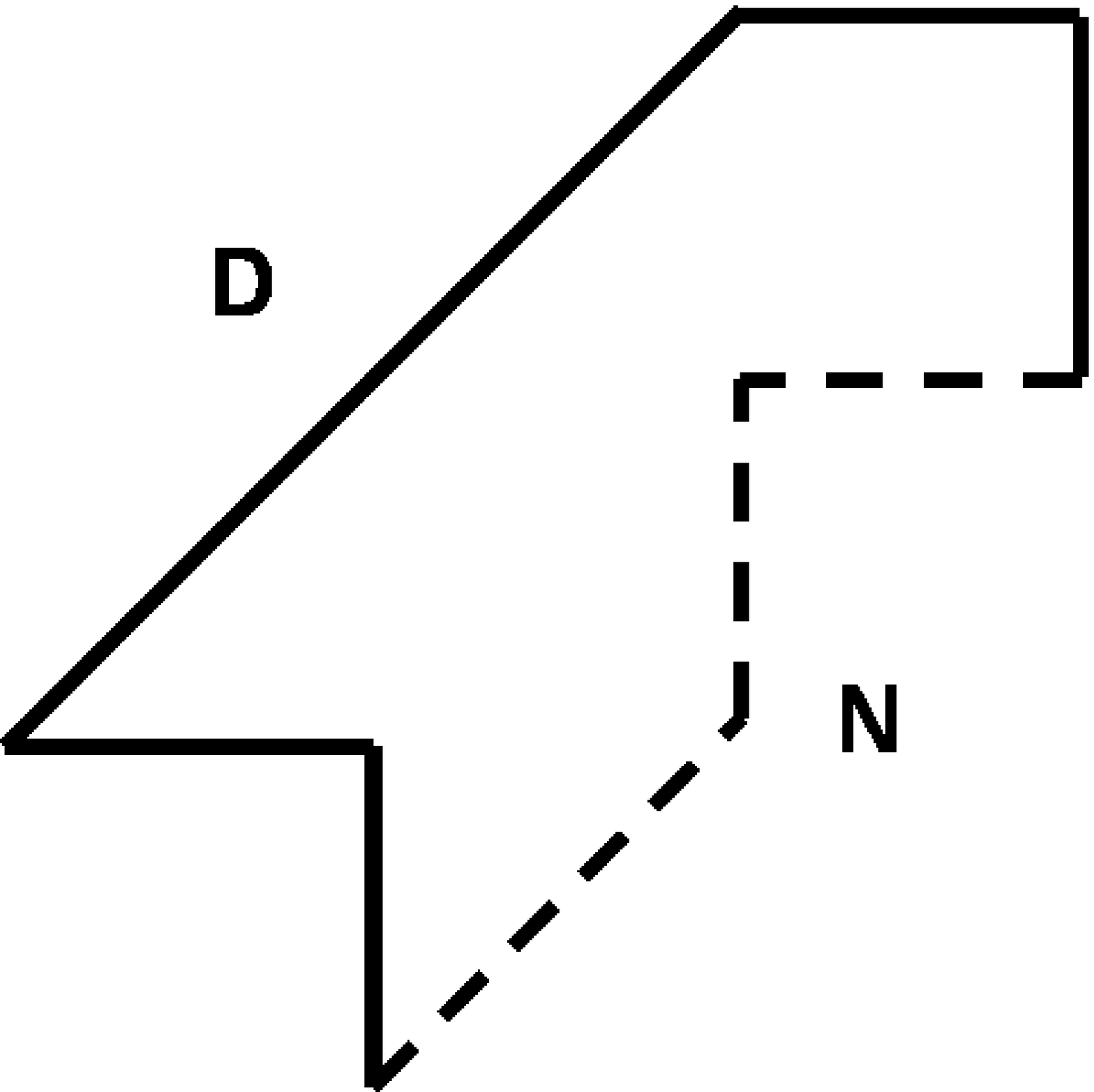}
\par\end{center}

\begin{center}
(a)
\par\end{center}%
\end{minipage}\hfill{}%
\begin{minipage}[t][1\totalheight]{0.4\columnwidth}%
\begin{center}
\includegraphics[width=0.5\textwidth]{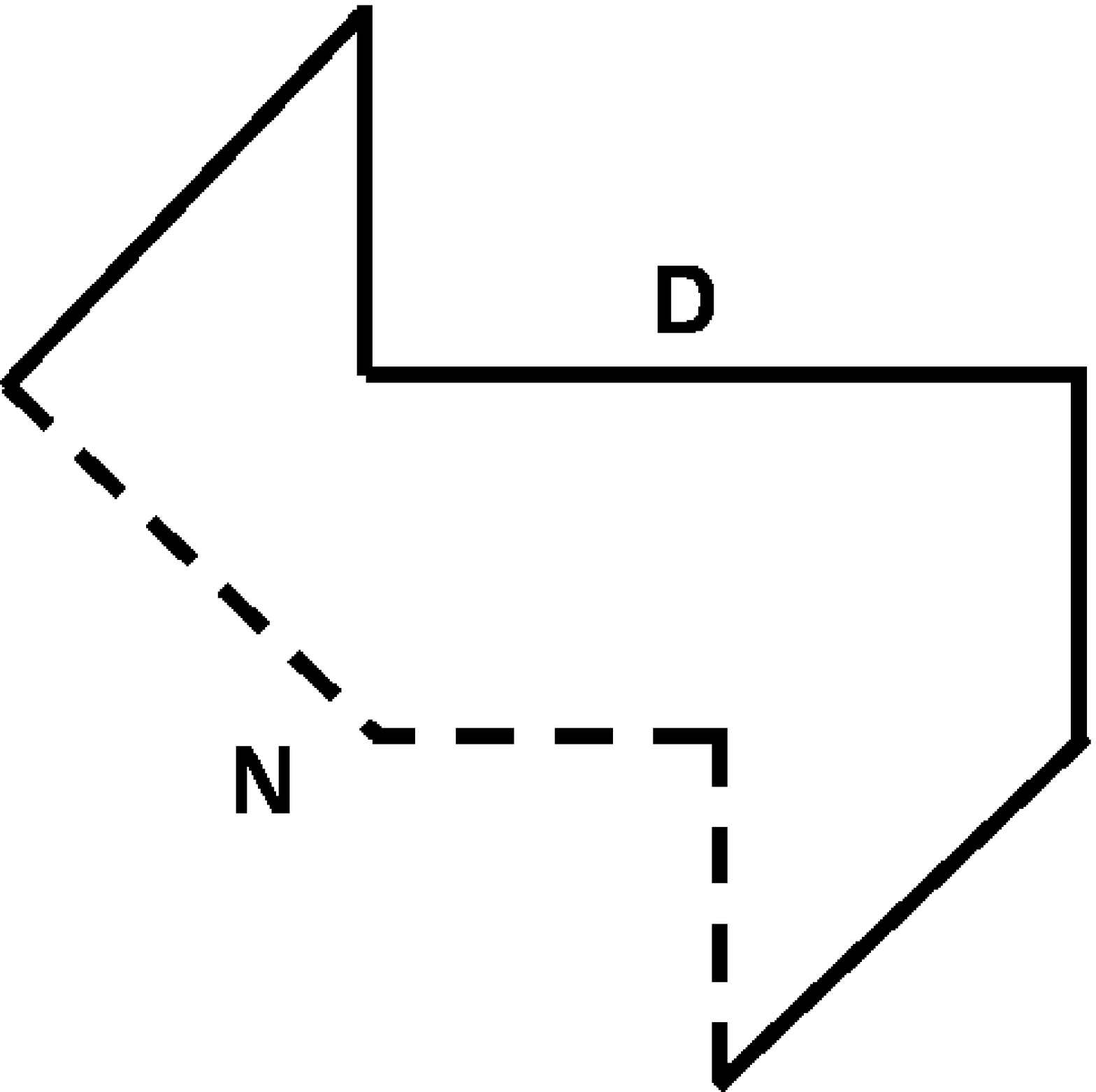}
\par\end{center}

\begin{center}
(b)
\par\end{center}%
\end{minipage}\hfill{}

\caption{The isospectral drums of Gordon et al.\ with new boundary
conditions.} \label{fig:new_Gordon}
\end{figure}

\par\end{center}

\subsection{Chapman's two piece band}
In $D_{4}$ there are no Sunada pairs, i.e., there are no
nonconjugate subgroups $A,B~\leq~D_{4}$ whose trivial
representations satisfy
$\mathrm{Ind}_{A}^{D_{4}}\mathbf{1}_{A}\cong\mathrm{Ind}_{B}^{D_{4}}\mathbf{1}_{B}$.
However, by simple arithmetic we can extract isospectrality even
from these {}``basic ingredients''.

We note that if $\left\{ R_{i}\right\} _{i=1}^{n}$ are
representations of $G$, then the disjoint union $\bigcup\limits
_{i=1}^{n}\nicefrac{\Gamma}{R_{i}}$ is isospectral to
$\nicefrac{\Gamma}{\bigoplus\limits _{i=1}^{n}\!\! R_{i}}$, as we
have $\Phi_{\bigcup\limits
_{i=1}^{n}\nicefrac{\Gamma}{R_{i}}}\left(\lambda\right)=\bigoplus\limits
_{i=1}^{n}\Phi_{\nicefrac{\Gamma}{R_{i}}}\left(\lambda\right)$, so
that \begin{equation*} \sigma_{\bigcup\limits
_{i=1}^{n}\nicefrac{\Gamma}{R_{i}}}\left(\lambda\right)=\sum\limits
_{i=1}^{n}\left\langle
\chi_{R_{i}},\chi_{\Phi_{\Gamma}\left(\lambda\right)}\right\rangle
_{G}=\left\langle \chi_{\bigoplus\limits _{i=1}^{n}\!\!
R_{i}},\chi_{\Phi_{\Gamma}\left(\lambda\right)}\right\rangle
_{G}=\sigma_{\nicefrac{\Gamma}{\bigoplus\limits _{i=1}^{n}\!\!
R_{i}}}\left(\lambda\right) \end{equation*} (this was implicitly
manifested in section \ref{subsec:holy_triangle}). Combining this
with theorem \ref{thm:mainthm}, we obtain the following: if $\left\{
H_{i}\right\} $,$\left\{ H'_{j}\right\} $ are finite sets of
subgroups of $G$, with corresponding representations $\left\{
R_{i}\right\} $, $\left\{ R_{j}'\right\} $, such that
$\bigoplus_{i}\mathrm{Ind}_{H_{i}}^{G}R_{i}\cong\bigoplus_{j}\mathrm{Ind}_{H'_{j}}^{G}R'_{j}$,
then $\bigcup_{i}\nicefrac{\Gamma}{R_{i}}$ is isospectral to
$\bigcup_{j}\nicefrac{\Gamma}{R'_{j}}$.

We recall the subgroups $H_1, H_2, H_3 \leq D_4$ from sections
\ref{sec:basic_example}, \ref{sec:extending_example} and consider
also $H_{4}=\left\{ e,\tau\sigma^{2}\right\} $ and $H_{5}=\left\{
e,\tau\sigma\right\}$. Even though no two inductions among $\left\{
\mathrm{Ind}_{H_{i}}^{D_{4}}\mathbf{1}_{H_{i}}\right\} _{i=1}^{5}$
are isomorphic, we do have that
$\mathrm{Ind}_{H_{1}}^{D_{4}}\mathbf{1}_{H_{1}}\oplus\mathrm{Ind}_{H_{5}}^{D_{4}}\mathbf{1}_{H_{5}}\cong\mathrm{Ind}_{H_{2}}^{D_{4}}\mathbf{1}_{H_{2}}\oplus\mathrm{Ind}_{H_{4}}^{D_{4}}\mathbf{1}_{H_{4}}$.
By the observation above, if $D_{4}$ acts on some object $\Gamma$,
then
$\nicefrac{\Gamma}{\mathbf{1}_{H_{1}}}\cup\nicefrac{\Gamma}{\mathbf{1}_{H_{5}}}$
is isospectral to
$\nicefrac{\Gamma}{\mathbf{1}_{H_{2}}}\cup\nicefrac{\Gamma}{\mathbf{1}_{H_{4}}}$.

\begin{center}
\begin{figure}[h]
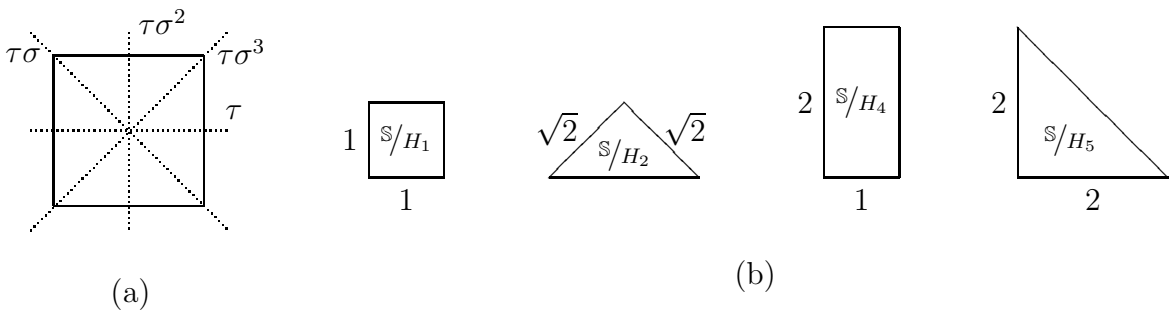

\begin{minipage}[c]{0.25\columnwidth}%
\begin{center}
$\xy(5,5)*{}="A";(5,25)*{}="B";(25,5)*{}="C";(25,25)*{}="D";"A";"B"**\dir{-};"A";"C"**\dir{-};"C";"D"**\dir{-};"B";"D"**\dir{-};(15,2)*{};(15,28)*{}**\dir{.};(2,15)*{};(28,15)*{}**\dir{.};(19.5,29.5)*{\tau\sigma^{2}};(29,17.5)*{\tau};(2,28)*{};(28,2)*{}**\dir{.};(2,2)*{};(28,28)*{}**\dir{.};(1,25)*{\tau\sigma};(30,25.5)*{\tau\sigma^{3}};\endxy$
\par\end{center}

\begin{center}
(a)~
\par\end{center}%
\end{minipage}\hfill{}%
\begin{minipage}[c]{0.7\columnwidth}%
\begin{center}
$\xy(5,5)*{};(15,5)*{}**\dir{-};(15,15)*{};(5,15)*{}**\dir{-};(15,5)*{};(15,15)*{}**\dir{-};(5,15)*{};(5,5)*{}**\dir{-};(2.5,10)*{1};(10,2)*{1};(10,10)*{\nicefrac{\mathbb{S}}{H_{1}}};\endxy$\hfill{}$\xy(5,5)*{};(25,5)*{}**\dir{-};(5,5)*{};(15,15)*{}**\dir{-};(25,5)*{};(15,15)*{}**\dir{-};(6,11.5)*{\sqrt{2}};(23,11.5)*{\sqrt{2}};(15,2)*{\vphantom{2}};(15,8)*{\nicefrac{\mathbb{S}}{H_{2}}};\endxy$\hfill{}$\xy(5,5)*{};(15,5)*{}**\dir{-};(15,25)*{};(5,25)*{}**\dir{-};(15,5)*{};(15,25)*{}**\dir{-};(5,25)*{};(5,5)*{}**\dir{-};(2.5,15)*{2};(10,2)*{1};(10,15)*{\nicefrac{\mathbb{S}}{H_{4}}};\endxy$\hfill{}$\xy(5,5)*{};(25,5)*{}**\dir{-};(25,5)*{};(5,25)*{}**\dir{-};(5,5)*{};(5,25)*{}**\dir{-};(2.5,15)*{2};(15,2)*{2};(12,10)*{\nicefrac{\mathbb{S}}{H_{5}}};\endxy$
\par\end{center}

\begin{center}
(b)
\par\end{center}%
\end{minipage}

\caption{\label{fig:Two-Piece}(a) A square $\mathbb{S}$, on which
$D_{4}$ acts, with the axes of reflection elements marked. (b)
Fundamental domains for the actions of $H_{1}=\left\langle
\tau,\tau\sigma^{2}\right\rangle $, $H_{2}=\left\langle
\tau\sigma,\tau\sigma^{3}\right\rangle $, $H_{4}=\left\langle
\tau\sigma^{2}\right\rangle $, $H_{5}=\left\langle
\tau\sigma\right\rangle $ on $\mathbb{S}$.}

\end{figure}

\par\end{center}

In figure \ref{fig:Two-Piece}, part (a) displays a square
$\mathbb{S}$, of side length 2, on which $D_{4}$ acts, and part (b)
displays fundamental domains for the actions of $D_{4}$'s subgroups
$\left\{ H_{i}\right\} _{i=1,2,4,5}$. If we supply $\mathbb{S}$ with
Neumann boundary conditions, we obtain that $\left\{
\nicefrac{\mathbb{S}}{\mathbf{1}_{H_{i}}}\right\} $ are the domains
shown in figure \ref{fig:Two-Piece}(b), also with Neumann boundary
conditions. The isospectrality of
$\nicefrac{\mathbb{S}}{\mathbf{1}_{H_{1}}}\cup\nicefrac{\mathbb{S}}{\mathbf{1}_{H_{5}}}$
and
$\nicefrac{\mathbb{S}}{\mathbf{1}_{H_{2}}}\cup\nicefrac{\mathbb{S}}{\mathbf{1}_{H_{4}}}$
is a known example constructed by Chapman \cite{Chapman}, which
shows that {}``one cannot hear the shape of a two-piece band''
\cite{Cipra}. Chapman obtains this isospectral example by
manipulating the drums of Gordon et al. $-$ he enlarges the number
of connected components by cutting the basic building block, and
then cancels out identical components in the two shapes. Chapman
shows that these domains are also isospectral if the Neumann
conditions at the boundaries are replaced by Dirichlet ones, and we
would like to establish this as well. A reasonable guess would be to
try $\mathbb{D}$, a square identical to $\mathbb{S}$ but with
Dirichlet boundary conditions. This, however, leads to
isospectrality of domains with mixed boundary conditions (figure
\ref{fig:Two-piece-mixed}).

\begin{center}
\begin{figure}[h]
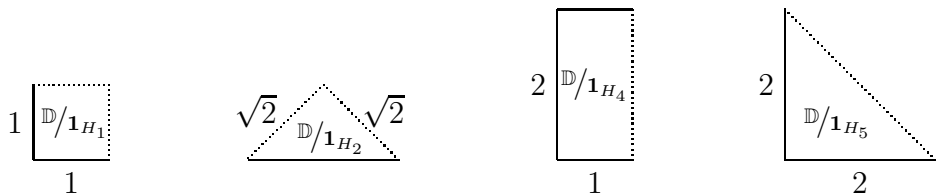

\begin{minipage}[t]{1\columnwidth}%
\noindent \begin{center}
\hfill{}$\xy(5,5)*{};(15,5)*{}**\dir{-};(15,15)*{};(5,15)*{}**\dir{.};(15,5)*{};(15,15)*{}**\dir{.};(5,15)*{};(5,5)*{}**\dir{-};(2.5,10)*{1};(10,2)*{1};(10.5,10)*{\nicefrac{\mathbb{D}}{\mathbf{1}_{H_{1}}}};\endxy$\hfill{}$\xy(5,5)*{};(25,5)*{}**\dir{-};(5,5)*{};(15,15)*{}**\dir{.};(25,5)*{};(15,15)*{}**\dir{.};(6,11.5)*{\sqrt{2}};(23,11.5)*{\sqrt{2}};(15,2)*{\vphantom{2}};(16,8)*{\nicefrac{\mathbb{D}}{\mathbf{1}_{H_{2}}}};\endxy$\hfill{}$\xy(5,5)*{};(15,5)*{}**\dir{-};(15,25)*{};(5,25)*{}**\dir{-};(15,5)*{};(15,25)*{}**\dir{.};(5,25)*{};(5,5)*{}**\dir{-};(2.5,15)*{2};(10,2)*{1};(10,15)*{\nicefrac{\mathbb{D}}{\mathbf{1}_{H_{4}}}};\endxy$\hfill{}$\xy(5,5)*{};(25,5)*{}**\dir{-};(25,5)*{};(5,25)*{}**\dir{.};(5,5)*{};(5,25)*{}**\dir{-};(2.5,15)*{2};(15,2)*{2};(12,10)*{\nicefrac{\mathbb{D}}{\mathbf{1}_{H_{5}}}};\endxy$\hfill{}
\par\end{center}%
\end{minipage}

\centering{}\caption{\label{fig:Two-piece-mixed}Various quotients of
$\mathbb{D}$, the square in figure \ref{fig:Two-Piece}(a) with
Dirichlet boundary (Solid lines stand for Dirichlet boundary
conditions and dashed ones for Neumann).
$\nicefrac{\mathbb{D}}{\mathbf{1}_{H_{1}}}\cup\nicefrac{\mathbb{D}}{\mathbf{1}_{H_{5}}}$
is isospectral to
$\nicefrac{\mathbb{D}}{\mathbf{1}_{H_{2}}}\cup\nicefrac{\mathbb{D}}{\mathbf{1}_{H_{4}}}$.}

\end{figure}

\par\end{center}

In figure \ref{fig:Two-piece-mixed}, the Dirichlet edges are the
remnants of $\mathbb{D}$'s boundary, whereas the Neumann edges are
the traces of the reflections by which the quotients were taken. One
is thus led to consider the following
representations:\begin{eqnarray*} R'_{1} & : & \left\{
\begin{array}{llll}
e\mapsto\left(1\right), & \tau\mapsto\left(-1\right), & \tau\sigma^{2}\mapsto\left(-1\right), & \sigma^{2}\mapsto\left(1\right)\end{array}\right\} \\
R'_{2} & : & \left\{ \begin{array}{llll}
e\mapsto\left(1\right), & \tau\sigma\mapsto\left(-1\right), & \tau\sigma^{3}\mapsto\left(-1\right), & \sigma^{2}\mapsto\left(1\right)\end{array}\right\} \\
R'_{4} & : & \left\{ \begin{array}{ll}
e\mapsto\left(1\right), & \tau\sigma^{2}\mapsto\left(-1\right)\end{array}\right\} \\
R'_{5} & : & \left\{ \begin{array}{ll} e\mapsto\left(1\right), &
\tau\sigma\mapsto\left(-1\right)\end{array}\right\}
\:.\end{eqnarray*} In $\left\{ R'_{i}\right\} $, the reflection
elements are sent to $\left(-1\right)$, so that the corresponding
quotients encode functions which are antisymmetric at the
corresponding axes, hence vanishing along them. The quotients
$\left\{ \nicefrac{\mathbb{D}}{R'_{i}}\right\} _{i=1,2,4,5}$ are
indeed the domains presented in figure \ref{fig:Two-Piece}(b) with
Dirichlet boundary conditions, and upon verifying that
$\mathrm{Ind}_{H_{1}}^{D_{4}}R'_{1}\oplus\mathrm{Ind}_{H_{5}}^{D_{4}}R'_{5}\cong\mathrm{Ind}_{H_{2}}^{D_{4}}R'_{2}\oplus\mathrm{Ind}_{H_{4}}^{D_{4}}R'_{4}$,
we obtain the isospectrality we have sought. Note that by taking
$\left\{ \nicefrac{\mathbb{S}}{R'_{i}}\right\} _{i=1,2,4,5}$, we
would have again obtained the mixed isospectral example in figure
\ref{fig:Two-piece-mixed}, but with the Dirichlet and Neumann
conditions swapped. We end by remarking that as usual, one can
extend the isospectral families above by considering other bases for
the representations
$\mathrm{Ind}_{H_{1}}^{D_{4}}\mathbf{1}_{H_{1}}\oplus\mathrm{Ind}_{H_{5}}^{D_{4}}\mathbf{1}_{H_{5}}$
and
$\mathrm{Ind}_{H_{1}}^{D_{4}}R'_{1}\oplus\mathrm{Ind}_{H_{5}}^{D_{4}}R'_{5}$,
but that in the general case the objects thus obtained will no
longer be planar domains.

\section{Summary and open questions}

\label{sec:summary} This paper describes a method which enables one
to construct isospectral objects, such as quantum graphs and drums.
The algebraic component of the underlying theory suggests theorem
\ref{thm:mainthm} and corollary \ref{cor:sunadapair} as our main
tools for producing isospectral objects. Their assumptions are less
strict than those of Sunada's theorem \cite{Sunada} and this allows
more degrees of freedom in the isospectral search. Another
ingredient is the assembly process of the so called quotient graphs,
whose construction accounts for yet more liberty. We found that for
a graph $\Gamma$ with a symmetry group $G$ and a representation $R$
of the group, we have a variety of choices to make for the
fundamental domain of the action of $G$ on $\Gamma$ and also for the
basis with respect to which $R$ is presented. These different
possibilities yield (possibly) different quotient graphs
$\nicefrac{\Gamma}{R}$, all isospecral to each other. We wish to
offer two perspectives on this dizzying freedom. On the one hand, it
invites us to test the strength of the method. Namely, given two
isospectral objects, can it be decided whether they arise as
quotients of some common object? We touched this question so far
merely by reconstructing some known isospectral objects in terms of
our method. On the other hand, it prompts one to classify these
sources of isospectrality and understand the interrelationships
between them. A fundamental question in this context is whether for
$R$ as above, there exists a choice of basis for $\ind_{H}^{G}R$
which makes $\nicefrac{\Gamma}{\ind_{H}^{G}R}$ the same as
$\nicefrac{\Gamma}{R}$ (rather than just isospectral). We saw a
demonstration of this in section \ref{sec:extending_example} where
$\nicefrac{\Gamma}{R_{1}}$ was the same as
$\nicefrac{\Gamma}{\ind_{H_{1}}^{G}R_{1}}$ with an appropriate
choice of basis. If this is always the case, it means that the
ability to change between different bases is a more fundamental
source of isospectrality. This should not cause one to abandon
theorem \ref{thm:mainthm} and corollary \ref{cor:sunadapair}. Their
role in such a case would be to indicate favorable bases for the
construction. First, they make the practical assembly of the
quotient easier by offering lower-dimensional representations to
divide by. Secondly, the quotient of a manifold by a
multidimensional representation is seldom a manifold, so one is led
to seek one-dimensional representations with isomorphic inductions
(see examples in section \ref{sec:drums}).

Pondering over the quotient graph, which stands at the heart of our
method, we are led to inquire how its various properties are
determined by the construction. Among these are topological ones,
such as the number of connected components and the number of
independent cycles of the graph. Others relate to the nature of the
boundary conditions, which in turn determine the qualities of the
differential operator on the graph. Of specific importance are
conditions which guarantee that the quotient graph has only Neumann
boundary conditions, or alternatively, boundary conditions which
ensure the self-adjointness of its Laplacian. This issue was
addressed in proposition \ref{prop:self-adj} and the example in
section \ref{subsec:holy_triangle}, but still awaits further
investigation.

Quantum graphs are the focus of this paper and they obtain a
thorough treatment. One reason for this, which was already
mentioned, is that under fairly mild conditions, the resulting
quotient object is also a quantum graph. The other reason is that it
is relatively simple to give a rigorous description of their
construction (see section \ref{sec:rigorous}). However, we have
demonstrated in section \ref{sec:drums} that the method is also
applicable to manifolds and drums, and it is desirable to examine
the possibility of obtaining other isospectral objects as well,
e.g., discrete graphs.

Another interesting application would be to relate the construction
method to the spectral trace formula for quantum graphs.
Specifically, we would like to show the equality of the trace
formulae for isospectral graphs just by examining the way in which
they are constructed. This includes a comparison of the total
lengths of the graphs and the lengths of their periodic orbits. A
similar question for isospectral planar domains is discussed in
\cite{Okada}. We propose to treat this issue by returning to the
origin of the spectral trace formula for quantum graphs \cite{KS97,
KS99}, which was developed by describing the boundary conditions in
terms of scattering matrices. Therefore, it may be worthwhile to
work out an isospectral theory, analogous to the one described in
this paper, but stated in terms of scattering matrices. Such an
approach may also pave the way for a similar isospectral theory for
discrete graphs, as a spectral trace formula for them was recently
developed using scattering matrices \cite{Smilansky07}.

We end by returning to Kac's question and asking what can one do
when hearing the shape of a graph (drum) is not possible. One answer
concerns the field of counting nodal domains of the Laplacian's
eigenfunctions. Some new works investigate the ability to resolve
the isospectrality of discrete graphs, quantum graphs and various
manifolds by counting their nodal domains \cite{Oren,
GSS05,BKP07,BOS08}. A specific method of doing so by relating the
nodal count of an isospectral pair to its transplantation was
developed in \cite{BSS06}.  The theory presented in this paper and
the transplantation it yields can perhaps lead to a general method
of isospectrality resolution. It may therefore be further asked
whether one can count whatever cannot be heard.

\ack

\label{sec:acknowledgments}It is an honour to acknowledge U.
Smilansky, who is the initiator of this work and an enthusiastic
promoter of it, and a pleasure to thank Z. Sela for his support and
encouragement. We are gratefull to M. Sieber for sharing with us his
notes, which led to the construction of the isospectral pair of
dihedral graphs. We are indebted to I.~Yaakov whose wise remark has
led us to examine inductions of representations. We thank D. Schüth
for the patient examination of the work and the generous support.
The comments and suggestions offered by J. Brüning, L. Friedlander,
S. Gnutzmann, O. Post, Z. Rudnick, M. Solomyak and T. Sunada are
highly appreciated. The work was supported by the Minerva Center for
non-linear Physics and the Einstein (Minerva) Center at the Weizmann
Institute, by an ISF fellowship, and by grants from the GIF (grant
I-808-228.14/2003) and BSF (grant 2006065). G.B. was financially
supported by D. Jakobson, KKISS, G.Gervais, NSERC, and Weizmann
Science Canada.

\appendix

\section{A short review of required elements of representation theory}

\label{appendix} Let $G$ be a finite group. A $d$-dimensional
\emph{representation} of $G$, denoted by $R$, consists of a vector
space $V^{R}$ of dimension $d$ equipped with an action of $G$, which
is described by a homomorphism $\rho_{R}:G\rightarrow GL(V^{R})$,
i.e., $\forall g_{1},g_{2}\in
G,\quad\rho_{R}(g_{1}g_{2})=\rho_{R}(g_{1})\rho_{R}(g_{2})$. Once a
basis for $V^{R}$ is chosen, one can also think of $\rho_{R}$ as a
homomorphism into $GL_{d}\left(\mathbb{C}\right)$. $\rho_{R}$ is
called the \emph{structure homomorphism} of $R$, and the vector
space $V^{R}$ the \emph{carrier space} of the representation. We
alternatingly use $R$, $\rho_{R}$ and $V^{R}$ when refering to
the representation.\\
 The \emph{character} of a representation $R$ is defined as $\chi_{R}:G\rightarrow\mathbb{C},\quad\chi_{R}(g)=tr(\rho_{R}(g))$.
We will also use the notation $\chi_{V}$ for the character of a
representation $R$ whose carrier space is $V$. There is an inner
product defined on the characters by \[
\langle\chi_{R_{1}},\,\chi_{R_{2}}\rangle_{G}:=\frac{1}{\left|G\right|}\sum_{g\in
G}\chi_{R_{1}}(g)\overline{\chi_{R_{2}}(g)}.\]
 A representation $R$ is called \emph{reducible} if there exists
a nontrivial subspace of the carrier space which is invariant under
the action of the group. Otherwise it is \emph{irreducible}. Up to
isomorphism, any finite group $G$ has a finite number $r$ of
irreducible representations, $\left\{ S_{i}\right\} _{i=1}^{r}$. We
often use $S$ to denote irreducible representations, and $R$ for
general ones. The characters of the irreducible representation obey
orthogonality relations $\forall\: i,j\in\left\{ 1..r\right\}
,\;\;\langle\chi_{S_{i}},\,\chi_{S_{j}}\rangle_{G}=\delta_{i,j}$.
\\
 Two important notions that are used throughout the paper are the
\emph{restriction} and the \emph{induction} of a representation. Let
$H$ be a subgroup of a group $G$. Let $R$ be a representation of
$G$. Then the restriction of $R$ from $G$ to $H$, denoted
$\mathrm{Res}_{H}^{G}R$, is described by
$V^{\mathrm{Res}_{H}^{G}R}=V^{R}$ and
$\rho_{\mathrm{Res}_{H}^{G}R}=\rho_{R}\Big|_{H}$. In particular, the
dimension of $\res_{H}^{G}R$ is equal to that
of $R$. \\
 Now, let $R$ be a representation of $H$. We describe the induction
of $R$ from $H$ to $G$, denoted $\ind_{H}^{G}R$ . We start with a
carrier space $W=V^{R}$ for the representation $R$ of $H$ and
construct a vector space $V=V^{\ind_{H}^{G}R}$ which carries a
representation of $G$. We choose representatives for the left cosets
of $H$ in $G$: $\left\{ t_{i}\right\} _{i=1}^{n}$, where
$n=\left[G:H\right]$. For each $i\in\left\{ 1..n\right\} $, we form
a space denoted $t_{i}W$, which is isomorphic to $W$. The prefix
$t_{i}$ of the elements $t_{i}v\in t_{i}W$ is currently only an
abstract notation without an actual meaning. However, as one may
expect, the vector $t_{i}v$ will obtain, via the described
construction, the meaning of the action of $t_{i}\in G$ on $v\in V$.
The desired carrier space $V$ is defined to be
$V=\bigoplus_{i=1}^{n}t_{i}W$ and we now equip it with an
appropriate action of $G$. Let $v\in V$, $g\in G$. There exist
unique $\left\{ v_{i}\right\} _{i=1}^{n}\subset W$ such that
$v=\sum_{i=1}^{n}t_{i}v_{i}$, and there exist unique $h_{i}\in H$
and $\sigma\left(i\right)\in\left\{ 1..n\right\} $ (which depend on
$g$) such that $gt_{i}=t_{\sigma\left(i\right)}h_{i}$ for
$i\in\left\{ i..n\right\} $. We are motivated by the {}``identity''
$gv=\sum_{i=1}^{n}gt_{i}v_{i}=\sum_{i=1}^{n}t_{\sigma\left(i\right)}h_{i}v_{i}$
to define \begin{equation} g\cdot
v=\sum_{i=1}^{n}t_{\sigma\left(i\right)}\left(h_{i}v_{i}\right).\end{equation}
 Where $h_{i}v_{i}$ is computed according to the representation $R$
from which we started. Note that the action of $g$ permutes the
subspaces $\left\{ t_{i}W\right\} _{i=1}^{n}$ among themselves and
in addition manifests the action of a corresponding $H$ element
(determined by $g$ and $t_{i}$) on each such subspace. It is left
for the reader to check that this indeed defines an action, i.e.,
$\forall g_{1},g_{2}~\in~G,\;\;(g_{1}g_{2})v=g_{1}(g_{2}v)$, and
that the obtained representation does not depend (up to isomorphism)
on the choice of representatives. Also, $V$ contains a subspace
which as a representation of $H$ is isomorphic to $R$, namely, the
$t_{i}W$ which corresponds to the trivial $H$-coset. In contrast to
the preservation of dimension of the restricted representation, we
have $\dim\ind_{H}^{G}R=\dim R\cdot\left[G:H\right]$.

It is known that the character completely identifies a
representation and it is therefore useful to describe the characters
of the restricted and the induced representations. Obviously,
$\chi_{\mathrm{Res}_{H}^{G}R}=\chi_{R}\at_{H}$, and explicit
calculation gives\begin{equation}
\chi_{\ind_{H}^{G}R}(g)=\sum_{i=1}^{n}\chi_{R}(t_{i}^{-1}gt_{i}),\label{eq:char_ind}\end{equation}
 where $\chi_{R}(t_{i}^{-1}gt_{i})=0$ for all $t_{i}^{-1}gt_{i}\notin H$.

Let us demonstrate this by referring to section
\ref{sec:basic_example} and calculating the induction of the
representation $R_{1}$ from $H_{1}$ to $G\cong D_{4}$. We choose the
representatives $\left\{ e,\sigma\right\} $ for the cosets of
$H_{1}$ in $D_{4}$. The character of the induced representation is
therefore \begin{eqnarray*}
\chi_{\ind_{H_1}^{G}R_1}(e) & = & \chi_{R_1}(e)+\chi_{R_1}(e)=2\\
\chi_{\ind_{H_1}^{G}R_1}(\sigma) & = & \chi_{R_1}(\sigma)+\chi_{R_1}(\sigma^{3}\sigma\sigma)=0\\
\chi_{\ind_{H_1}^{G}R_1}(\sigma^{2}) & = & \chi_{R_1}(\sigma^{2})+\chi_{R_1}(\sigma^{3}\sigma^{2}\sigma)=-2\\
\chi_{\ind_{H_1}^{G}R_1}(\tau) & = & \chi_{R_1}(\tau)+\chi_{R_1}(\sigma^{3}\tau\sigma)=0\\
\chi_{\ind_{H_1}^{G}R_1}(\tau\sigma) & = &
\chi_{R_1}(\tau\sigma)+\chi_{R_1}(\sigma^{3}\tau\sigma\sigma)=0\end{eqnarray*}
 In the above we have calculated the character only for chosen representatives
of the conjugacy classes of $G$ (which is obviously enough). We
indeed obtain the character of the two-dimensional irreducible
representation of $D_{4}$ (see for example (\ref{eq:repform1})).

We end by stating a fundamental theorem, known as \emph{Frobenius
reciprocity} (\cite{Curtis}, section 10A): Let $G$ be a group with a
representation $R_{1}$. Let $H$ be a subgroup of $G$ and $R_{2}$ a
representation of $H$. Then
\[
\langle\chi_{\res_{H}^{G}R_{1}},\chi_{R_{2}}\rangle_{H}=\langle\chi_{R_{1}},\chi_{\ind_{H}^{G}R_{2}}\rangle_{G}.\]
 This is the main element that stands behind the proof of theorem
\ref{thm:mainthm}.

\end{document}